%% file: ms.tex
\documentclass[manuscript]{acmart}
\usepackage{subcaption}
\usepackage{graphicx}
\usepackage[export]{adjustbox}

\usepackage{pifont}
\usepackage{soul} 
\usepackage{amsmath}
\usepackage{graphicx}
\usepackage[linesnumbered,ruled]{algorithm2e}
 \newcommand{\namePaper}{NERO\xspace} 
\usepackage{algpseudocode}
\usepackage{titlesec}
\usepackage{lscape}
\usepackage{xcolor}
\usepackage{textcomp}
\usepackage{tikz}
\usetikzlibrary{chains,shapes,matrix,backgrounds,positioning}
\newtoggle{quickdecim}
\usetikzlibrary{decorations}
\usetikzlibrary{decorations.pathreplacing}
\usetikzlibrary{calc}
\usetikzlibrary{arrows}
\newcommand\MemoryLayout[1]{
 \begin{scope}
     \draw[thick](0,0)--++(0,3)node[above]{$0$};
     \foreach \pt/\col/\lab [remember=\pt as \tp (initially 0)] in {#1} {
       \foreach \a in {\tp,...,\pt-1} {
          \draw[fill=\col](-\a,0) rectangle ++(-1,2);
       }
       \draw[thick](-\pt,0)--++(0,3)node[above]{$\pt$};
       \if\lab\relax\relax\else
         \draw[thick,decorate, decoration={brace,amplitude=4mm}]
            (-\tp,-0.2)--node[below=4mm]{\lab} (-\pt,-0.2);
       \fi
     }
       \end{scope}
}

\newcommand*\circled[1]{\tikz[baseline=(char.base)]{
            \node[shape=circle,draw,inner sep=0pt,fill=black, text=white] (char) {#1};}}
\newcommand*\circledWhite[1]{\tikz[baseline=(char.base)]{
            \node[shape=circle,draw,inner sep=0pt,fill=white, text=black] (char) {#1};}}

\include{pythonlisting}
\usepackage{array}
\usepackage{xcolor}
 \makeatletter
 \newcommand*{\@rowstyle}{}
\newcommand*{\rowstyle}[1]{
 \gdef\@rowstyle{#1}%
 \@rowstyle\ignorespaces%
}
\newcolumntype{=}{
>{\gdef\@rowstyle{}}%
}
\newcolumntype{+}{
>{\@rowstyle}%
}

 \newcolumntype{C}[1]{>{\centering\arraybackslash}p{#1}}
 \makeatother

\newcommand{\etal}{\textit{et al.}}
\newcommand{\myGlobalTransformation}[2]
{
    \pgftransformcm{1}{0}{0.4}{0.5}{\pgfpoint{#1cm}{#2cm}}
}

\newcommand{\gridThreeD}[3]
{
    \begin{scope}
        \myGlobalTransformation{#1}{#2};
        \draw [#3] grid (5,5);
    \end{scope}
}

\newcommand{\gridThreeDSecond}[3]
{
    \begin{scope}
        \myGlobalTransformation{#1}{#2};
        \draw [#3] grid (3,3);
    \end{scope}
}

\newcommand{\gridThreeDThird}[3]
{
    \begin{scope}
        \myGlobalTransformation{#1}{#2};
        \draw [#3] grid (1,1);
    \end{scope}
}

\tikzstyle myBG=[line width=3pt,opacity=1.0]

\newcommand{\dotesUpper}[2]
{
    \begin{scope}
        \myGlobalTransformation{#1}{#2};
   
     \node at (2.5,0.5) [circle,fill=black] {};
     \node at (1.5,1.5) [circle,fill=black] {};
    \node at (3.5,1.5) [circle,fill=black] {};

     \node at (0.5,2.5) [circle,fill=black] {};
      \node at (2.5,2.5) [circle,fill=black] {};
      \node at (4.5,2.5) [circle,fill=black] {};
      \node at (1.5,3.5) [circle,fill=black] {};
         \node at (3.5,3.5) [circle,fill=black] {};
           \node at (2.5,4.5) [circle,fill=black] {};

    \end{scope}
}

\newcommand{\lowerDots}[2]
{
    \begin{scope}
        \myGlobalTransformation{#1}{#2};
 
     \node at (0.5,1.5) [circle,fill=black] {};
       \node at (-1.5,1.5) [circle,fill=black] {};
          \node at (-0.5,0.5) [circle,fill=black] {};
           \node at (-0.5,2.5) [circle,fill=black] {};

    \end{scope}
}

\newcommand{\lowestDot}[2]
{
    \begin{scope}
        \myGlobalTransformation{#1}{#2};
 
    \node at (0.5,1.5) [circle,fill=black] {};

    \end{scope}
}
\definecolor{dred}{rgb}{0.75,0.00, 0.00}
\definecolor{dpink}{rgb}{0.75,0.00, 0.75}
\definecolor{ddpink}{rgb}{1.0, 0.20, 1.0}
\definecolor{dgreen}{rgb}{0.0, 0.50, 0.25}
\definecolor{dblack}{rgb}{0.00, 0.0, 0.00}
\definecolor{dblue}{rgb}{0.00, 0.00, 0.75}
\definecolor{feedb}{rgb}{0.75, 0.00, 0.75}

\newcommand{\vadvc}{\texttt{vadvc}\xspace} 
\newcommand{\hdiff}{\texttt{hdiff}\xspace} 
\newcommand{\cop}{\texttt{copy}\xspace} 

\newcommand{\gagan}[1]{{\color{dblack}#1}}
\newcommand{\gagann}[1]{{\color{dblack}#1}}
\newcommand{\gagannn}[1]{{\color{dblack}#1}}
\newcommand{\gonur}[1]{{\color{dblack}#1}}
\newcommand{\garxiv}[1]{{\color{dblack}#1}}
\newcommand{\sr}[1]{{\color{dblack}#1}}
\newcommand{\dd}[1]{{\color{dblack}#1}}

\newcommand{\gtrets}[1]{{\color{black}#1}}

\newcommand{\juan}[1]{{\color{dblack}#1}}
\newcommand{\juang}[1]{{\color{dblack}#1}}
\newcommand{\juangg}[1]{{\color{dblack}#1}}
\newcommand{\juanggg}[1]{{\color{dblack}#1}}

\newcommand{\juann}[1]{{\color{black}#1}}

\newcommand{\grev}[1]{{\color{black}#1}}
\newcommand{\gcamera}[1]{{\color{black}#1}}
\newcommand{\gcameraa}[1]{{\color{black}#1}}

\newcommand{\om}[1]{{\color{black}#1}}

\AtBeginDocument{%
  \providecommand\BibTeX{{%
    \normalfont B\kern-0.5em{\scshape i\kern-0.25em b}\kern-0.8em\TeX}}}

\begin{document}

\title{\namePaper: A Near High-Bandwidth Memory Stencil Accelerator for \\ Weather Prediction Modeling }
\title{\namePaper: Near-Memory FPGA-Based Acceleration of \\Weather Prediction Application}
\title{\namePaper: Near-Memory Weather Modeling on FPGAs}
\title{\namePaper: Near-Memory FPGA-Based Acceleration of \\Weather Prediction Application}
\title{\namePaper: Solving Weather Prediction Partial Differential Equations Using a Near-Memory Reconfigurable Fabric}
\title{\namePaper: Solving Complex Weather Prediction Stencils Using a Near-Memory Reconfigurable Fabric}
\title{\fontsize{16}{15}\selectfont Accelerating Weather Prediction using\\ Near-Memory Reconfigurable Fabric}
\title{\fontsize{16}{15}\selectfont Accelerating Weather Prediction using Near-Memory Reconfigurable Fabric}
\rtitle{Accelerating Weather Prediction using\\ Near-Memory Reconfigurable Fabric}
\author{Gagandeep Singh}
\email{gagan.gagandeepsingh@safari.ethz.ch}
\affiliation{%
  \institution{ETH Zürich}
   \country{Switzerland}
}

\author{Dionysios Diamantopoulos}
\email{did@zurich.ibm.com}
\affiliation{%
  \institution{IBM Research Europe, Zürich Lab}
  \country{Switzerland}}
\author{Juan G{\'o}mez-Luna}
 \email{juan.gomez@safari.ethz.ch}
\affiliation{%
 \institution{ETH Zürich}
 \country{Switzerland}
}

\author{Christoph Hagleitner}
  \email{hle@zurich.ibm.com}
\affiliation{%
  \institution{IBM Research Europe, Zürich Lab}
  \country{Switzerland}}

\author{Sander Stuijk}
 \email{s.stuijk@tue.nl}
\affiliation{%
  \institution{Eindhoven Univesity of Technology}
  \country{The Netherlands}
 }

\author{Henk Corporaal}
\email{h.corporaal@tue.nl}
\affiliation{%
  \institution{Eindhoven Univesity of Technology}
  \country{The Netherlands}
}

\author{Onur Mutlu}
\email{omutlu@ethz.ch}
\affiliation{%
  \institution{ETH Zürich}
  \country{Switzerland}
}


\renewcommand{\shortauthors}{Singh, \emph{et al.}}


\input{sections/abstract}



\maketitle



\input{sections/introduction}
\input{sections/background}
\input{sections/accelerator.tex}

\input{sections/results}

\input{sections/takeaways}

\input{sections/relatedWork}

\input{sections/conclusion}
\section*{Acknowledgments} 
\label{sec:acknowledgment}
This work was performed in the framework of the Horizon 2020 program for the project ``Near-Memory Computing (NeMeCo)''. It is funded by the European Commission under Marie Sklodowska-Curie Innovative Training Networks European Industrial Doctorate (Project ID: 676240). Special thanks to Florian Auernhammer and Raphael Polig for providing support with the IBM systems. We appreciate valuable discussions with Kaan Kara and Ronald Luijten. \gagannn{We would also like to thank Bruno Mesnet and Alexandre Castellane from IBM France for help with the SNAP and OC-Accel framework.} This work was partially supported by the H2020 research and innovation programme under grant agreement No 732631, project OPRECOMP.
\juang{We also thank {Google, Huawei, Intel, Microsoft, SRC, and VMware} for their funding support \om{to the SAFARI Research Group}.}
\bibliographystyle{ACM-Reference-Format}
\bibliography{ms}
\end{document}

%% file: sections/abstract.tex
\begin{abstract}
Ongoing climate change calls for \gagan{fast and accurate} 
weather and climate \gagan{modeling}. However, \gagan{when solving large-scale \gagann{weather prediction}
simulations,} state-of-the-art CPU and GPU implementations suffer from limited performance and high energy consumption. \juan{These implementations are \gagan{dominated by} complex irregular memory access patterns and low arithmetic intensity \gagan{that} pose fundamental challenge\gagannn{s} to acceleration}. To overcome these challenges, we propose \gagan{and evaluate} the use of near-memory acceleration using a reconfigurable fabric with high-bandwidth memory (HBM). \gagan{We focus on} \juan{compound stencils \gagan{that} are fundamental kernels} in weather prediction models. \gagan{By using} high-level synthesis techniques, \gagan{we} develop \juang{\namePaper,} an \juang{FPGA+HBM-based} accelerator connected through OCAPI (Open Coherent Accelerator Processor Interface) to an IBM POWER9 host system. \gagan{Our} experimental results show that \juang{\namePaper} outperforms a 16-core POWER9 \gagann{system} by $5.3\times$ and $12.7\times$  when running two \juanggg{different} compound stencil kernels. \juang{\namePaper} \gagann{reduces the energy consumption by} $12\times$ and $35\times$ \juanggg{for the same two kernels \gagannn{over the POWER9 system}} \gagann{with an energy efficiency of 1.61 GFLOPS/Watt and 21.01 GFLOPS/Watt}. \gagan{We conclude that} employing near-memory acceleration solutions for weather prediction \gagan{modeling}~\gagan{is~promising} \juanggg{as a means to achieve \gagannn{both} high performance and \gagannn{high} energy efficiency}.
        
\end{abstract}

%% file: sections/introduction.tex
\section{Introduction} 
{Accurate weather prediction \om{and climate modeling} using detailed weather models is essential to \gagan{make} weather-dependent \om{and climate-related} decisions in a timely manner.}
These models are based on physical laws that describe various components of the atmosphere~\cite{schar2020kilometer}. 
The Consortium for Small-Scale Modeling (COSMO)~\cite{doms1999nonhydrostatic} 
\juang{built} one such weather model 
to meet the high-resolution forecasting requirements of weather services. The COSMO model is a non-hydrostatic atmospheric prediction model currently being used by a dozen nations for meteorological purposes and research applications. 

The central part of the COSMO model (\juan{called \emph{dynamical core} or \emph{dycore}}) solves the Euler equations on a curvilinear grid and applies implicit discretization in the vertical dimension \gagann{(i.e., parameters are dependent on each other at the same time instance~\cite{bonaventura2000semi})}  and 
explicit discretization in the horizontal dimension \gagan{(i.e., 
\juanggg{a solution is dependent on the previous system state}~\cite{bonaventura2000semi})}. The use of different discretizations leads to three computational patterns~\cite{cosmo_knl}: \juang{1)} horizontal stencils, \juang{2)} tridiagonal solvers in the vertical dimension, and \juang{3)} point-wise computation. These computational kernels are compound stencil kernels that operate on a three-dimensional grid~\cite{gysi2015modesto}.  
\emph{Vertical advection} (\gagan{\texttt{vadvc}}) and  \emph{horizontal diffusion} (\gagan{\texttt{hdiff}})  are such compound kernels found in the \emph{dycore} of the COSMO \gagannn{weather prediction} model. {These kernels} are representative \gagan{of} the data access {patterns and algorithmic} complexity of the entire COSMO model. \gagan{They} are similar to the kernels used in other weather and climate models~\cite{kehler2016high,neale2010description,doi:10.1175/WAF-D-17-0097.1}.  Their performance is dominated by memory-bound operations with unique irregular memory access patterns 
\juang{and} low arithmetic intensity that often results in $<$10\% sustained floating-point performance on current CPU-based systems~\cite{chris}. 

\begin{figure}[h]
  \centering
  \includegraphics[width=0.75\textwidth,trim={1cm 0.8cm 1.5cm 0.6cm},clip]{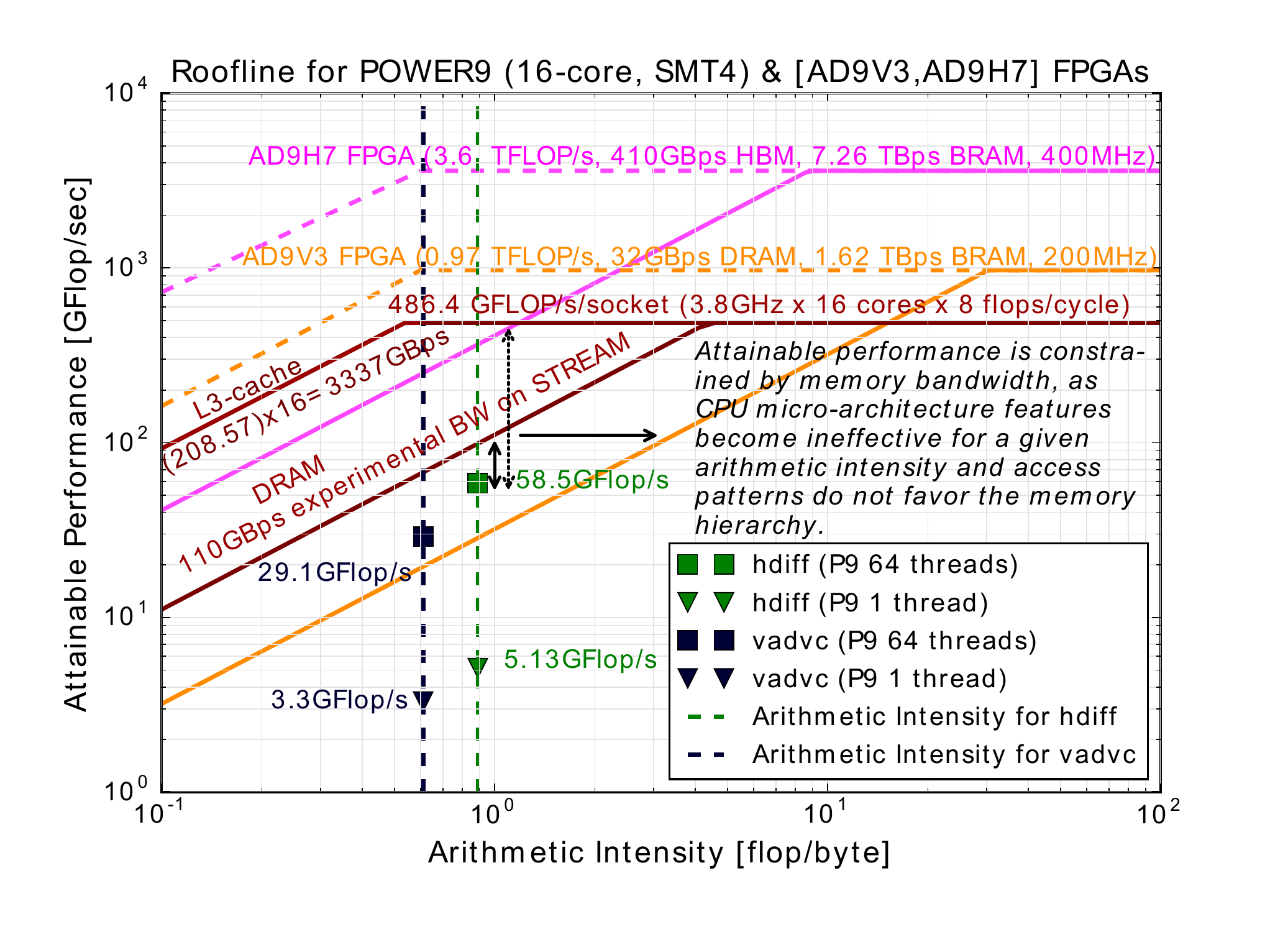}
    \caption{Roofline~\cite{williams2009roofline} for POWER9 (1-socket) showing vertical advection {(\texttt{vadvc})} and horizontal diffusion {(\texttt{hdiff})} kernels for single-thread and 64-thread 
    implementations. 
    \juang{The plot shows also the rooflines} 
    of the FPGAs used in our work with peak DRAM and on-chip BRAM bandwidth.}
 \label{fig:roofline}
\end{figure}

Figure~\ref{fig:roofline} shows the roofline plot\gagan{~\cite{williams2009roofline}} for {an} IBM 16-core POWER9 CPU (IC922).\footnote{IBM and POWER9 are registered trademarks or common law marks of International Business Machines Corp., registered in many jurisdictions worldwide. Other product and service names might be trademarks of IBM or other companies.} 
After optimizing the \texttt{vadvc} and \texttt{hdiff} kernels for the POWER architecture\footnote{\grev{We use single instruction, multiple data (SIMD)~\cite{flynn1966very,illiac_simd_1968}, and simultaneous multithreading (SMT)~\cite{smt} techniques to fill the hardware pipelines. We use the same tiling size for both CPU and FPGA-based designs. While compiling these kernels, we use the IBM XLC~\cite{ibmxlc} 16 C/C++ compiler that is optimized for IBM POWER~\cite{POWER9} machines with the following flags: qarch=pwr9, qtune=pwr9, O3, q64,  qprefetch=aggressive, qsmp=omp, and qsimd=auto.}} by following the approach in~\cite{xu2018performance}, 
\juan{they} achieve {29.1}~GFLOP/s and 58.5~GFLOP/s, \juan{respectively}, \juang{for 64 threads}.  {Our roofline analysis 
\juang{indicates} that these kernels are constrained by the host DRAM bandwidth.} 
\juan{Their} low arithmetic intensity limits 
\juang{their performance, which is one order of magnitude smaller than the peak performance,} and results in a fundamental memory bottleneck that 
standard CPU-based optimization techniques \juan{cannot overcome}.

\gtrets{{Heterogeneous computing} has emerged as an answer to improve the system performance in an energy-efficient way. Heterogeneous computing entails complementing processing elements with different compute capabilities, each to perform the tasks to which it is best suited. In the HPC domain, coupling specialized compute units with general-purpose cores can meet the high-performance computing demands with the ability to realize exascale systems  needed to process data-intensive workloads~\cite{nair2015active}. The graphics processing unit (GPU) is one of the most popular acceleration platforms. GPUs have been used to accelerate workloads like computer graphics and linear algebra~\cite{volkov2008benchmarking} because of their many-core architecture. However, GPUs are power-hungry due to high transistor density and, depending on the power constraints, may not always be the ideal platform for implementation. Recently, the use of field-programmable gate array (FPGA) in accelerating machine learning workloads with high energy efficiency has inspired researchers to explore the use of FPGAs instead of GPUs for various high-performance computing applications~\cite{caulfield2016cloud,duarte2018fast}.  FPGAs provide a unique combination of flexibility and performance without the cost, complexity, and risk of developing custom application-specific integrated circuits (ASICs). } Modern FPGAs show four key trends. 
\begin{itemize}
    \item The advancements in the stacking technology with  high-bandwidth memory ({HBM})~\cite{hbm,xilinx_utlra,hbm_specs,intel_altera,lee2016smla} blends DRAM on the same package as an FPGA. This integration allows us to implement our accelerator logic in close proximity to the memory with a lower latency and much higher memory bandwidth 
    than the traditional DDR4-based FPGA boards. Memory-bound applications on FPGAs are limited by \juang{the relatively} low DDR4 bandwidth (72 GB/s 
\gagan{for four independent dual-rank DIMM interfaces}\gagann{~\cite{VCU}}). 
HBM-based FPGAs can overcome this limitation with
 a  peak bandwidth of 410~GB/s~\cite{kara2020hbm}. 
   
    \item New cache-coherent interconnects, such as  Open Coherent Accelerator Processor Interface (OCAPI)~\cite{openCAPI}, Cache Coherent Interconnect for Accelerators (CCIX)~\cite{benton2017ccix}, and Compute Express Link (CXL)~\cite{sharma2019compute},  allow tight integration of FPGAs with CPUs at high bidirectional bandwidth {({on} the order of tens of GB/s)}. This integration {reduces programming effort and} enables us to coherently access the host system's memory through a pointer rather than having multiple copies of the data.   
     \item The introduction of UltraRAM (URAM)~\cite{uram} along with the BlockRAM (BRAM) that offers massive \emph{scratchpad}-based on-chip memory next to the logic. \gtrets{URAM is more denser than BRAM, but is not as distributed in the FPGA layout as the BRAM.} 
    \item FPGAs are being manufactured with an advanced technology node of 7-14nm FinFET technology~\cite{gaide2019xilinx} that offers higher performance. 
 
\end{itemize}

These above trends suggest that {modern FPGA architectures with near-memory compute capabilities can alleviate the \emph{memory bottleneck} of real-world data-intensive applications~\cite{singh2021fpga}.  
However, a study of their advantages for 
real-world memory-bound applications is still missing. In this work, our goal is to overcome the memory} bottleneck {of weather prediction} kernels by {exploiting near-memory {computation} capability on} FPGA accelerators with high-bandwidth memory (HBM)~\cite{6757501,hbm,lee2016smla} {that are attached} {to the host CPU}. Figure~\ref{fig:roofline} shows the roofline model{s} of the two FPGA cards {(AD9V3~{\cite{ad9v3}} and AD9H7~{\cite{ad9h7}})} used in this work. 
FPGAs \juang{can} \gagan{handle} 
irregular memory access patterns \juang{efficiently} and 
offer significantly 
\juang{higher} memory bandwidth \juang{than the host CPU with} 
their on-chip URAMs \gagan{(UltraRAM)}, BRAMs \gagan{(block RAM)}, and \gcamera{on-package} HBM (\gagan{high-bandwidth memory} for the AD9H7 card). 
However, taking full advantage of 
FPGAs for accelerating a workload is not a trivial task. To compensate \gagan{for} the higher clock frequency of the \gagan{baseline} CPUs, 
\gagan{our} FPGAs must exploit at least one order of magnitude more parallelism in a target workload. 
\juang{This is challenging, as it requires sufficient FPGA programming skills to \gagan{map the workload and} optimize the design for the FPGA microarchitecture.
}

\juan{We aim} to answer the following \juan{research question}: \textbf{Can FPGA-based \gagan{accelerators} with HBM \gagan{mitigate} the performance bottleneck of \juang{memory-bound} compound weather \gagan{prediction} kernels 
in an energy-efficient way?} As an answer to this question\gagann{,} we present \namePaper, a \underline{ne}ar-HBM accelerator for weathe\underline{r} predicti\underline{o}n. 
\juan{We design and implement} \namePaper~
\juan{on an FPGA with HBM to} optimize two 
kernels (vertical advection and horizontal diffusion), 
\juan{which notably} represent 
\juan{the} spectrum of {computational diversity} found in the COSMO \gagann{weather prediction} application. 
\juang{We co-design a hardware-software framework and provide an optimized API to interface efficiently with the rest of the COSMO model, which runs on the CPU}.
Our \gagan{FPGA-based} solution \gagan{for \texttt{hdiff} and \texttt{vadvc}} leads to 
\gagan{performance improvements of $5.3\times$ and $12.7\times$ and}
\juang{energy reductions} of $12\times$ and $35\times$, \juanggg{respectively,} \juang{with respect to optimized CPU implementations~\cite{xu2018performance}}. 

\gagan{T}he major contributions of \namePaper are \juang{as follows}:
\begin{itemize}
\item \juang{We perform \gagan{a detailed} roofline analysis to show that representative weather \gagan{prediction} kernels are constrained by memory bandwidth on \gagannn{state-of-the-art} CPU systems.}

\item {We propose \namePaper, \gagan{the first} near-HBM FPGA-based accelerator} for representative kernels from a real-world weather prediction application.

\item \juang{We optimize \namePaper~with} a data-centric caching scheme with precision-optimized tiling for a heterogeneous memory hierarchy \gagan{(consisting of URAM, BRAM, and HBM)}.

\item \juan{We evaluate the performance and energy consumption of our accelerator and perform a scalability analysis.} 
We show that an FPGA+HBM-based design 
outperforms a complete 16-core POWER9 \gagann{system}~(
\juang{running} 64~threads)~by~$5.3\times$ for 
\juang{the} {vertical advection} \juanggg{(\texttt{vadvc})} and $12.7\times$ for 
\juang{the} {horizontal diffusion} \juanggg{(\texttt{hdiff})} kernels with energy \gagann{reductions of $12\times$ and $35\times$, respectively. Our design provides an energy efficiency of 1.61 GLOPS/Watt and 21.01 GFLOPS/Watt for \texttt{vadvc} and \texttt{hdiff} kernels,} \juanggg{respectively}.

\end{itemize}

\gcamera{This work extends our previous work~\cite{singh2020nero} as follows. First, we add new results using a state-of-the-art OpenCAPI (OCAPI) interface~\om{\cite{openCAPI}}. OCAPI provides two key opportunities compared to our previous CAPI2 implementation: (1) OCAPI has double the bitwidth (1024-bit) of our previously used CAPI2 interface, and (2) the memory coherency logic has been moved to the host CPU side, which provides more area and allows us to run our design at a higher frequency. Our implementation and evaluation with state-of-the-art OCAPI improves the performance for our two main workloads from weather modeling (vadvc and hdiff) by 37\% and 44\%, respectively, compared to a CAPI-based  HBM  design~\cite{singh2020nero}. All the functions in our design now operate on a 1024-bit \gcameraa{per clock} rather than 512-bit \gcameraa{per clock, utilizing the maximum processing throughput of OCAPI  (POWER9 cache line is 1024-bit). Thus, we can build a dataflow accelerator with wide AXI  streams of 1024-bit.} Second, we develop \om{\texttt{HBM\_multi+OCAPI}}-based versions of our workloads that make use of multiple channels per processing element (PE). This multi-channel implementation allows the PEs to exploit significantly higher bandwidth. As a result, the workloads achieve an average speedup of 1.5x over the single-channel version for a single PE. 
Third, we provide a discussion section (Section~\ref{sec:discussion}) that provides various insights and takeaways while designing HBM-based FPGA accelerators, which we believe would be useful for future FPGA architects \om{and programmers}. Fourth, we implement and evaluate \om{the \cop} stencil~\cite{cosmo_knl} \om{(Figure~\ref{fig:copy_stencil})}, a stencil from the COSMO model to benchmark the peak performance on a platform. Fifth, we use the OC-Accel framework\footnote{https://github.com/OpenCAPI/oc-accel} instead of the SNAP framework for OCAPI accelerator development and deployment. Sixth, we provide a comparison of state-of-the-art works in stencil acceleration (Table~\ref{tab:compare_works}).

}

%% file: sections/background.tex
\section{Background} 
\label{sec:background}
\juang{In this section, we first provide} 
an overview of the \juang{\texttt{vadvc} and \texttt{hdiff}} {compound stencils, 
\juang{which represent} a large fraction of the overall computational load of the COSMO \gagannn{weather prediction} model}. 
\gtrets{Second, we introduce the OC-Accel \gagan{(OpenCAPI Acceleration)} framework} that we use to connect our \namePaper~accelerator to an IBM POWER9~system.

\subsection{\juangg{Representative} COSMO Stencils}
A stencil operation updates values in a structured multidimensional grid based on the values of a fixed local neighborhood of grid points. \gtrets{ In weather and climate modeling, several stencil operations are compounded together that operate on multiple input elements to generate one or more output results.}
Vertical advection (\texttt{vadvc}) and horizontal diffusion (\texttt{hdiff}) from the COSMO model are two such compound \juang{stencil} kernels, which represent the typical code patterns found in the \emph{dycore} of COSMO.  
Algorithm~\ref{algo:hdiffKernel} shows \gagan{the} pseudo-code for \texttt{vadvc} and \texttt{hdiff}  kernels. 
The horizontal diffusion kernel iterates over a 3D grid, performing \textit{Laplacian} and \textit{flux} to calculate different grid points, as shown in Figure~\ref{fig:hdiffMemory}.
\gtrets{A single \textit{Laplacian} stencil accesses
the input grid at five memory offsets, the result of which is used to calculate the \textit{flux} \gonur{stencil}. \hdiff has purely horizontal access patterns and does not have dependencies in the vertical dimension. {Thus,} it can be fully parallelized in the vertical dimension. Figure~\ref{fig:hdiffMemory_layout} shows the memory layout for the horizontal diffusion kernel. We observe that the indirect memory accesses of the input grid domain can {severely} impact cache efficiency on our current CPU-based systems.}

\input{images/hdiff}
\gtrets{Vertical advection has a higher degree of complexity since it uses the Thomas algorithm~\cite{thomas} to solve a tridiagonal matrix of \gonur{weather data (called \emph{fields}, such as air pressure, wind velocity, and temperature)} along the vertical axis. \vadvc consists of a forward sweep that is followed by a backward sweep along the vertical dimension. \vadvc requires access to \gonur{the weather data,} which are stored as array structures while performing forward and sweep computations.
Unlike the conventional stencil kernels, vertical advection has dependencies in the vertical direction, which leads to limited available parallelism and irregular memory access patterns.  For example, when the input grid is stored by \emph{row}, accessing data elements in the \emph{depth} dimension typically results in {many} cache {misses}~\cite{xu2018performance}. }

\input{algorithms/cosmo.tex}

\begin{figure}[h]
  \includegraphics[width=0.45\linewidth,trim={0cm 0cm 0cm 0cm},clip]{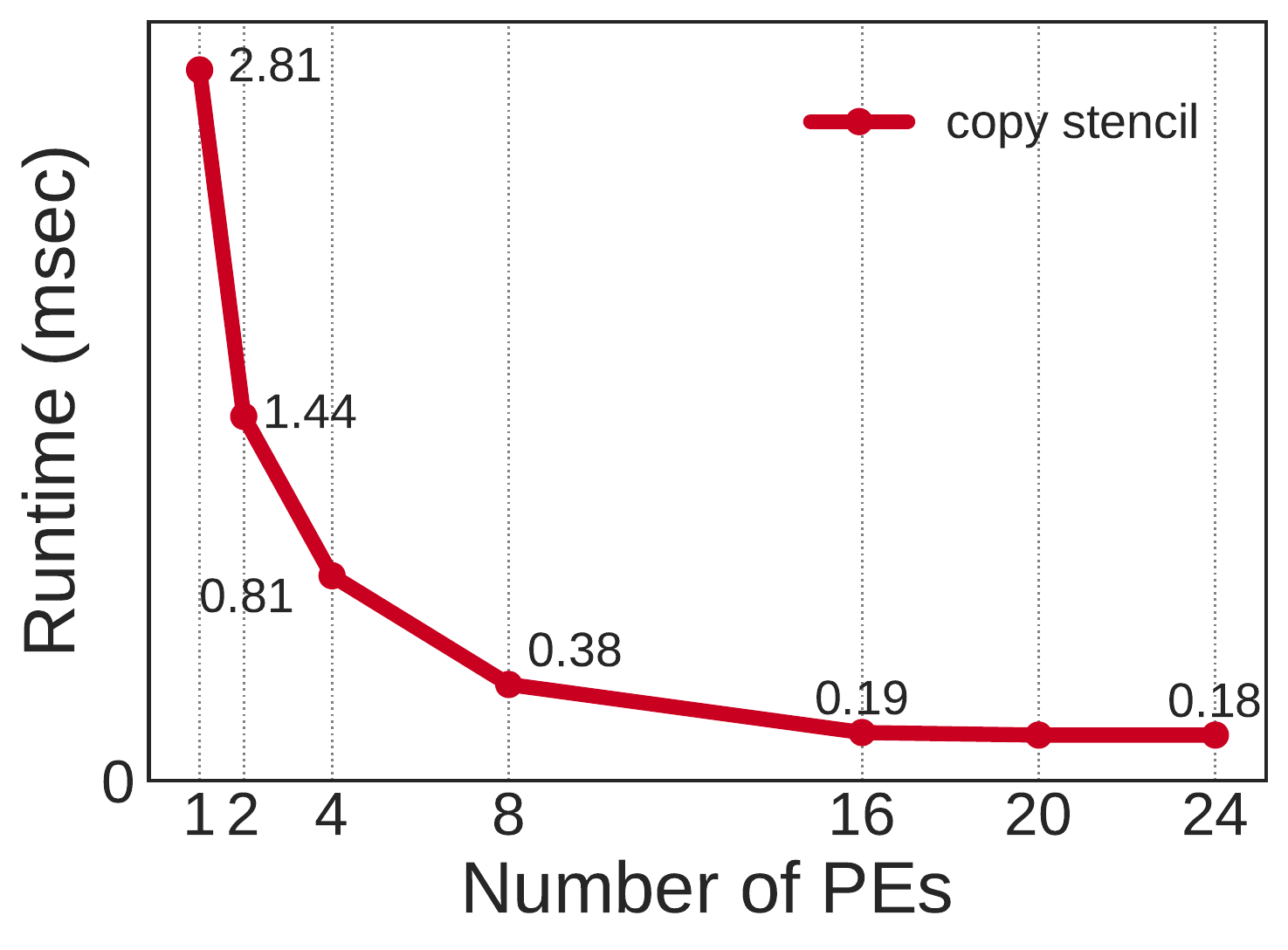}
  \vspace{-11pt}
     \caption{  \gtrets{Performance of copy stencil on our HBM-based FPGA platform.}   \label{fig:copy_stencil}}
\end{figure}

Such compound kernels are dominated by memory-bound operations with complex memory access patterns and low arithmetic intensity. This poses a fundamental challenge to acceleration. \juanggg{CPU implementations of these kernels~\cite{xu2018performance} suffer from limited data locality and inefficient memory usage, as our roofline analysis in Figure~\ref{fig:roofline} exposes}. 
\gtrets{In Figure~\ref{fig:copy_stencil} we implement a \emph{copy stencil} from the COSMO weather model to evaluate the performance potential of our HBM-based FPGA platform for the weather prediction application. A \emph{\cop stencil} performs an element-wise copy operation over the complete input grid. It is the simplest stencil found in the COSMO model and, hence, serves as a benchmark to characterize the achievable peak performance on a platform for weather kernels. {To implement a \cop stencil, we divide the 3D grid data among the processing elements (PEs), where each PE performs an element-wise copy operation.} We were able to enable only 24 HBM memory channels because adding more HBM channels leads to {timing constraint violations}. From the figure, we make two observations. First, as we increase the number of processing elements (PEs), we can exploit data-level parallelism because of dedicated HBM channels serving data to a PE. Second, the maximum achievable performance tends to saturate after 16 PEs. Since we implement our design in a dataflow pipeline manner, all the functions run in parallel, and the overall latency is equal to the maximum latency out of all the functions. After 16 PEs, for \cop, \grev{we observe that most of the time is spent in the FPGA \gcameraa{computation} logic rather than the transfer of data from an HBM memory channel.   }}

\subsection{OC-Accel {Framework}}
The OpenPOWER Foundation Accelerator Workgroup~\cite{open_power} created the OC-Accel framework, 
\juanggg{an open-source environment for FPGA programming productivity}. 
OC-Accel \gagan{provides three key benefits}~\cite{wenzel2018getting}: (i) it enables an improved developer productivity for FPGA acceleration and eas\gagan{es the} use of CAPI's cache-coheren\gagan{ce mechanism},  (ii) it places 
\juanggg{FPGA-}accelerated compute engines, 
\juanggg{also known as} FPGA \textit{actions}, closer to \gagan{relevant} data to achieve better performance, and \gtrets{(iii) access to FPGA memory via user-level DMA (Direct Memory Access) semantics.} OC-Accel provides a simple API to invoke an accelerated \emph{action} and provides programming methods to instantiate customized accelerated \emph{actions} on the FPGA side.  These accelerated \emph{actions} can be specified in C/C++ code that is then compiled to the FPGA target using the Xilinx Vivado High-Level Synthesis (HLS) tool~\cite{hls}. 

\gtrets{The benefits of employing such cache-coherent interconnect links for attaching FPGAs to CPUs, as opposed to the traditional DMA-like communication protocols (e.g., PCIe), are not only the ultra lower-latency and the higher bandwidth of the communication, but most importantly, the ability of the accelerator to access the entire memory space of the CPU coherently, without consuming excessive CPU cycles. Traditionally, the host processor has a shared memory space across its cores with coherent caches. Attached devices such as FPGAs, GPUs, network, and storage controllers are memory-mapped and use a DMA to transfer data between local and system memory across an interconnect such as PCIe. The attached devices can not see the entire system memory but only a part of it. Communication between the host processor and attached devices requires an inefficient software stack, including user-space software, drivers, and kernel-space modules, in comparison to the communication scheme between CPU cores using shared memory. Especially when DRAM memory bandwidth becomes a constraint, requiring extra memory-to-memory copies to move data from one address space to another is cumbersome \om{and low performance}~\cite{seshadri2013rowclone,Fang2020}. This is the driving force of the industry to push for coherency and shared memory across CPU cores and attached devices, like FPGAs. This way, the accelerators act as peers to the processor cores. \grev{Note that CAPI2 is built on top of PCIe. However, CAPI2 provides the following two advantages. First,  a CAPI-attached device, unlike a PCIe device, can perform Direct Memory Access (DMA) to application memory \emph{without} calls to a device driver or underlying operating system kernel, resulting in a reduction in latency. Avoiding these unnecessary memory calls improves performance significantly compared to the traditional PCIe I/O model~\cite{stuecheli2015capi}. Second, CAPI2 provides cache-coherent access to the CPU memory, allowing the FPGA to \emph{directly} access the host memory. Such direct cache-coherent access reduces the FPGA developer's burden and debugging time by overcoming read after write (RAW) and write and read (WAR) dependencies, which is typically the FPGA developer’s responsibility. OCAPI is a new technology built from the ground up. It includes a faster PHY layer \gcameraa{(BlueLink 25Gb/s x8 lanes~\cite{openCAPI}) } than its CAPI predecessors, providing double the bitwidth between the host and an accelerator.}}

%% file: images/hdiff.tex
\begin{figure*}[h]
\begin{subfigure}[h]{0.4\textwidth}
    \resizebox{\textwidth}{!}{
  \begin{tikzpicture}[rotate=90,transform shape]

    \gridThreeD{0}{8.25}{black};
    \node at (8,10.5)[ rotate=270] {\huge \textit{Laplacian}};
    \node at (7.35,10.5)[ rotate=270] {\huge \textit{Stencil}};
      \dotesUpper{0}{8.25};
    \gridThreeDSecond{1}{4}{black};
    \node  at (8,5.5)[ rotate=270] {\huge \textit{Flux}};
    \node at (7.35,5.5)[ rotate=270] {\huge \textit{Stencil}};
     \lowerDots{3}{4};
     
    \gridThreeDThird{2}{0}{black};
     \node  at (7.5,0.25) [ rotate=270] {\huge \textit{Output}};
     \lowestDot{1.6}{-0.45}

      \draw [thick,->,blue] (2.75,4.25) -- (2.7,8.4);
       \draw [thick,->,blue] (2.75,4.25) -- (2.1,8.9);
       \draw [thick,->,blue] (2.75,4.25) -- (4.1,8.9);
       \draw [thick,->,blue] (2.75,4.25) -- (3.5,9.4);

       \draw [thick,->,orange] (2.1,4.7) -- (3.5,9.4);
       \draw [thick,->,orange] (2.1,4.7) -- (1.5,9.4);
       \draw [thick,->,orange] (2.1,4.7) -- (2.1,8.9);
       \draw [thick,->,orange] (2.1,4.7) -- (2.9,9.9);
     
       \draw [thick,->,red] (4.1,4.7) -- (3.5,9.4);
       \draw [thick,->,red] (4.1,4.7) -- (5.5,9.4);
       \draw [thick,->,red] (4.1,4.7) -- (4.1,8.9);
       \draw [thick,->,red] (4.1,4.7) -- (4.9,9.9);

           \draw [thick,->,brown] (3.5,5.25) -- (3.5,9.4);
           \draw [thick,->,brown] (3.5,5.25) -- (2.9,9.9);
           \draw [thick,->,brown] (3.5,5.25) -- (4.9,9.9);
           \draw [thick,->,brown] (3.5,5.25) -- (4.3,10.4);
         \draw [thick,->,violet] (2.7,0.25) -- (3.5,5.15);
         \draw [thick,->,violet] (2.7,0.25) -- (4.1,4.65);
         \draw [thick,->,violet] (2.7,0.25) -- (2.1,4.65);
         \draw [thick,->,violet] (2.7,0.25) -- (2.75,4.15);

    
   
\end{tikzpicture}
}
 \vspace{-32pt}
   \caption{ \label{fig:hdiffMemory}}
\end{subfigure}%
\begin{subfigure}[h]{0.4\textwidth}
\vspace{20pt}
  \includegraphics[width=0.95\linewidth,trim={0cm 0cm 0cm 0.5cm},clip]{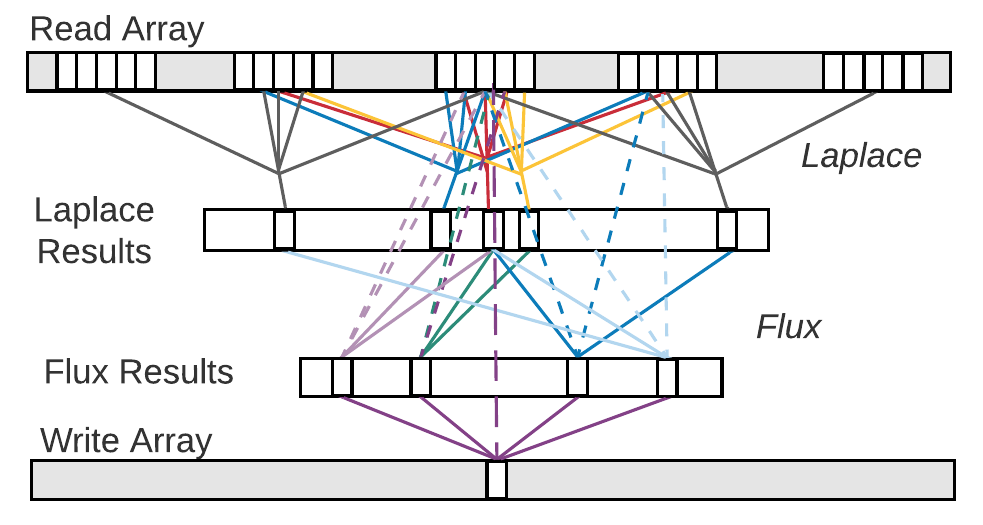}
   \caption{ \label{fig:hdiffMemory_layout}}
\end{subfigure}
\vspace{-10pt}
\caption[Two numerical solutions]{(a) Horizontal diffusion kernel composition {using Laplacian and flux stencils} in a two dimensional plane~\cite{narmada}. (b)    Memory layout of horizontal diffusion from 3D grid onto 1D array. \label{fig:hdiff_access}}
\end{figure*}

%% file: algorithms/cosmo.tex
\begin{algorithm}[h]
\SetAlgoLined
\DontPrintSemicolon
\SetNoFillComment
\caption{Pseudo-code for vertical advection and horizontal diffusion \gagan{kernels}
used by the COSMO~\cite{doms1999nonhydrostatic} \gagannn{weather prediction} model.}
\label{algo:hdiffKernel}
\SetKwFunction{FMain}{verticalAdvection}
   \SetKwFunction{FTest}{forwardSweep}
    \SetKwFunction{FBack}{backwardSweep}
\SetKwProg{Fn}{Function}{}{end}
  
  \Fn{\FMain{float* ccol,
                float* dcol,
                float* wcon,
                float* ustage,
                float* upos,
                float* utens,
                float* utensstage
           }}{
       
             \For{$c\gets2$ \KwTo $column-2$}{
                    \For{$r\gets2$ \KwTo row-2}{
                    
               \Fn{\FTest{float* ccol,
                float* dcol,
                float* wcon,
                float* ustage,
                float* upos,
                float* utens,
                float* utensstage}} { 
                    \For{$d\gets1$ \KwTo $depth$}{
                       \tcc*[l]{forward sweep calculation}
                       }
                     }   
                       
                \Fn{\FBack{float* ccol,
                float* dcol,
                float* wcon,
                float* ustage,
                float* upos,
                float* utens,
                float* utensstage}} { 
                    \For{$d\gets depth-1$ \KwTo$ 1$}{
                       \tcc*[l]{backward sweep calculation}}}
                    }
                  }

}


\hrule
\SetKwFunction{FMain}{horizontalDiffusion}
\SetKwProg{Fn}{Function}{}{end}
  \Fn{\FMain{float* src, float* dst}}{
       \For{$d\gets1$ \KwTo $depth$}{
             \For{$c\gets2$ \KwTo $column-2$}{
                    \For{$r\gets2$ \KwTo row-2}{
                        \tcc*[l]{\gagan{L}aplacian calculat\gagan{ion}}
                        $lap_{CR}=laplaceCalculate(c,r)$
                        \tcc{row-laplacian}
                        $lap_{CRm}=laplaceCalculate(c,r-1)$\;      
                        $lap_{CRp}=laplaceCalculate(c,r+1)$
                        \tcc{column-laplacian}
                        $lap_{CmR}=laplaceCalculate(c-1,r)$\;
                        $lap_{CpR}=laplaceCalculate(c+1,r)$
                        \tcc{column-flux calculat\gagan{ion}}\
                        $flux_{C} = lap_{CpR} - lap_{CR}$\;
                        $flux_{Cm} = lap_{CR} - lap_{CmR}$\;
                        \tcc{row-flux calculat\gagan{ion}}
                        $flux_{R} = lap_{CRp} - lap_{CR}$\;
                        $flux_{Rm} = lap_{CR} - lap_{CmR}$\;
                        \tcc{output calculat\gagan{ion}}
                        $dest[d][c][r] = src[d][c][r] -
                            c1 * (flux_{CR}- flux_{CmR}) + (flux_{CR}- flux_{CRm})$
                        }
            
                }
   
         }
}
\end{algorithm}

%% file: sections/accelerator.tex
\section{Design Methodology}
\label{sec:design}
\subsection{NERO, A Near HBM Weather Prediction Accelerator}
The low arithmetic intensity of real-world weather \gagan{prediction} kernels limits the attainable performance on current multi-core systems. This sub-optimal performance is due to \gagan{the kernels'} complex memory \juanggg{access} patterns and their inefficiency in exploiting a rigid cache hierarchy, \gagan{as quantified in \juangg{the} roofline plot in} Figure~\ref{fig:roofline}. 
These kernels \gagannn{cannot} fully utilize the available memory bandwidth, which leads to 
high 
\juanggg{data movement overheads} in terms of latency and energy consumption. We address 
\juanggg{these inefficiencies} by developing an architecture \gagan{that} combines fewer off-chip data \gagan{accesses} with higher throughput for the loaded data. \gagan{To this end}, our accelerator design \gagan{takes} a data-centric approach~\cite{mutlu2019,ghose2019processing,teserract,singh2019near,NAPEL,hsieh2016accelerating,7551394,ahn2015pim,googleWorkloads,kim2018grim,mutlu2021intelligent,boroumand2021google,mutlu2020modern} \gagan{that exploits} near high-bandwidth memory acceleration.

\begin{figure*}[h]
\begin{subfigure}[t]{0.5\textwidth}
 \includegraphics[width=1\linewidth,trim={0.5cm 1.2cm 0.5cm 1cm},clip]{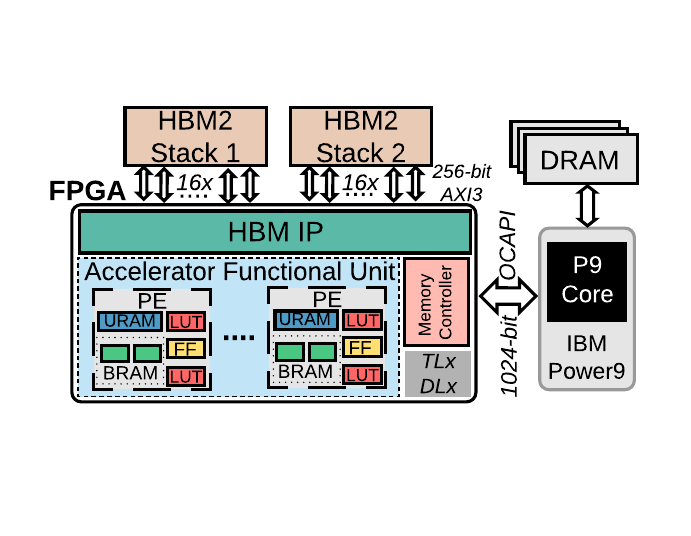}
   \caption{
  \label{fig:system}}
\end{subfigure}%
\begin{subfigure}[t]{0.4\textwidth}
  \includegraphics[width=0.95\linewidth,trim={0cm 0cm 0cm 0cm},clip]{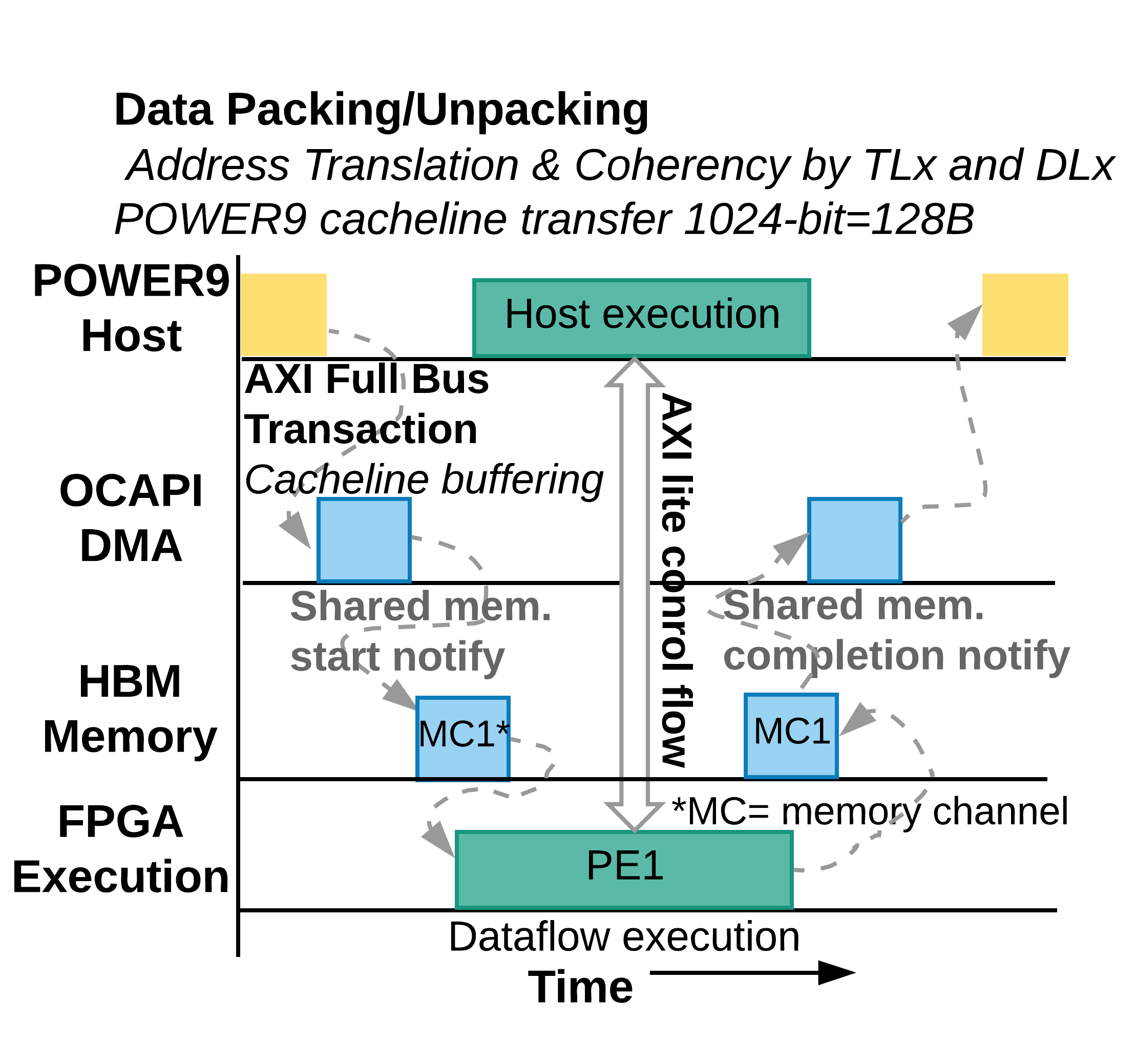}
  \caption{
  \label{fig:execution}}
\end{subfigure}
\vspace{-10pt}
\caption[Two numerical solutions]{(a) Heterogeneous platform with an IBM POWER9 system connected to an HBM-based FPGA board via~OCAPI. We also show components of an FPGA: flip-flop (FF), lookup table (LUT), UltraRAM (URAM), and Block RAM (BRAM).
 (b)  Execution timeline with data flow sequence from the  host  DRAM  to the onboard  FPGA  memory.\label{fig:complete_hbm_flow}}
\end{figure*}

\begin{figure}[h]
 \centering
  \includegraphics[width=0.95\linewidth,trim={0.5cm 0.8cm 0cm 0cm},clip]{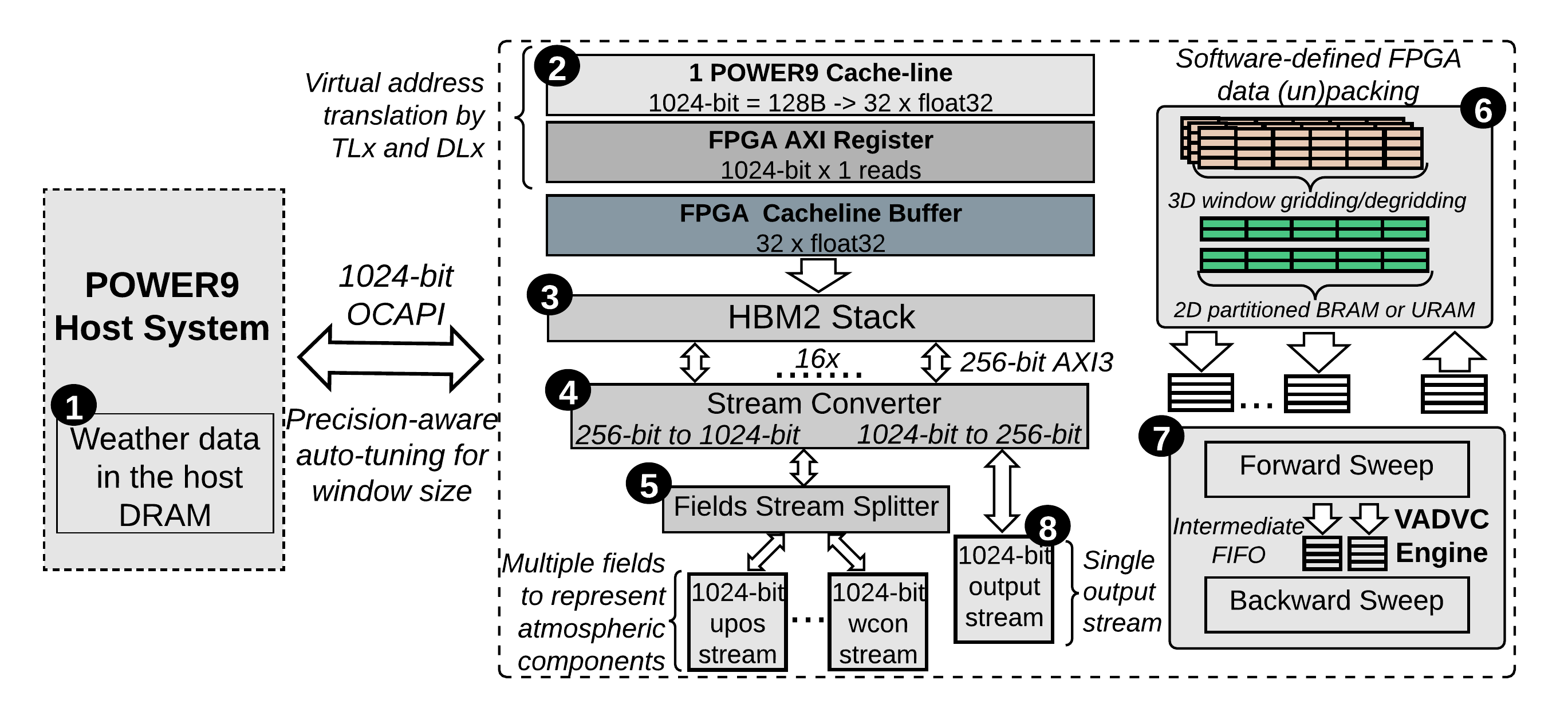}
  \vspace{-2pt}
   \caption{Architecture overview of~\namePaper~with data flow sequence from the host DRAM to the on-board FPGA memory via POWER9 cachelines. We {depict a} single {processing element (PE)} fetching data from a dedicated HBM port. The number of HBM ports scales linearly with the number of {PEs}. Heterogeneous partitioning of on-chip memory blocks \juangg{reduces} read \juangg{and} write latencies across the FPGA memory hierarchy.
  \label{fig:complete_flow}}
\end{figure}

Figure~\ref{fig:system} shows a high-level overview of our integrated system. \gagan{An HBM-based} FPGA is connected to a server system based on an IBM POWER9 processor using the \juangg{Open Coherent Accelerator Processor Interface}  (OCAPI). 
The FPGA consists of two HBM stacks\footnote{In this work, we enable only a single stack based on our resource and power consumption analysis \gagan{for \gagannn{the} \texttt{vadvc} kernel.}}, each with 16 \emph{pseudo-memory channels}~\cite{axi_hbm}. \gagan{A} channel is exposed to the FPGA as a 256-bit wide port, and in total, 
\juanggg{the FPGA} has 32 such ports. The HBM IP provides 8 memory controllers \juanggg{(per stack)} to handle the data transfer to and from the HBM memory ports. Our design consists of an \emph{accelerator functional unit} (AFU) that interacts with the host system through the TLx (Transaction Layer) and the DLx (Data Link Layer), which are the OCAPI endpoint on the FPGA. An AFU comprises of multiple \emph{processing elements} (PEs) that perform compound stencil computation. Figure~\ref{fig:complete_flow} shows the architecture overview of \namePaper. As vertical advection is the most complex kernel, we depict \gagan{our architecture design flow for} vertical advection. We use a similar design for the \gagan{horizontal diffusion} kernel.

The weather data, based on the atmospheric model resolution grid, is stored in the DRAM of the host system \gagan{(\circled{1} in Figure~\ref{fig:complete_flow})}.  We employ \juangg{the} double buffering technique  between the CPU and the FPGA to hide the PCIe \gagan{(Peripheral Component Interconnect Express~\cite{pcie})} \gagannn{transfer} {latency}.  {By configuring a buffer of 64 cache lines, between the AXI4 interface of OCAPI/TLx-DLx and the AFU, we can reach the theoretical peak bandwidth of OCAPI (\juanggg{i.e.,} 32 GB/s).} We create a specialized memory hierarchy from the heterogeneous FPGA memories \gagan{(i.e., URAM, BRAM, and HBM)}. By using a greedy algorithm, we determine the best-suited hierarchy for our kernel. The memory controller (shown in Figure~\ref{fig:system}) handles the data placement to the appropriate memory type \gagan{based on the programmer's directives}.

\gagan{On the FPGA, following the initial buffering (\circled{2}), the \gagan{transferred} grid data is mapped onto the HBM memory (\circled{3}).  As \gagan{the} FPGA ha\gagan{s} limited resources, we propose a 3D window-based grid transfer from the host DRAM to the FPGA, facilitating a smaller, less power-hungry deployment. The window size represents the portion of the grid a processing element ({PE} in Figure~\ref{fig:system}) would process. \gagan{Most FPGA developers manually optimize for the right window size. However, manual optimization is tedious because of the huge design space, and \juanggg{it} requires expert guidance. 
Further, selecting an inappropriate window size lead\gagan{s} to sub-optimal results. 
\juangg{Our experiments (in Section~\ref{subsection:evaluation}) show that}: (1) finding the \juangg{best} window size is critical in terms of \juangg{the} area vs. performance trade-off, and (2) the \juangg{best window} size depends on the datatype precision. Hence, instead of pruning the design space manually, we formulate 
\juanggg{the search for} the \juangg{best} window size as a multi-objective auto-tuning problem taking into account the datatype precision. We make use of OpenTuner~\cite{opentuner}, 
\juanggg{which} uses machine-learning techniques to guide the design-space~exploration~\cite{singh2021modeling}}.}


Our design consists of multiple PEs (shown in Figure~\ref{fig:system}) that exploit data-level parallelism in COSMO \gagannn{weather prediction} kernels. \gagan{A dedicated HBM memory port is assigned to a specific PE}; therefore, we enable as many HBM ports as the number of PEs. This allows us to use the high \gagan{HBM} bandwidth \gagannn{effectively} because each PE fetches from an independent port. 
In our design, we use a switch, which provides the capability to 
\juanggg{bypass} the HBM, \juangg{when the} grid size \juangg{is small}, and map the data directly onto the FPGA's URAM \gagan{and BRAM}. The HBM port provides 256-bit data, \gagannn{which} is a quarter of the size of \juangg{the} OCAPI \gagannn{bitwidth} (1024-bit). Therefore, to match \gagannn{the} OCAPI bandwidth, we introduce a stream converter logic (\circled{4}) that converts \juangg{a} 256-bit HBM stream to \juangg{a} 1024-bit stream (OCAPI compatible) or vice versa.
From HBM, \gagan{a PE} reads a single stream of data that consists of all the fields\footnote{Fields represent atmospheric components like wind, pressure, velocity, etc. that are required for weather calculation.} \juangg{that} are needed for a specific COSMO kernel computation. \gagan{The PEs} use a fields stream splitter logic (\circled{5}) that splits a single HBM stream to multiple streams (1024-bit each), one for each field.
  
\gagann{To optimize a PE\gagan{,} we apply various optimization strategies. First, we exploit the inherent parallelism in \gagann{a given} algorithm through hardware pipelining. \gagann{Second}, we partition on-chip memory to avoid the stalling of our pipelined design, since the on-chip BRAM/URAM has only two read/write ports. \gagann{Third}, all the tasks execute in a dataflow manner that enables task-level parallelism. \texttt{vadvc} is more \gagan{ computationally complex} than \texttt{hdiff} because it involves forward and backward sweeps \gagan{with dependencies in \juangg{the} z-dimension}. While \texttt{hdiff} performs \gagan{only} \juangg{Laplacian} and flux calculations with dependencies in \juangg{the} x- and y-dimensions. 
Therefore, we demonstrate our design flow by means of \gagannn{the} \texttt{vadvc} kernel (Figure~\ref{fig:complete_flow}). 
Note \juangg{that} we show \juangg{only} a single port-based PE operation.  \juanggg{However}, for multiple PEs, we enable multiple HBM ports.} 

We make use of memory reshaping techniques to  configure our memory space with multiple parallel BRAMs or URAMs~\cite{dioFPT}. \gagann{We form an intermediate memory hierarchy by decomposing (or slicing) 3D window data into a 2D grid. } This allows us to bridge the latency gap between the HBM memory and our accelerator. Moreover, it allows us to exploit the available FPGA resources efficiently. Unlike traditionally-fixed CPU memory hierarchies, which perform poorly with irregular access patterns and suffer from cache pollution effects \om{and cache miss latency}, application-specific memory hierarchies \juangg{are} shown to improve \juangg{energy and latency} by tailoring the cache levels and cache sizes to \juangg{an} application's memory access patterns~\cite{jenga}.
  
The main computation pipeline (\circled{7}) \gagann{consists of a forward and a backward sweep logic}. The forward sweep \gagannn{results} are stored in \juangg{an} intermediate buffer to allow for backward sweep calculation. \juangg{Upon} completion of \juangg{the} backward sweep, results are \gagan{placed in} an output buffer \gagan{that is followed by a degridding logic (\circled{6}). The degridding logic} converts the calculated results to a 1024-bit wide output stream (\circled{8}). As there is only a single output stream (both in \texttt{vadvc} and \texttt{hdiff}), we do not need extra logic to merge the streams. The 1024-bit \gagan{wide} stream goes through an HBM stream converter logic (\circled{4}) that converts the stream bitwidth to HBM port size (256-bit).  
  
\gagan{Figure~\ref{fig:execution} shows the execution timeline from our host system to the FPGA board for a single PE. The host offloads the processing to an FPGA and 
\juangg{transfers} the required data \juangg{via DMA (direct memory access)} over \juangg{the} OCAPI interface. The OC-Accel framework allows for parallel execution of the host and our FPGA PEs while exchanging control signals over the AXI lite interface\gagan{~\cite{axilite}}. On task completion, the AFU notif\gagannn{ies} the host system via \gagannn{the} AXI lite interface and 
\juangg{transfers} back the results \juangg{via DMA}. }

\subsection{\namePaper~Application Framework}
\begin{figure}[h]
  \centering
 \includegraphics[width=0.7\linewidth,trim={0cm 0cm 0.1cm 0cm},clip]{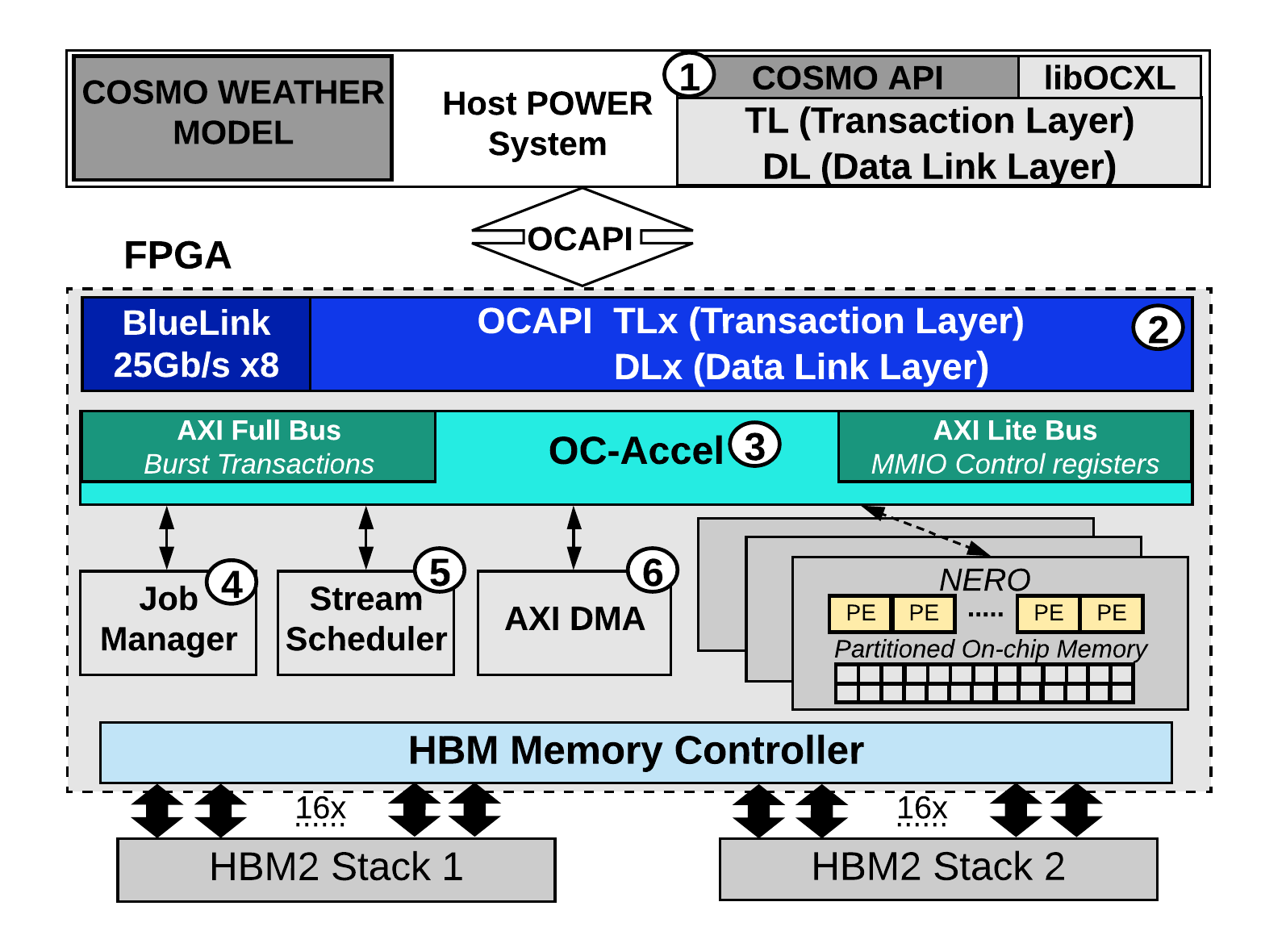}
 \vspace{-10pt}
  \caption{
  \juangg{\namePaper~application framework.} 
  We co-design our software and hardware using \gagannn{the} OC-Accel framework. COSMO API allows the host to offload kernels to our FPGA platform.
 \label{fig:snap_api}}
 \end{figure}

Figure~\ref{fig:snap_api} shows the \namePaper~application framework \gagan{to support our architecture}. \gcamera{Our previous work~\cite{singh2020nero,narmada} describes the corresponding application framework using SNAP-CAPI2.}  A software-defined COSMO API (\circledWhite{1}) handles offloading jobs to \namePaper~with an interrupt-based queuing mechanism.  This allows for minimal CPU usage (and, hence, power \gagan{usage}) during FPGA \gagan{operation}. \namePaper~employs an array of processing elements to compute COSMO kernels, such as vertical advection or horizontal diffusion. Additionally, we pipeline our PEs to exploit the available spatial parallelism. 
By accessing the host memory through 
\juanggg{the OCAPI} cache-coherent link, \namePaper~acts as a peer to the CPU.  This is enabled \gagannn{through} \gagan{the} TLx (Transaction Layer) and the DLx (Data Link Layer) (\circledWhite{2}). \gagan{ OC-Accel (\circledWhite{3}) allows for seamless \juangg{integration} of the COSMO API with our OCAPI-based accelerator.} The job manager (\circledWhite{4}) dispatches jobs to streams, which are managed in the stream scheduler (\circledWhite{5}). The execution of a job is done by streams that determine which data is to be read from the host memory and sent to the PE array through DMA transfers (\circledWhite{6}). The pool of heterogeneous on-chip memory is used to store the input data from the main memory and the intermediate data generated by \juangg{each} PE.  

%% file: sections/results.tex
\section{Results}
\label{sec:results}


\subsection{Experimental Setup}
{We evaluate our accelerator designs for  \vadvc, and \hdiff  in terms of performance, energy consumption, and FPGA resource utilization on two different FPGAs, and two different external data communication interfaces between the CPU and the FPGA board.}
We implement our accelerator designs for \vadvc, and \hdiff 
on  \sr{both 1)} \gonur{an} Alpha-Data ADM-PCIE-9H7 card~\garxiv{\cite{ad9h7} }featuring the Xilinx Virtex Ultrascale+ XCVU37P-FSVH2892-2-e~\garxiv{\cite{vu37p}} with 8GiB HBM2~\cite{hbm} \sr{and 2)} \gonur{an} Alpha-Data ADM-PCIE-9V3 card~\garxiv{\cite{ad9v3}} featuring the Xilinx Virtex Ultrascale+ XCVU3P-FFVC1517-2-i with 8GiB DDR4~\garxiv{\cite{vu37p}}, connected to an IBM POWER9 host system. For the external data communication interface, we use both CAPI2~\garxiv{\cite{stuecheli2015capi}} and the state-of-the-art OCAPI (OpenCAPI)~\garxiv{\cite{openCAPI}} interface. 
We compare these implementations to execution on a POWER9 CPU with 16 cores \juan{({using all} 64 hardware threads).} 
Table~\ref{tab:systemparameters}
provides our system parameters.
\grev{We co-design our hardware and software interface around the OC-Accel framework~\cite{ocaccel} while using the HLS design flow. Our development machine is x86 Intel\textsuperscript{\sffamily\textregistered} Xeon\textsuperscript{\sffamily\textregistered}
7.9.2009~\cite{centos} distribution with GNU Compiler Collection (GCC) version 4.8.5~\cite{gnu}. We use Xilinx Vivado 2019.2~\cite{vivado} suite to develop our accelerator designs.  }

\input{tables/system.tex}

\subsection{Performance Tuning}

\begin{figure}[h]
  \centering
  \includegraphics[width=1\linewidth,trim={0.4cm 0.3cm 0.35cm 0.2cm},clip]{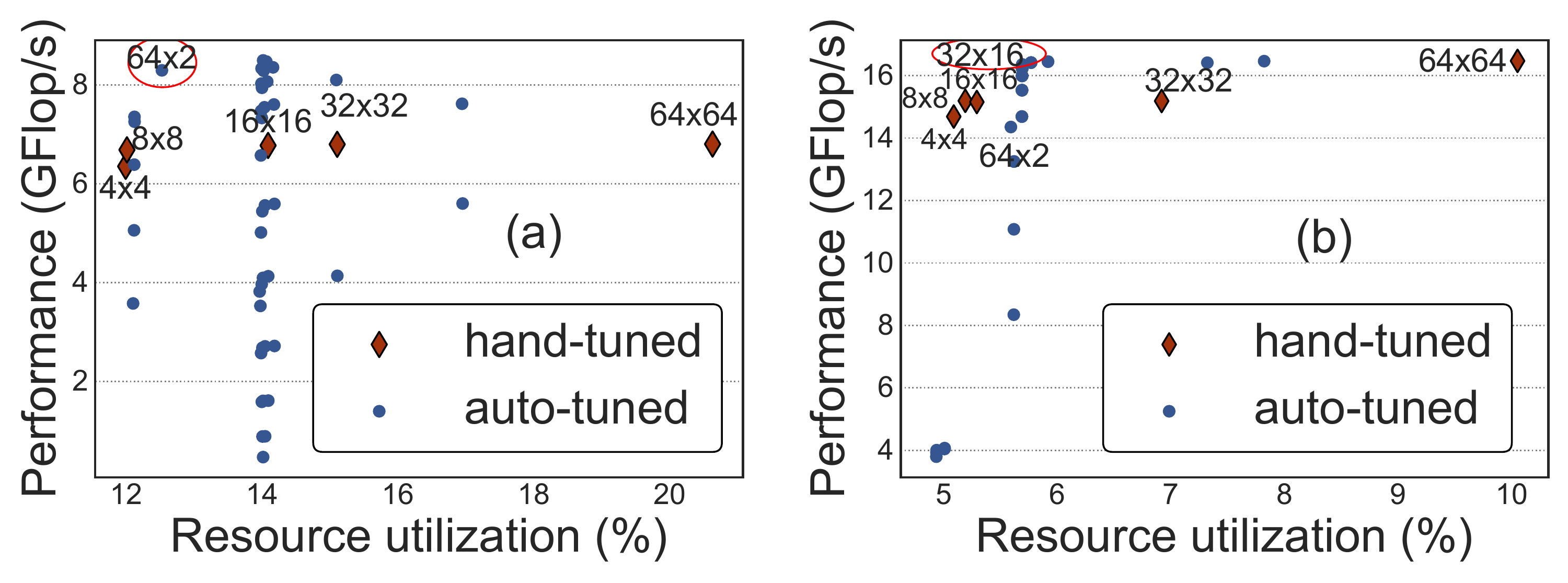}
  \vspace{-10pt}
      \caption{Performance \gagann{and FPGA resource utilization} \gagannn{of} single \texttt{vadvc} PE\gagann{, as a function of tile-size,} using hand-tuning and auto-tuning for (a) single-precision (32-bit) and (b) half-precision (16-bit). We highlight the Pareto-optimal solution that we use for our \texttt{vadvc} accelerator \gagan{(with a red circle). Note that }the Pareto-optimal solution changes with precision.
  \label{fig:single_afu}}
 \end{figure}

We 
\juanggg{run} our experiments using a $256\times256\times64$-point domain similar to the grid domain used by the COSMO \gagannn{weather prediction} model.  We employ an auto-tuning technique to determine a Pareto-optimal solution (in terms of performance and resource utilization) for our 3D window dimensions.
{The auto-tuning \juanggg{with OpenTuner} exhaustively searches for every tile size in the x- and y-dimensions for \texttt{vadvc}.\footnote{\texttt{vadvc} has dependencies in \gagann{the} z-dimension; therefore, it cannot be parallelized in the z-dimension.} For \texttt{hdiff}, we consider sizes in all three dimensions. We define our auto-tuning as a multi-objective optimization with the goal \gagannn{of maximizing} performance with  minimal resource utilization.} Section~\ref{sec:design} provides further details on our design. \grev{We evaluate frequency values between 50-400 MHz, with an increment of 50MHz, utilizing the complete spectrum of compatible frequency configurations supported by the OC-Accel framework~\cite{ocaccel}. }
Figure~\ref{fig:single_afu} shows hand-tuned and auto-tuned \gagann{performance and FPGA resource utilization} results for {\texttt{vadvc}}\gagann{, as a function of the chosen tile size. From the figure, we draw two observations.} 

First, by using the auto-tuning approach and our careful FPGA microarchitecture design, we can get \gagan{P}areto-optimal results with a tile size of $64\times2\times64$ for single-precision {\texttt{vadvc}}\gagan{, which} gives us a peak performance of 8.49 GFLOP/s. For half-precision, we use a tile size of $32\times16\times64$ to achieve a peak performance of 16.5 GFLOP/s. We employ a similar strategy for {\texttt{hdiff}} to attain a single-precision performance of 30.3 GFLOP/s with a tile size of $16\times64\times8$ and a \gagan{half-precision} performance of 77.8 GFLOP/s \gagan{with} a tile size of $64\times8\times64$. 

Second, in FPGA acceleration, designers usually rely on expert judgement to find the appropriate tile-size and often adapt the design to \gagannn{use} homogeneous tile sizes. However, as shown in Figure~\ref{fig:single_afu}\gagan{,} such hand-tuned implementations lead to sub-optimal results in terms of either resource utilization~or~performance. 

\gagann{We conclude that} the \gagann{Pareto-optimal} tile size depends on the data precision used\gagan{:} a \gagan{good} tile-size for single-precision might lead to poor results when used with half-precision.
\subsection{Performance Analysis}
\label{subsection:evaluation}

\begin{figure}[h]
  \centering
  \includegraphics[width=1\linewidth,trim={13.4cm 0.3cm 0.35cm 0.4cm},clip]{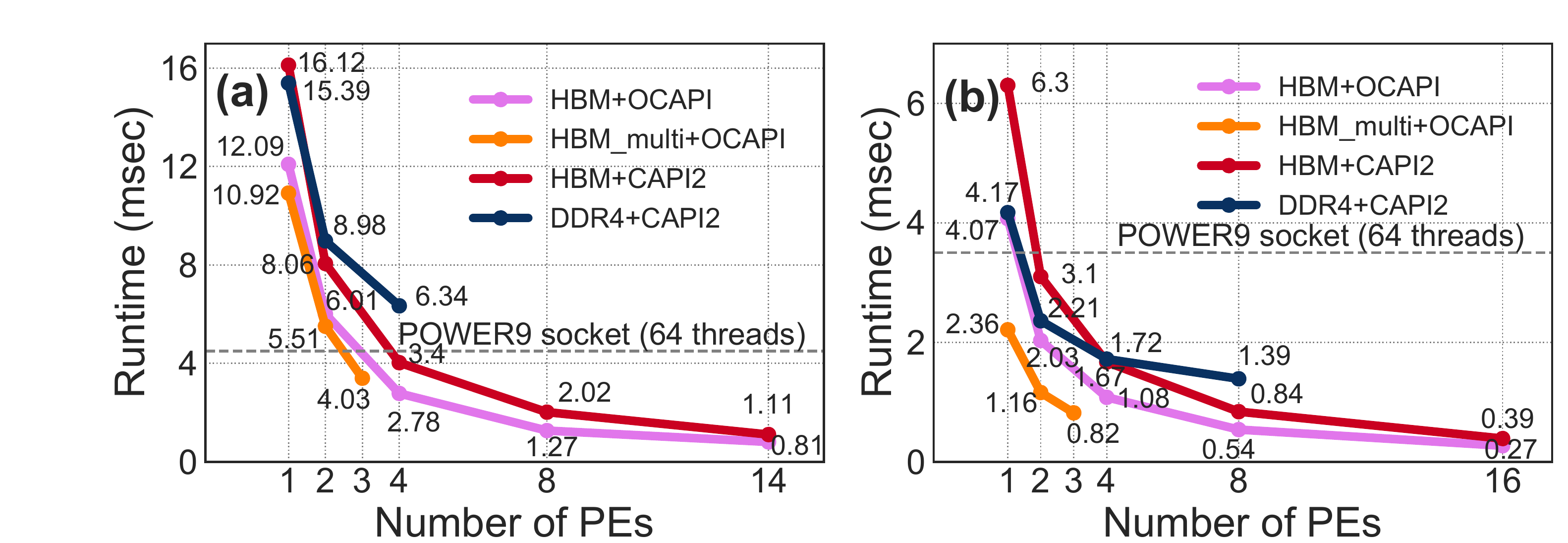}
  \vspace{-10pt}
   \caption{ \gagann{Single-precision} performance for (a) \texttt{vadvc} and (b) \texttt{hdiff}\gagann{, as a function of accelerator PE count on the HBM- and DDR4-based FPGA boards}. 
   \gagan{We also show \gagann{the}} single socket (64 threads) performance of \gagan{an} IBM POWER9 host system for both \texttt{vadvc} and \texttt{hdiff}. \gtrets{For HBM-based design, we implement our accelerator with \gonur{both the} CAPI2 interface and the state-of-the-art OpenCAPI (OCAPI) interface (with both single channel and multiple \sr{channels} per PE).} 
  \label{fig:perf}}
 \end{figure}

Figure~\ref{fig:perf} shows \gagann{single-precision} performance results for the (a) vertical advection \juann{(\texttt{vadvc})} and (b) horizontal diffusion kernels \juann{(\texttt{hdiff})}. 
For both kernels, we implement our design \juanggg{on} an HBM- and \juanggg{a} DDR4-based FPGA board. 
\juann{For the DDR4-based design, we use CAPI2 (\texttt{DDR4+CAPI2} in Figure~\ref{fig:perf}. 
For the HBM-based design, we use CAPI2 (\texttt{HBM+CAPI2}) and OCAPI. We evaluate two versions of the HBM-based design with OCAPI: (1) one with a single channel per PE (\texttt{HBM+OCAPI}), and (2) one with multiple channels (i.e., 4 HBM pseudo channels) per PE (\texttt{HBM\_multi+OCAPI}).}
To compare the performance 
\juann{of these four versions}, we scale the number of PEs and analyze the change in execution time.  \gtrets{ We also tested different domain sizes, varying from $64\times64\times64$-point to $1024\times1024\times64$-point and observe that the runtime scales linearly and the overall performance (GLOP/s) remain constant. This shows the scalability of our accelerator design. }

We draw \juann{five} observations from the figure.

\juann{First, the maximum number of PEs that we can fit on the FPGA boards varies for different versions of our design. 
For the DDR4-based design, we can accommodate only 4 PEs/8 PEs for \texttt{vadvc}/\texttt{hdiff} on the 9V3 board. 
For the HBM-based design, we can fit 14 PEs/16 PEs for \texttt{vadvc}/\texttt{hdiff} for both \texttt{HBM+CAPI2} and \texttt{HBM+OCAPI} versions before exhausting the on-board resources. 
The \texttt{HBM\_multi+OCAPI} version can only fit 3 PEs (i.e., 12 HBM channels) for both \texttt{vadvc} and \texttt{hdiff} because adding more HBM channels leads to timing constraint violations.}


\juann{Second, the full-blown \texttt{HBM+OCAPI} versions (i.e., with the maximum number of PEs) of \texttt{vadvc} and \texttt{hdiff} outperform the 64-thread IBM POWER9 CPU version by $5.3\times$, and $12.7\times$, respectively. We achieve 37\% and 44\% higher performance for \texttt{vadvc} and \texttt{hdiff}, respectively, with \texttt{HBM+OCAPI} than \texttt{HBM+CAPI2} due to the following two reasons: 
(1) OCAPI provides double the bitwidth (1024-bit) of the CAPI2 interface (512-bit), which provides a higher bandwidth to the host CPU, i.e., 22.1/22.0 GB/s R/W versus 13.9/14.0 GB/s; and 
(2) with OCAPI, memory coherency logic is moved onto the IBM POWER CPU, which provides more FPGA area and allows us to run our accelerator logic at a higher clock frequency (250MHz for OCAPI versus 200MHz for CAPI2).} \grev{We observe that when implementing our accelerator designs with targets above those frequencies, the respective timing report results in the worst negative slack (WNS) higher than 200ps, which OC-Accel developers regard as dangerous for system stability. At lower frequencies, we achieved lower performance, regardless of the number of PEs.}  \gtrets{ Our single-precision \texttt{HBM+OCAPI}-based FPGA implementations {provide} 157.1~GFLOP/s and 608.4~GFLOP/s for \vadvc and \hdiff, respectively. For half-precision, if we use
the same amount of {PE}s as in single precision, {our accelerator} 
{reaches} a performance of 329.9~GFLOP/s for \texttt{vadvc} ($2.1\times$ {the} single-precision performance) and 1.5~TFLOP/s for \texttt{hdiff} ($2.5\times$ {the} single-precision performance).}


\juann{Third, for a single PE, \texttt{DDR4-CAPI2} is faster than \texttt{HBM-CAPI2} for both \texttt{vadvc} and \texttt{hdiff}. This  higher performance is because the HBM-based design uses one HBM channel per PE, and the bus width of the DDR4 channel (512 bits) is larger than that of an HBM channel (256 bits). 
Therefore, the HBM channel has a lower transfer rate of 0.8-2.1 GT/s (Gigatransfers per second) than a DDR4 channel (2.1-4.3 GT/s), resulting in a theoretical bandwidth of 12.8 GB/s and 25.6 GB/s per channel, respectively. 
One way to match the DDR4 bus width is to have a single PE fetch data from multiple HBM channels in parallel. In Figure~\ref{fig:perf}, our multi-channel setting (\texttt{HBM\_multi+OCAPI}) uses 4 HBM pseudo channels per PE to match the bitwidth of the OCAPI interface. 
We observe that by fetching more data from multiple channels, compared to the single-channel-single PE design (\texttt{HBM+OCAPI}), \texttt{HBM\_multi+OCAPI} achieves $1.2\times$ and $1.8\times$ performance improvement for \texttt{vadvc} and \texttt{hdiff}, respectively.}


\juann{Fourth, as we increase the number of PEs, we divide the workload evenly across PEs. 
As a result, we observe linear scaling in the performance of HBM-based designs, where each PE reads and writes through a dedicated HBM channel. 
For multi-channel designs, we observe that the best-performing multi-channel-single PE design (i.e., using 3 PEs with 12 HBM channels for both workloads) has $4.7\times$ and $3.1\times$ lower performance than the best-performing single-channel-single PE design (i.e., 14 PEs for \texttt{vadvc} and 16 PEs for \texttt{hdiff}, respectively). 
This observation shows that there is a tradeoff between (1)~enabling more HBM pseudo channels to provide each PE with more bandwidth, and (2)~implementing more PEs in the available area. 
For both \texttt{vadvc} and \texttt{hdiff}, data transfer and computation take a comparable amount of time. Therefore, we are able to achieve a linear execution time reduction with the number of PEs.}


\juann{Fifth, the performance of the DDR-based designs scales non-linearly for \texttt{vadvc} and \texttt{hdiff} with the number of PEs, as all PEs access memory through the same channel. 
Multiple PEs compete for a single memory channel, which causes frequent memory stalls due to contention in the memory channel.} 


\subsection{Energy Efficiency Analysis}
\begin{figure}[h]
  \centering
  \includegraphics[width=1\linewidth,trim={0cm 0cm 0cm 0cm},clip]{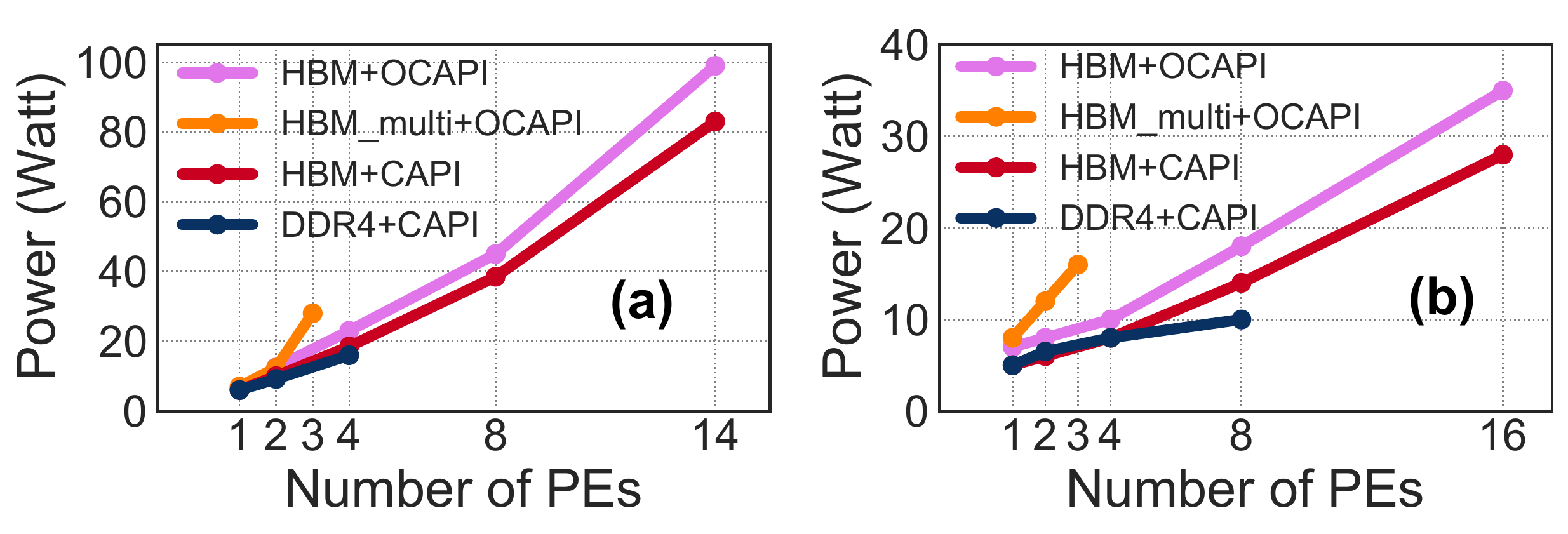}
  \vspace{-10pt}
     \caption{Active Power Consumption for (a) \texttt{vadvc} and (b) \texttt{hdiff} on HBM- and DDR4-based FPGA boards. \gtrets{For HBM-based design, we implement our accelerator with \gonur{both the} CAPI2 interface and the state-of-the-art OpenCAPI (OCAPI) interface (with both single channel and multiple \sr{channels} per PE).} 
  \label{fig:power_consumption}}
 \end{figure}

\begin{figure}[h]
  \centering
  \includegraphics[width=1\linewidth,trim={13.4cm 0.3cm 0.35cm 0.4cm},clip]{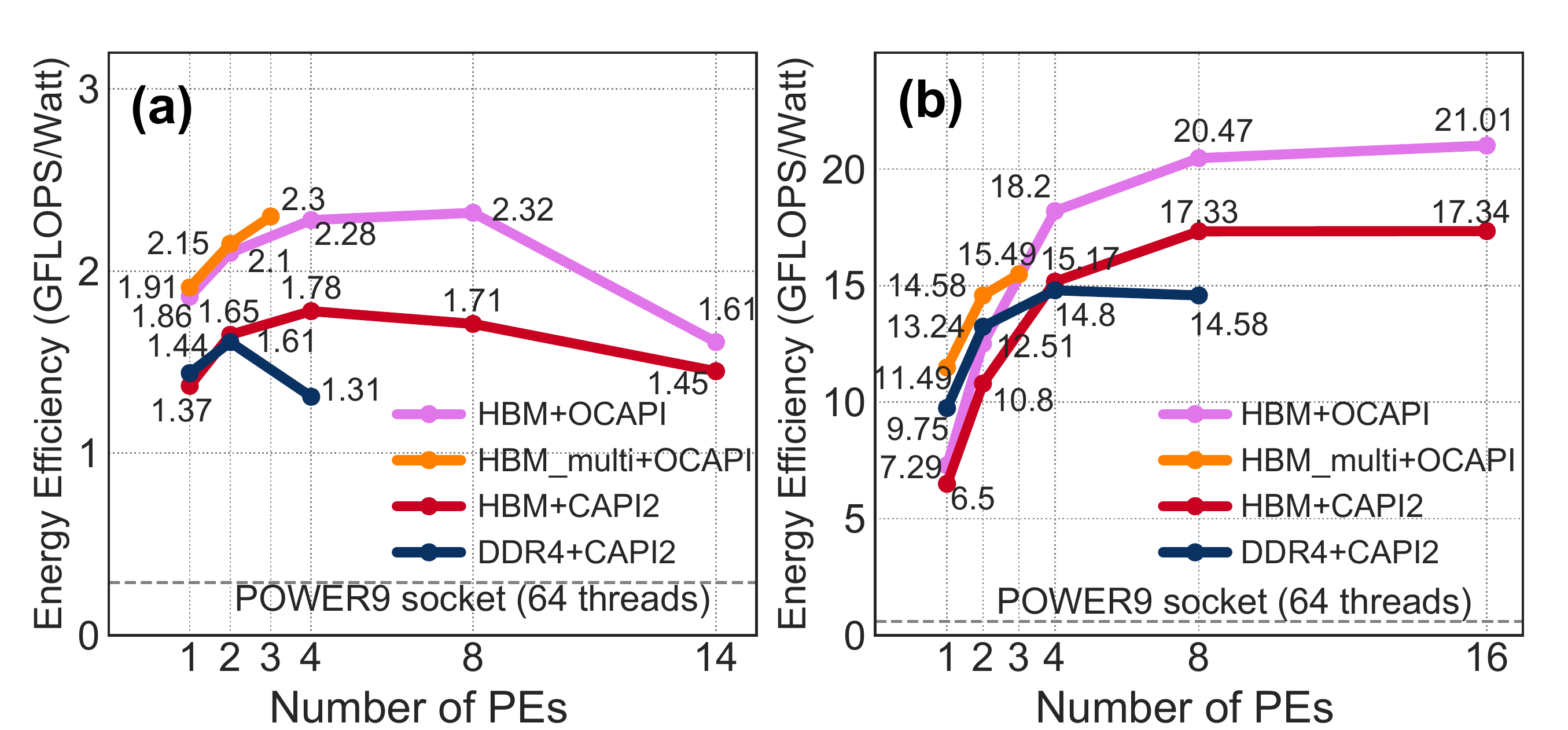}
  \vspace{-10pt}
     \caption{Energy efficiency for (a) \texttt{vadvc} and (b) \texttt{hdiff} on HBM- and DDR4-based FPGA boards. \gagann{We also show \gagann{the}} single socket (64 threads) energy efficiency of \gagan{an} IBM POWER9 host system for both \texttt{vadvc} and \texttt{hdiff}. \gtrets{For HBM-based design, we implement our accelerator with \gonur{both the} CAPI2 interface and the state-of-the-art OpenCAPI (OCAPI) interface (with both single channel and multiple \sr{channels} per PE).} 
  \label{fig:energy_eff}}
 \end{figure}

We compare the energy consumption of our accelerator to \gagan{a} \juanggg{16-core} POWER9 host system.  We use the AMESTER\footnote{https://github.com/open-power/amester} tool to measure the active power\footnote{Active power denotes the difference between the total power of a complete socket (including CPU, memory, fans, I/O, etc.) when an application is running compared to when it is idle.} consumption. 
\juanggg{We measure}~99.2~Watt\gagan{s} for \texttt{vadvc} and ~97.9~Watt\gagan{s} for \texttt{hdiff} by monitoring \gagannn{built} power sensors in 
\juanggg{the POWER9}~system. 
\grev{For {\texttt{vadvc}} and {\texttt{hdiff}} 
\juanggg{on the HBM-} and DDR4-based designs, Figure~\ref{fig:power_consumption} and Figure~\ref{fig:energy_eff} shows the active power consumption and the energy efficiency (GFLOPS per Watt), respectively.} 

We make \juann{five} observations from Figure~\ref{fig:power_consumption} and Figure~\ref{fig:energy_eff}.


\juann{First, the full-blown \texttt{HBM+OCAPI} designs (i.e., 14 PEs for \texttt{vadvc} and 16 PEs for \texttt{hdiff}) achieve energy efficiency values of 1.61~GFLOPS/Watt and 21.01~GFLOPS/Watt for \texttt{vadvc} and \texttt{hdiff}, respectively. 
These represent improvements of $12\times$ and $35\times$ compared to the IBM POWER9 system for \texttt{vadvc} and \texttt{hdiff}, respectively.}


\juann{Second, the \texttt{DDR4-CAPI2} designs for \texttt{vadvc} and \texttt{hdiff} are slightly more energy-efficient ($1.1\times$ to $1.5\times$) than the \texttt{HBM-CAPI2} designs when the number of PEs is small. 
This observation is in line with our discussion about performance with small PE counts in Section~\ref{subsection:evaluation}. 
However, as we increase the number of PEs, the \texttt{HBM-CAPI2} designs provide higher energy efficiency since they make use of multiple HBM channels.}

\juann{Third, the energy efficiency of the HBM-based designs (\texttt{HBM+CAPI2}, \texttt{HBM+OCAPI}) for \texttt{hdiff} increases with the number of PEs until a saturation point (8 PEs). 
This trend is because every additional HBM channel increases power consumption by $\sim$1 Watt (for the HBM AXI3 interface operating at 250MHz with a logic toggle rate of $\sim$12.5\%).}

\juann{Fourth, \texttt{HBM+OCAPI}, \texttt{HBM+CAPI2}, and \texttt{DDR4+CAPI2} versions of \texttt{vadvc} achieve their highest energy efficiency at a number of PEs that is smaller than the maximum possible. 
There is a large amount of control flow in \texttt{vadvc}, which leads to large resource utilization. As a result, as shown in Figure~\ref{fig:power_consumption}, increasing the PE count increases power consumption dramatically, causing lower energy efficiency.}


\juann{Fifth, the multi-channel-single PE designs (\texttt{HBM\_multi+OCAPI}) are more energy-efficient than the single-channel-single PE designs (\texttt{HBM+OCAPI}) for the same number of PEs. 
However, \texttt{HBM+OCAPI} designs achieve higher energy efficiency for higher numbers of PEs, which are not affordable for \texttt{HBM\_multi+OCAPI} designs.}



\subsection{\gagann{FPGA Resource Utilization}}
Table~\ref{tab:utilization} shows the \gagan{resource} utilization \gagan{of \texttt{vadvc}} and \texttt{hdiff} on the AD9H7 board. \gagann{We draw two observations. First, there is a high BRAM consumption compared to other FPGA resources. This is because we implement input, field, and output signals as \texttt{hls::streams}. In \gagannn{high-level synthesis}, by default, streams are implemented as FIFOs that make use of BRAM. Second,  \texttt{vadvc} has a much larger resource consumption  than \texttt{hdiff} because \texttt{vadvc} has higher computational complexity and requires \gagannn{a larger} number of fields to \gagan{perform} the \gagan{compound} stencil calculation. \gagan{Note that for} \texttt{hdiff}, we can \gagan{accommodate} more {PE}s, but in this work, we  make use of only \gagan{a} single HBM stack. Therefore\gagan{,} we \gagan{use} 16 {PE}s \gagan{because} a single \gagan{HBM} stack offers} \juann{up to 16 memory channels.}

\input{tables/utilization}

%% file: tables/system.tex
  \vspace{0.2cm}
\begin{table}[h]
  \caption{\gagan{S}ystem parameters and hardware configuration for the CPU and the FPGA board.}
  \vspace{-0.4cm}
    \label{tab:systemparameters}
      \begin{center}
      \small
\resizebox{0.65\linewidth}{!}{%
\begin{tabular}{|=l@{\hspace{0.1\tabcolsep}}|=p{5.8cm}|}

\hline
\textbf{Host CPU} 

 & 16-core IBM POWER9 AC922~\cite{POWER9} \\&@3.2 GHz, 4\gagann{-way} SMT~\cite{smt}\\
 \hline
 \textbf{Cache-Hierarchy}&16$\times$32 KiB L1-I/D, 256 KiB L2, 10 MiB L3 \\
 \hline
 \textbf{System Memory}&32GiB RDIMM DDR4 2666 MHz~\cite{rdimm} \\
\hline

 \begin{tabular}[c]{@{}l@{}}
 \textbf{HBM-based} \\\textbf{FPGA Board} \end{tabular}  &
  \begin{tabular}[c]{@{}l@{}}
 Alpha Data
ADM-PCIE-9H7~\cite{ad9h7}\\
 Xilinx Virtex Ultrascale+ XCVU37P-2~\cite{vu37p}\\
  8GiB (HBM2~\cite{hbm}) with PCIe Gen4 x8~\cite{pcie}\\
 \end{tabular}\\
\hline
 \begin{tabular}[c]{@{}l@{}}
 \textbf{DDR4-based} \\\textbf{FPGA Board} \end{tabular}  & \begin{tabular}[c]{@{}l@{}}
 Alpha Data
ADM-PCIE-9V3~\cite{ad9v3}\\
 Xilinx Virtex Ultrascale+ XCVU3P-2~\cite{vu37p}\\
 8GiB (DDR4) with PCIe Gen4 x8~\cite{pcie}
  \end{tabular}\\

\hline
 \textbf{OS details}& Ubuntu 20.04.3 LTS~\cite{ubuntu}, GNU Compiler Collection (GCC) version 9.3.0~\cite{gnu}, IBM XL C/C++ 16~\cite{ibmxlc}\\
 \hline
\end{tabular}
}
  \end{center}
 \end{table}

%% file: tables/utilization.tex
  \vspace{0.2cm}
\begin{table}[h]
  \caption{FPGA resource utilization \gagannn{in} \gagan{our highest-performing HBM-based design\gagannn{s for}} \texttt{vadvc} and \texttt{hdiff}. }
  \vspace{-0.4cm}
    \label{tab:utilization}
          \begin{center}
\resizebox{0.5\linewidth}{!}{%
\begin{tabular}{llllllc}
\hline
\textbf{Algorithm} & \textbf{BRAM} & \textbf{DSP} & \textbf{FF} & \textbf{LUT} & \textbf{URAM} \\ \hline
\texttt{vadvc}              & 94\%            & 39\%           & 37\%          & 55\%           & 53\%                         \\ 
\texttt{hdiff}              & 96\%            & 4\%            & 10\%           & 15\%           & 8\%                           \\ \hline
\end{tabular}
}
  \end{center}
\vspace{-0.2cm}
\end{table}

%% file: sections/takeaways.tex
\gtrets{
\section{Discussion and Key Takeaways}
\label{sec:discussion}
A wide range of application domains have emerged with the ubiquity of computing platforms in every aspect of our daily lives. These modern workloads (e.g., machine
learning, graph processing, and bioinformatics) demand high compute capabilities within strict power constraints~\cite{ghose2019processing}. However, today's computing systems are getting constrained by current technological capabilities, making them incapable of delivering the required performance. \juan{This paper presents our \gonur{recent} efforts to leverage \gonur{near-memory computing capable FPGA-based accelerators} \gagan{to accelerate} two major kernels from the weather prediction application in an energy-efficient way. We summarize the most important insights and takeaways as follows.}

\juan{First, our evaluation shows that \gonur{High-Bandwidth Memory}-based \gonur{near-memory} \gagann{FPGA} accelerator designs can improve performance by \gonur{ 5.3$\times$-12.7$\times$ and energy efficiency by 12$\times$-35$\times$ over} a \gcamera{single-socket} high-end \gagann{16-core} \gonur{IBM} POWER9 CPU.}

\juan{Second, our HBM-based \gagann{FPGA accelerator} designs employ a dedicated HBM channel per PE. Such a design avoids memory access congestion, which is typical in DDR4-based FPGA designs, and ensures memory bandwidth scaling with the number of PEs. As a result, \gagann{in most of the data-parallel applications,} performance scales linearly with the number of PEs. \gtrets{Therefore, HBM provides an attractive solution for scale-out computation.}}

Third, the data needs to be adequately mapped to each HBM channel's address space.   A data mapping scheme should map data in such a way that the data required by the processing unit is readily available in the vicinity (data and code co-location).  An inefficient data mapping mechanism can severely hamper the benefits of processing close to memory. 

Fourth, we make use of OCAPI in a coarse-grained way, since we offload the entire application to the FPGA. In this case, OCAPI ensures that the FPGA accelerators access the entire CPU memory with the minimum number of memory copies between the host and the FPGA, e.g., avoiding the intermediate buffer copies that a traditional PCIe-based DMA invokes~\cite{10.1145/2897937.2897972}. However, depending on the application, the CAPI protocol can be employed in finer-grained algorithm-hardware co-design, like the \textit{ExtraV}  \cite{10.14778/3137765.3137776}, where the authors aggressively utilize the fine-grained communication capability of OCAPI to boost graph analytics performance.

\juan{Fifth, \sr{the} maximum performance of our HBM-based design is \gonur{reached using} the maximum PE count that we can fit in the reconfigurable fabric, with each PE having a dedicated HBM channel.} \grev{ However, adding more PEs could lead to timing constraint violations for HBM-based designs. As shown with our multi-channel setting (Section~\ref{subsection:evaluation}), where we can fit only 3 PEs for both \vadvc and \hdiff, enabling more HBM channels leads to timing constraint violations.} HBM-based FPGAs consist of multiple super-logic regions (SLRs)~\garxiv{\cite{hbm_slr}}, where an SLR represents a single FPGA die. All HBM channels are connected only to SLR0, while other SLRs have indirect connections to the HBM channels. Therefore, if a PE is implemented in a non-SLR0 region for a large design, it might make timing closure difficult. A possible way to alleviate timing issues is by running the AFU at a lower frequency, which eases the place and route.

\gtrets{
Sixth, the energy efficiency of our HBM-based designs tends to saturate (or even \gonur{reduces}) as we increase the number of PEs \gonur{beyond some point}. The highest energy efficiency is achieved with a PE count that is smaller than the highest-performing PE count. The \gonur{major} reason \gonur{for a decrease in the energy efficiency is the increase in power consumption with} every additional HBM channel.

Seventh, the emerging cache-coherent  interconnects standards like CXL~\cite{sharma2019compute}, CCIX~\cite{benton2017ccix}, and OCAPI~\cite{openCAPI} could be vital in improving the performance and energy efficiency of big data workloads running on FPGA-based devices because they avoid having multiple data copies. However, a very \om{small number} of works, such as \cite{10.14778/3137765.3137776}, leverage the \emph{coherency} aspect of these interconnects. More quantitative exploration is required to analyze the advantages and disadvantages of using these interconnects.}

Eighth, we are witnessing an enormous amount of data being generated across multiple application domains~\cite{nair2015active,singh2021modeling} like weather prediction modeling, radio astronomy, bioinformatics, material science, chemistry, health sciences, etc. The processing of the sheer amount of generated data is one of the biggest challenges to overcome. In this paper, we demonstrate the capabilities of near-memory reconfigurable accelerators in the domain of weather prediction, however, there are many other high-performance computing applications  where such near-memory architectures can alleviate the data movement bottleneck.}

%% file: sections/relatedWork.tex
\section{Related Work}
\label{sec:relatedWork}
\juang{To our knowledge, this is the first work to evaluate the benefits of using FPGAs equipped with \gagann{high-bandwidth memory (HBM) to accelerate} {real-world weather \om{modeling} stencils.}} \gagannn{We exploit the near-memory capabilities of such FPGAs to accelerate important weather prediction kernels.} \om{Exploiting the high-bandwidth memory in FPGAs,} we answer the following questions with our work. First, do real-world weather prediction applications benefit from HBM-based FPGAs? Second, how can we scale the processing in terms of not only run-time but also energy efficiency? Third, what does the system look like regarding computation and data movement with an HBM-enabled FPGA when integrating accelerators for real-world weather prediction workloads? 

\gagannn{Modern workloads exhibit limited locality and operate on large amounts of data, which causes frequent data movement between the memory subsystem and the processing units~\cite{ghose2019processing,mutlu2019,googleWorkloads,mutlu2019enabling,mutlu2021intelligent,boroumand2021google,mutlu2020modern}.  This frequent data movement has a severe impact on overall system performance and energy efficiency. \gtrets{ For example, in the domain of climate and weather modeling,  there is a data avalanche due to large atmospheric simulations~\cite{schar2020kilometer}. Major efforts are currently underway towards refining the resolution grid of climate models that would generate \emph{zettabytes} of data~\cite{schar2020kilometer}. These high-resolution simulations are useful to predict and address events like severe storms. However, the sheer amount of generated data is one of the biggest challenges to overcome. We find another relevant example in radio astronomy. The first phase of the Square Kilometre Array (SKA) aims to process over 100 terabytes of raw data samples per second, yielding of the order of 300 petabytes of SKA data produced annually~\cite{6898703,singh2018review}. { Recent biological disciplines such as genomics have also emerged as one of the most data-intensive workloads across all different sciences wherein just a single human genome sequence produces hundreds of gigabytes of raw data. With the rapid advancement in sequencing technology, the data volume in genomics is projected to surpass the data volume in all other application domains~\cite{navarro2019genomics}.}  }

A way to alleviate this \emph{data movement bottleneck}~\cite{singh2019near, mutlu2019, ghose2019processing,mutlu2019enabling,googleWorkloads,mutlu2020modern,hajinazar2021simdram,seshadri2019dram,mutlu2021intelligent,boroumand2021google,mutlu2020modern} is \emph{near-memory computing} (NMC), which consists of placing processing units closer to memory. 
NMC is enabled by new memory technologies, such as 3D-stacked memories~\cite{7477494,6757501,6025219,hbm,lee2016smla,fimdram2021ISSCC,fimdram2021ISCA,mutlu2021intelligent,boroumand2021google,mutlu2020modern}, and also by cache-coherent interconnects~\cite{openCAPI, benton2017ccix, sharma2019compute},  which allow close integration of processing units and memory units. 
Depending on the applications \gagannn{and systems} of interest (e.g.,~\cite{nai2017graphpim,7056040,lee2018application,kang2013enabling,hashemi2016continuous,akin2015data,babarinsa2015jafar,lee2015bssync,chi2016prime,kim2016neurocube,asghari2016chameleon,boroumand2016lazypim,seshadri2015gather,liu2017concurrent,gao2015practical,morad2015gp,googleWorkloads,teserract,ahn2015pim,hsieh2016accelerating,hashemi2016accelerating,mutlu2020modern,senolcalimicro2020,oliveira2021pimbench,upmem2021,ke2020recnmp,fernandez2020natsa,ConTutto_2017_MICRO,simon2020blade,giannoula2021syncron,gao2019computedram,seshadri2017simple,angizi2019aligns,mutlu2021intelligent,boroumand2021google,mutlu2020modern,gomez2021benchmarking,seshadri2017ambit,seshadri2015fast,li2016pinatubo,seshadri2016buddy,besta2021sisa,park2021trim,wu2021sieve}), prior works propose different types of near-memory processing units, such as general-purpose CPU cores~\cite{lee2018application,alian2018application,de2017mondrian,koo2017summarizer,7927081,teserract,nair2015active,6844483,googleWorkloads,boroumand2016lazypim,boroumand2019conda,gu2016biscuit,liu2018processing,oliveira2021pimbench,upmem2021}, GPU cores~\cite{zhang2014top,7756764,7551394,ghose2019demystifying}, reconfigurable units~\cite{7446059, jun2015bluedbm,istvan2017caribou,narmada}, or fixed-function units~\cite{ahn2015pim,hsieh2016accelerating,gu2016biscuit,nai2017graphpim,liu2018processing,kim2018grim,hashemi2016accelerating,hashemi2016continuous,boroumand2021google}}.

FPGA accelerators are promising to enhance overall system performance \gagann{with low power consumption.}
Past works~\cite{jun2015bluedbm, kara2017fpga, alser2019shouji, giefers2015accelerating,8373077,10.14778/3137765.3137776,alser2017,chai_icpe19,chang2017collaborative,jiang2020,alser2019sneakysnake,alser2020accelerating,alser2020technology} show that FPGAs can be employed {effectively} for a wide range of applications. \gtrets{FPGAs provide a unique combination of flexibility and performance without the cost, complexity, and risk of developing custom application-specific integrated circuits (ASICs). 
The researchers at CERN, for example, are using FPGAs to accelerate physics workload in CERN's \dd{exploration} of dark matter~\cite{duarte2018fast}. Microsoft's Project Catapult
~\cite{caulfield2016cloud} is another example of how FPGAs can be used in the data center infrastructure. \dd{Driven by Catapult's promising research results, Microsoft further deployed the architecture on the Azure cloud marketplace~\cite{FPGA_in_azure}. } Such integration {for certain workloads} can even offer more energy efficiency than CPU or GPU-based systems.} The recent addition of HBM to \juang{FPGAs} 
\juang{presents an opportunity to exploit} 
high \juang{memory} bandwidth with \gagannn{the} low-power FPGA fabric. 
The potential of high-bandwidth memory~\cite{hbm,lee2016smla} has been explored in many-core processors~\cite{hbm_joins,ghose2019demystifying} and GPUs~\cite{hbm_gpu_data_intensive,ghose2019demystifying}. 
\juang{Recent benchmarking works~\cite{wang2020, kara2020hbm} show the potential of HBM for FPGAs.}

\juang{\namePaper} is the first work to accelerate a real-world HPC weather prediction application using \juang{the} FPGA+HBM fabric. {Compared to a previous work~\cite{narmada} that \juang{optimizes only the} horizontal diffusion kernel \juang{on an FPGA with DDR4 memory}, our analysis reveals \juang{that the} vertical advection kernel has a much lower compute intensity with little to no regularity.
Therefore, our work \juang{accelerates} both kernels that together represent the algorithmic diversity of the entire COSMO \gagannn{weather prediction} model.
\grev{Our current work differs \om{from~\cite{singh2020nero}} in the following aspects. First, we design and evaluate both horizontal diffusion and vertical advection stencils. Vertical advection  is the most complex stencil in the entire COSMO application. Second, we integrate and implement our accelerator design with an HBM-based FPGA. The bus width of the DDR4 channel (512 bits) is larger than that of an HBM channel (256 bits). Therefore, the HBM channel has a lower transfer rate of 0.8-2.1 GT/s (Gigatransfers per second) than a DDR4 channel (2.1-4.3 GT/s), resulting in a theoretical bandwidth of 12.8 GB/s and 25.6 GB/s per channel, respectively. However, HBM exposes 32 memory channels that provide 4x more bandwidth (410 GB/s for HBM~\cite{kara2020hbm}) compared to traditional DDR4 bandwidth (72 GB/s for four independent dual-rank DIMM interfaces~\cite{VCU}). Therefore, the use of HBM imposes an architectural shift. We evaluate and demonstrate the use of HBM for scaling an accelerator design with different channels provided by HBM. Third, we use an auto-tuning framework to find the right window size (Figure 6) that demonstrates the importance of finding the right window size. Fourth, we provide new results using a state-of-the-art OpenCAPI (OCAPI) interface with the OC-Accel framework. OCAPI provides two key opportunities compared to our previous CAPI2 implementation: (1) OCAPI has double the bitwidth of our previously used CAPI2 interface, (2) a major component of the memory coherency logic is moved to the host CPU side, which provides more FPGA area and enables designs with higher frequency. Due to the above optimizations, we improve the performance for horizontal diffusion by 1.2x on a DDR4-based board and 4.7x on an HBM-based board compared to our previous work NARMADA~\cite{narmada}. }}


\gagan{Enabling higher performance for stencil computations has been a subject of optimizations 
across the \juang{whole} computing stack~\cite{sano2014multi,7582502,chi2018soda,de2018designing,christen2011patus,  datta2009optimization, meng2011performance, henretty2011data,fu2011eliminating, strzodka2010cache,zohouri2018combined,waidyasooriya2019multi, tang2011pochoir,gonzalez1997speculative,armejach2018stencil,gysi2015modesto}}. \gtrets{Stencil computation is essential for numerical simulations of finite difference methods~(FDM)~\cite{ozicsik2017finite} and is applied in iterative solvers of linear equation systems. We use stencil computation in a wide range of applications, including computational fluid dynamics~\cite{huynh2014high}, image processing~\cite{hermosilla2008non}, weather prediction modeling~\cite{doms1999nonhydrostatic}, etc.}

\grev{
Unlike stencils found in the literature~\cite{waidyasooriya2019multi,singh2019low,de2021stencilflow,sano2014multi,7582502,chi2018soda,de2018designing}, real-world compound stencils consist of a collection of stencils that perform a sequence of element-wise computations with complex interdependencies. Such compound kernels have complex memory access patterns and low arithmetic intensity because they have limited operations per loaded value. Our work is the first work to accelerate both horizontal diffusion and vertical advection stencils, which are representative of data access patterns and the algorithmic complexity found in the entire COSMO weather model.  }

Table~\ref{tab:compare_works} lists recent works (including \namePaper) that use FPGA to accelerate stencil-based application. We also mention works that accelerate elementary stencils (\texttt{7-point}, \texttt{25-point} Jacobi, \texttt{Hotspot}, and \texttt{Diffusion}). \grev{We make the following three observations. First, the elementary stencils can achieve much higher performance on comparable FPGA devices than complex weather stencils (such as \hdiff) even without using HBM. This high performance is because elementary stencils have a higher arithmetic intensity than weather stencils. Due to their data-parallel nature, these elementary stencils can further take advantage of the increased bandwidth provided by HBM in an energy-efficient way. Second, weather stencils can reach only 2\%-17\% of the peak theoretical performance of an FPGA board.  This low peak performance is because weather stencils have several elementary stencils cascaded together with data interdependencies that lead to complex memory access patterns. Third, compared to NARMADA~\cite{narmada}, which uses a DDR4-based design, our HBM-based design achieves $4.7\times$ performance improvement by exploiting the high bandwidth provided by the HBM. }

\input{tables/relatedWork}

\juang{Szustak~\etal} accelerate the MPDATA advection scheme on multi-core CPU~\cite{szustak2013using} and computational fluid dynamics kernels on FPGA~\cite{mpdata}. \gtrets{
Singh~\etal~\cite{singh2019low} explore the applicability of different number formats and exhaustively search for the appropriate bit-width for memory-bound stencil kernels to improve performance and energy efficiency with minimal loss in the accuracy.}  Bianco~\etal~\cite{bianco2013gpu} \juan{optimize the COSMO \gagannn{weather prediction} model} for GPUs.  \grev{ Thaler~\etal~\cite{cosmo_knl}, in a collaboration work between the Swiss National Supercomputing Centre (CSCS) and the Federal Institute of Meteorology and Climatology (MeteoSwiss),  discuss the importance of horizontal diffusion and vertical advection kernels in the entire COSMO model. These kernels together represent the algorithmic diversity of the entire COSMO weather prediction model~\cite{cosmo_knl,gysi2015modesto,bianco2013gpu}. They {port} COSMO to a many-core system.} \grev{Compared to their Intel KNL~\om{\cite{intelknl}} (or NVIDIA P100~\om{\cite{nvidiap100}}) implementation, we observe that our FPGA-based \vadvc and \hdiff design provides $1.5\times$ (or $1.4\times$) and $3.2\times$ (or $2.1\times$) performance improvements, respectively. } Several works~\cite{de2021stencilflow,lai2019heterocl, li2020heterohalide,wang2017comprehensive} propose frameworks for generating optimized stencil code for FPGA-based platforms. 
Wahib~\etal~\cite{wahib2014scalable} \juang{develop} an analytical performance model for choosing an optimal GPU-based execution strategy for various scientific applications, including COSMO. 
Gysi~\etal~\cite{gysi2015modesto} provide guidelines for optimizing stencil kernels for CPU--GPU systems.


%% file: tables/relatedWork.tex
\begin{table}[h]
  \caption{Overview of the state-of-art stencil implementations on FPGAs. \grev{For each work, we mention the technology node (Tech. node), DRAM memory technology (Mem. Tech.), theoretical peak floating-point  performance (Peak Perf. (TFLOPS)), available peak memory bandwidth (Peak B/W (GB/s)), frequency of the accelerator logic (Freq. (MHz)), overall logic utilization (Logic Util.), overall memory utilization (Mem. Util.), achieved performance (Perf. (GOp/s)), and the percentage of achieved peak roofline performance (Ach. Roof.). }  }
  \vspace{-0.4cm}
    \label{tab:compare_works}
      \begin{center}
      \small
\resizebox{1\linewidth}{!}{%
\begin{tabular}{c|c|c|l|l|l|l|l|l|l|l|l|l}
\textbf{Stencil} & Work & Year & Device & Tech. node & Mem. Tech. & Peak Perf. (TFLOPS) & Peak B/W (GB/s)   & Freq. (MHz)& Logic Util. & Mem. Util. & Perf. (GOp/s) & Ach. Roof.  \\
\hline
\texttt{Diffusion 3D} &\cite{waidyasooriya2019multi}  & 2019 & Arria 10 & TSMC 20nm   &DDR3 &1.4 &34& 276& 32\% & 47\% & 628.0  &44.9\%\\
\texttt{Hotspot 3D} &\cite{waidyasooriya2019multi}  & 2019 & Arria 10  & TSMC 20nm &DDR3 & 1.4& 34& 240& 34\% & 81\% & 630 &45.0\% \\
\texttt{7-point 3D} &\cite{singh2019low} & 2019 & XCVU3P & TSMC 16FF+ & DDR4&0.97 &25.6 &180& 23.5\% & 39\% & 228.4& 23.7\% \\  %
\texttt{25-point 3D } & \cite{singh2019low}  & 2019 & XCVU3P & TSMC 16FF+ &DDR4 &0.97 & 25.6&190 & 49\% & 39\% & 327.7 & 34.1\%  \\ %
\texttt{3D Jacobi} &\cite{de2021stencilflow}  & 2021 & Stratix 10 & Intel 14nm FinFet &DDR4& 9.2 &76.8&292-317  &  - & - & 568.2 & 6.2\% \\

\hline
\hdiff & \cite{narmada} & 2019 & XCVU3P & TSMC 16FF+ & DDR4&0.97 &25.6 &200 & 64.5\% & 64.1\% & 129.9& 13.3\% \\%
\hdiff & \cite{de2021stencilflow} & 2021 &  Stratix 10 & Intel 14nm FinFet & DDR4& 9.2 &76.8&292-317 & 26.0\% & 20\% & 145.0 [513.0$^\dagger$] & 1.6\% [5.5\%] \\ %

\hdiff & [Ours] & 2021 & XCVU37P & TSMC 16FF+ & HBM & 3.6 & 204.8$^\mathsection$ [410] & 250& 12.5\% & 52\% & 608.4 & 16.9\%  \\%

\hline
\end{tabular}
}
\end{center}
{\footnotesize$^\dagger$ When simulated using an infinite memory bandwidth. \quad \quad
\footnotesize$^\mathsection$ Note that we use only a single HBM stack due to resource limitations.}
\end{table}


%% file: sections/conclusion.tex
\newpage
\section{Conclusion}
\label{sec:conclusion}
\juang{We introduce} \namePaper, the first design and implementation on a reconfigurable fabric \juang{with \gagann{high-bandwidth memory} (HBM)} to accelerate representative weather prediction kernels, i.e., vertical advection (\texttt{vadvc}) and horizontal diffusion (\texttt{hdiff}), from a real-world weather prediction application. 
These kernels are compound stencils that are found in various weather prediction applications, including the COSMO  model. \gagannn{We show that c}ompound kernels do not perform well on conventional architectures due to their complex data access patterns and low data reusability, which \gagann{make} them memory-bounded. 
\juang{Therefore, they greatly benefit from our near-memory computing solution \gagannn{that} takes advantage of the high \gagann{data transfer} bandwidth of HBM.} \gtrets{We use a heterogeneous system comprising of IBM POWER9 CPU with field-programmable gate array (FPGA) as our target platform. 
We create a heterogeneous domain-specific memory hierarchy using on-chip {URAM}s and  {BRAM}s, \gcamera{and on-package HBM} on an FPGA. Unlike \om{conventional} fixed CPU memory hierarchies, which perform poorly with irregular access patterns and suffer from cache pollution effects, application-specific memory hierarchies are shown to improve \om{both} energy and latency by tailoring the cache levels and cache sizes to an application's memory access patterns.  }


\juang{\namePaper's implementations of \texttt{vadvc} and \texttt{hdiff} outperform the optimized software implementations on a 16-core POWER9 with \gagann{4-way multithreading} by $5.3\times$ and $12.7\times$, with $12\times$ and $35\times$ less energy consumption, respectively. 
We conclude that hardware acceleration on \gagann{an} FPGA+HBM fabric is a promising solution for compound stencil\gagann{s} present in weather prediction applications. \gagann{We hope that o}ur \gagann{reconfigurable near\gagannn{-}memory} accelerator inspire\gagann{s} developers of different high-performance computing applications that suffer from \gagann{the} memory bottleneck.
}

%% file: ms.bbl

\begin{thebibliography}{179}


\ifx \showCODEN    \undefined \def \showCODEN     #1{\unskip}     \fi
\ifx \showDOI      \undefined \def \showDOI       #1{#1}\fi
\ifx \showISBNx    \undefined \def \showISBNx     #1{\unskip}     \fi
\ifx \showISBNxiii \undefined \def \showISBNxiii  #1{\unskip}     \fi
\ifx \showISSN     \undefined \def \showISSN      #1{\unskip}     \fi
\ifx \showLCCN     \undefined \def \showLCCN      #1{\unskip}     \fi
\ifx \shownote     \undefined \def \shownote      #1{#1}          \fi
\ifx \showarticletitle \undefined \def \showarticletitle #1{#1}   \fi
\ifx \showURL      \undefined \def \showURL       {\relax}        \fi
\providecommand\bibfield[2]{#2}
\providecommand\bibinfo[2]{#2}
\providecommand\natexlab[1]{#1}
\providecommand\showeprint[2][]{arXiv:#2}

\bibitem[\protect\citeauthoryear{??}{ad9}{[n.d.]a}]%
        {ad9h7}
\bibinfo{booktitle}{\emph{{ADM-PCIE-9H7}-{H}igh-{S}peed {C}ommunications {H}ub,
  \url{https://www.alpha-data.com/dcp/products.php?product=adm-pcie-9h7}}}.
\newblock


\bibitem[\protect\citeauthoryear{??}{ad9}{[n.d.]b}]%
        {ad9v3}
\bibinfo{booktitle}{\emph{{ADM-PCIE-9V3}-{H}igh-{P}erformance {N}etwork
  {A}ccelerator,
  \url{https://www.alpha-data.com/dcp/products.php?product=adm-pcie-9v3}}}.
\newblock


\bibitem[\protect\citeauthoryear{??}{axi}{[n.d.]a}]%
        {axi_hbm}
\bibinfo{booktitle}{\emph{{AXI} {H}igh {B}andwidth {M}emory {C}ontroller v1.0,
  \url{https://www.xilinx.com/support/documentation/ip_documentation/hbm/v1_0/pg276-axi-hbm.pdf}}}.
\newblock


\bibitem[\protect\citeauthoryear{??}{axi}{[n.d.]b}]%
        {axilite}
\bibinfo{booktitle}{\emph{{AXI} {R}eference {G}uide,
  \url{https://www.xilinx.com/support/documentation/ip_documentation/ug761_axi_reference_guide.pdf}}}.
\newblock


\bibitem[\protect\citeauthoryear{??}{cen}{[n.d.]}]%
        {centos}
\bibinfo{booktitle}{\emph{{CentOS-7 (2009) Release Notes,
  \url{https://wiki.centos.org/Manuals/ReleaseNotes/CentOS7.2009} }}}.
\newblock


\bibitem[\protect\citeauthoryear{??}{gnu}{[n.d.]}]%
        {gnu}
\bibinfo{booktitle}{\emph{{GCC, the GNU Compiler Collection,
  \url{https://gcc.gnu.org/} }}}.
\newblock


\bibitem[\protect\citeauthoryear{??}{hbm}{[n.d.]a}]%
        {hbm}
\bibinfo{booktitle}{\emph{{H}igh {B}andwidth {M}emory {(HBM) DRAM} ({JESD}235),
  \url{https://www.jedec.org/document_search?search_api_views_fulltext=jesd235}}}.
\newblock


\bibitem[\protect\citeauthoryear{??}{hbm}{[n.d.]b}]%
        {hbm_specs}
\bibinfo{booktitle}{\emph{High {B}andwidth {M}emory ({HBM}) {DRAM},
  \url{https://www.jedec.org/sites/default/files/JESD235B-HBM_Ballout.zip}}}.
\newblock


\bibitem[\protect\citeauthoryear{??}{ibm}{[n.d.]}]%
        {ibmxlc}
\bibinfo{booktitle}{\emph{{IBM XL C/C++ for Linux},
  \url{https://www.ibm.com/products/xl-cpp-linux-compiler-power}}}.
\newblock


\bibitem[\protect\citeauthoryear{??}{int}{[n.d.]a}]%
        {intel_altera}
\bibinfo{booktitle}{\emph{{I}ntel {S}tratix 10 {MX FPGA}s,
  \url{https://www.intel.com/content/www/us/en/products/programmable/sip/stratix-10-mx.html}}}.
\newblock


\bibitem[\protect\citeauthoryear{??}{int}{[n.d.]b}]%
        {intelknl}
\bibinfo{booktitle}{\emph{{Intel® Xeon Phi™ Processor 7230 (16GB, 1.30 GHz,
  64 core)},
  \url{https://www.intel.com/content/www/us/en/products/sku/94034/intel-xeon-phi-processor-7230-16gb-1-30-ghz-64-core/specifications.html}}}.
\newblock


\bibitem[\protect\citeauthoryear{??}{nvi}{[n.d.]}]%
        {nvidiap100}
\bibinfo{booktitle}{\emph{{NVIDIA® TESLA® P100 GPU ACCELERATOR},
  \url{https://images.nvidia.com/content/tesla/pdf/nvidia-tesla-p100-PCIe-datasheet.pdf}}}.
\newblock


\bibitem[\protect\citeauthoryear{??}{oca}{[n.d.]}]%
        {ocaccel}
\bibinfo{booktitle}{\emph{{OC-Accel,
  \url{https://opencapi.github.io/oc-accel-doc/} }}}.
\newblock


\bibitem[\protect\citeauthoryear{??}{ope}{[n.d.]}]%
        {open_power}
\bibinfo{booktitle}{\emph{{O}pen{POWER} {W}ork {G}roups,
  \url{https://openpowerfoundation.org/technical/working-groups}}}.
\newblock


\bibitem[\protect\citeauthoryear{??}{rdi}{[n.d.]}]%
        {rdimm}
\bibinfo{booktitle}{\emph{{RDIMM},
  \url{https://www.micron.com/products/dram-modules/rdimm}}}.
\newblock


\bibitem[\protect\citeauthoryear{??}{ubu}{[n.d.]}]%
        {ubuntu}
\bibinfo{booktitle}{\emph{{Ubuntu 20.04.3 LTS (Focal Fossa)},
  \url{https://releases.ubuntu.com/20.04/}}}.
\newblock


\bibitem[\protect\citeauthoryear{??}{ura}{[n.d.]}]%
        {uram}
\bibinfo{booktitle}{\emph{{UltraScale Architecture Memory Resources},
  \url{https://www.xilinx.com/support/documentation/user_guides/ug573-ultrascale-memory-resources.pdf}}}.
\newblock


\bibitem[\protect\citeauthoryear{??}{hbm}{[n.d.]c}]%
        {hbm_slr}
\bibinfo{booktitle}{\emph{{V}irtex {U}ltra{S}cale+ {HBM FPGA}: {A}
  {R}evolutionary {I}ncrease in {M}emory {P}erformance,
  \url{https://www.xilinx.com/support/documentation/white_papers/wp485-hbm.pdf}}}.
\newblock


\bibitem[\protect\citeauthoryear{??}{vu3}{[n.d.]}]%
        {vu37p}
\bibinfo{booktitle}{\emph{{V}irtex {U}ltra{S}cale+,
  \url{https://www.xilinx.com/products/silicon-devices/fpga/virtex-ultrascale-plus.html}}}.
\newblock


\bibitem[\protect\citeauthoryear{??}{hls}{[n.d.]}]%
        {hls}
\bibinfo{booktitle}{\emph{{V}ivado {H}igh-{L}evel {S}ynthesis,
  \url{https://www.xilinx.com/products/design-tools/vivado/integration/esl-design.html}}}.
\newblock


\bibitem[\protect\citeauthoryear{??}{VCU}{[n.d.]}]%
        {VCU}
\bibinfo{booktitle}{\emph{Xilinx {VCU}1525,
  \url{https://www.xilinx.com/products/boards-and-kits/ vcu1525-a.html}}}.
\newblock


\bibitem[\protect\citeauthoryear{??}{xil}{[n.d.]}]%
        {xilinx_utlra}
\bibinfo{booktitle}{\emph{Xilinx {V}irtex {U}ltraScale+,
  \url{https://www.xilinx.com/products/silicon-devices/fpga/virtex-ultrascale-plus.html}}}.
\newblock


\bibitem[\protect\citeauthoryear{??}{viv}{[n.d.]}]%
        {vivado}
\bibinfo{booktitle}{\emph{Xilinx Vivado,
  \url{https://www.xilinx.com/support/download.html}}}.
\newblock


\bibitem[\protect\citeauthoryear{Ahn, Hong, Yoo, Mutlu, and Choi}{Ahn
  et~al\mbox{.}}{2015a}]%
        {teserract}
\bibfield{author}{\bibinfo{person}{Junwhan Ahn}, \bibinfo{person}{Sungpack
  Hong}, \bibinfo{person}{Sungjoo Yoo}, \bibinfo{person}{Onur Mutlu}, {and}
  \bibinfo{person}{Kiyoung Choi}.}
\newblock \showarticletitle{{A Scalable Processing-in-Memory Accelerator for
  Parallel Graph Processing}}. In \bibinfo{booktitle}{\emph{ISCA}}
  \bibinfo{year}{2015}\natexlab{a}.
\newblock


\bibitem[\protect\citeauthoryear{Ahn, Yoo, Mutlu, and Choi}{Ahn
  et~al\mbox{.}}{2015b}]%
        {ahn2015pim}
\bibfield{author}{\bibinfo{person}{Junwhan Ahn}, \bibinfo{person}{Sungjoo Yoo},
  \bibinfo{person}{Onur Mutlu}, {and} \bibinfo{person}{Kiyoung Choi}.}
\newblock \showarticletitle{{PIM-Enabled Instructions: A Low-Overhead,
  Locality-Aware Processing-in-Memory Architecture}}. In
  \bibinfo{booktitle}{\emph{ISCA}} \bibinfo{year}{2015}\natexlab{b}.
\newblock


\bibitem[\protect\citeauthoryear{Akin, Franchetti, and Hoe}{Akin
  et~al\mbox{.}}{2015}]%
        {akin2015data}
\bibfield{author}{\bibinfo{person}{Berkin Akin}, \bibinfo{person}{Franz
  Franchetti}, {and} \bibinfo{person}{James~C Hoe}.}
\newblock \showarticletitle{{Data Reorganization in Memory Using 3D-stacked
  DRAM}}. In \bibinfo{booktitle}{\emph{ISCA}} \bibinfo{year}{2015}\natexlab{}.
\newblock


\bibitem[\protect\citeauthoryear{{Alian}, {Min}, {Asgharimoghaddam}, {Dhar},
  {Wang}, {Roewer}, {McPadden}, {O'Halloran}, {Chen}, {Xiong}, {Kim}, {Hwu},
  and {Kim}}{{Alian} et~al\mbox{.}}{2018}]%
        {alian2018application}
\bibfield{author}{\bibinfo{person}{M. {Alian}}, \bibinfo{person}{S.~W. {Min}},
  \bibinfo{person}{H. {Asgharimoghaddam}}, \bibinfo{person}{A. {Dhar}},
  \bibinfo{person}{D.~K. {Wang}}, \bibinfo{person}{T. {Roewer}},
  \bibinfo{person}{A. {McPadden}}, \bibinfo{person}{O. {O'Halloran}},
  \bibinfo{person}{D. {Chen}}, \bibinfo{person}{J. {Xiong}},
  \bibinfo{person}{D. {Kim}}, \bibinfo{person}{W. {Hwu}}, {and}
  \bibinfo{person}{N.~S. {Kim}}.}
\newblock \showarticletitle{{Application-Transparent Near-Memory Processing
  Architecture with Memory Channel Network}}. In
  \bibinfo{booktitle}{\emph{MICRO}} \bibinfo{year}{2018}\natexlab{}.
\newblock


\bibitem[\protect\citeauthoryear{Alser, Bing{\"o}l, Cali, Kim, Ghose, Alkan,
  and Mutlu}{Alser et~al\mbox{.}}{2020a}]%
        {alser2020accelerating}
\bibfield{author}{\bibinfo{person}{Mohammed Alser}, \bibinfo{person}{Z{\"u}lal
  Bing{\"o}l}, \bibinfo{person}{Damla~Senol Cali}, \bibinfo{person}{Jeremie
  Kim}, \bibinfo{person}{Saugata Ghose}, \bibinfo{person}{Can Alkan}, {and}
  \bibinfo{person}{Onur Mutlu}.}
\newblock \showarticletitle{{Accelerating Genome Analysis: A Primer on an
  Ongoing Journey}}. In \bibinfo{booktitle}{\emph{IEEE Micro}}
  \bibinfo{year}{2020}\natexlab{a}.
\newblock


\bibitem[\protect\citeauthoryear{Alser, Hassan, Kumar, Mutlu, and Alkan}{Alser
  et~al\mbox{.}}{2019}]%
        {alser2019shouji}
\bibfield{author}{\bibinfo{person}{Mohammed Alser}, \bibinfo{person}{Hasan
  Hassan}, \bibinfo{person}{Akash Kumar}, \bibinfo{person}{Onur Mutlu}, {and}
  \bibinfo{person}{Can Alkan}.}
\newblock \showarticletitle{{Shouji: A Fast and Efficient Pre-Alignment Filter
  for Sequence Alignment}}. In \bibinfo{booktitle}{\emph{Bioinformatics}}
  \bibinfo{year}{2019}\natexlab{}.
\newblock


\bibitem[\protect\citeauthoryear{Alser, Hassan, Xin, Ergin, Mutlu, and
  Alkan}{Alser et~al\mbox{.}}{2017}]%
        {alser2017}
\bibfield{author}{\bibinfo{person}{Mohammed Alser}, \bibinfo{person}{Hasan
  Hassan}, \bibinfo{person}{Hongyi Xin}, \bibinfo{person}{Oğuz Ergin},
  \bibinfo{person}{Onur Mutlu}, {and} \bibinfo{person}{Can Alkan}.}
\newblock \showarticletitle{{GateKeeper: A New Hardware Architecture for
  Accelerating Pre-Alignment in {DNA} Short Read Mapping}}. In
  \bibinfo{booktitle}{\emph{Bioinformatics}} \bibinfo{year}{2017}\natexlab{}.
\newblock


\bibitem[\protect\citeauthoryear{Alser, Rotman, Taraszka, Shi, Baykal, Yang,
  Xue, Knyazev, Singer, Balliu, et~al\mbox{.}}{Alser et~al\mbox{.}}{2020b}]%
        {alser2020technology}
\bibfield{author}{\bibinfo{person}{Mohammed Alser}, \bibinfo{person}{Jeremy
  Rotman}, \bibinfo{person}{Kodi Taraszka}, \bibinfo{person}{Huwenbo Shi},
  \bibinfo{person}{Pelin~Icer Baykal}, \bibinfo{person}{Harry~Taegyun Yang},
  \bibinfo{person}{Victor Xue}, \bibinfo{person}{Sergey Knyazev},
  \bibinfo{person}{Benjamin~D Singer}, \bibinfo{person}{Brunilda Balliu},
  {et~al\mbox{.}}}
\newblock \showarticletitle{{Technology Dictates Algorithms: Recent
  Developments in Read Alignment}}. In \bibinfo{booktitle}{\emph{Genome
  Biology}} \bibinfo{year}{2020}\natexlab{b}.
\newblock


\bibitem[\protect\citeauthoryear{Alser, Shahroodi, Gomez-Luna, Alkan, and
  Mutlu}{Alser et~al\mbox{.}}{2020c}]%
        {alser2019sneakysnake}
\bibfield{author}{\bibinfo{person}{Mohammed Alser}, \bibinfo{person}{Taha
  Shahroodi}, \bibinfo{person}{Juan Gomez-Luna}, \bibinfo{person}{Can Alkan},
  {and} \bibinfo{person}{Onur Mutlu}.}
\newblock \showarticletitle{{SneakySnake: A Fast and Accurate Universal Genome
  Pre-Alignment Filter for CPUs, GPUs, and FPGAs}}. In
  \bibinfo{booktitle}{\emph{Bioinformatics}} \bibinfo{year}{2020}\natexlab{c}.
\newblock


\bibitem[\protect\citeauthoryear{Angizi, Sun, Zhang, and Fan}{Angizi
  et~al\mbox{.}}{2019}]%
        {angizi2019aligns}
\bibfield{author}{\bibinfo{person}{Shaahin Angizi}, \bibinfo{person}{Jiao Sun},
  \bibinfo{person}{Wei Zhang}, {and} \bibinfo{person}{Deliang Fan}.}
\newblock \showarticletitle{{AlignS: A Processing-In-Memory Accelerator for DNA
  Short Read Alignment Leveraging SOT-MRAM}}. In
  \bibinfo{booktitle}{\emph{DAC}} \bibinfo{year}{2019}\natexlab{}.
\newblock


\bibitem[\protect\citeauthoryear{Ansel, Kamil, Veeramachaneni, Ragan-Kelley,
  Bosboom, O'Reilly, and Amarasinghe}{Ansel et~al\mbox{.}}{2014}]%
        {opentuner}
\bibfield{author}{\bibinfo{person}{Jason Ansel}, \bibinfo{person}{Shoaib
  Kamil}, \bibinfo{person}{Kalyan Veeramachaneni}, \bibinfo{person}{Jonathan
  Ragan-Kelley}, \bibinfo{person}{Jeffrey Bosboom}, \bibinfo{person}{Una-May
  O'Reilly}, {and} \bibinfo{person}{Saman Amarasinghe}.}
\newblock \showarticletitle{{Open{T}uner: {A}n Extensible Framework for Program
  Autotuning}}. In \bibinfo{booktitle}{\emph{PACT}}
  \bibinfo{year}{2014}\natexlab{}.
\newblock


\bibitem[\protect\citeauthoryear{Armejach, Caminal, Cebrian,
  Gonz{\'a}lez-Alberquilla, Adeniyi-Jones, Valero, Casas, and
  Moret{\'o}}{Armejach et~al\mbox{.}}{2018}]%
        {armejach2018stencil}
\bibfield{author}{\bibinfo{person}{Adri{\`a} Armejach}, \bibinfo{person}{Helena
  Caminal}, \bibinfo{person}{Juan~M Cebrian}, \bibinfo{person}{Rekai
  Gonz{\'a}lez-Alberquilla}, \bibinfo{person}{Chris Adeniyi-Jones},
  \bibinfo{person}{Mateo Valero}, \bibinfo{person}{Marc Casas}, {and}
  \bibinfo{person}{Miquel Moret{\'o}}.}
\newblock \showarticletitle{{Stencil Codes on a Vector Length Agnostic
  Architecture}}. In \bibinfo{booktitle}{\emph{PACT}}
  \bibinfo{year}{2018}\natexlab{}.
\newblock


\bibitem[\protect\citeauthoryear{Asghari-Moghaddam, Son, Ahn, and
  Kim}{Asghari-Moghaddam et~al\mbox{.}}{2016}]%
        {asghari2016chameleon}
\bibfield{author}{\bibinfo{person}{Hadi Asghari-Moghaddam},
  \bibinfo{person}{Young~Hoon Son}, \bibinfo{person}{Jung~Ho Ahn}, {and}
  \bibinfo{person}{Nam~Sung Kim}.}
\newblock \showarticletitle{{Chameleon: Versatile and Practical Near-DRAM
  Acceleration Architecture for Large Memory Systems}}. In
  \bibinfo{booktitle}{\emph{MICRO}} \bibinfo{year}{2016}\natexlab{}.
\newblock


\bibitem[\protect\citeauthoryear{Babarinsa and Idreos}{Babarinsa and
  Idreos}{2015}]%
        {babarinsa2015jafar}
\bibfield{author}{\bibinfo{person}{Oreoluwatomiwa~O Babarinsa} {and}
  \bibinfo{person}{Stratos Idreos}.}
\newblock \showarticletitle{{JAFAR: Near-Data Processing for Databases}}. In
  \bibinfo{booktitle}{\emph{SIGMOD}} \bibinfo{year}{2015}\natexlab{}.
\newblock


\bibitem[\protect\citeauthoryear{Barnes, Brown, Kato, Kuck, Slotnick, and
  Stokes}{Barnes et~al\mbox{.}}{1968}]%
        {illiac_simd_1968}
\bibfield{author}{\bibinfo{person}{George~H Barnes}, \bibinfo{person}{Richard~M
  Brown}, \bibinfo{person}{Maso Kato}, \bibinfo{person}{David~J Kuck},
  \bibinfo{person}{Daniel~L Slotnick}, {and} \bibinfo{person}{Richard~A
  Stokes}.}
\newblock \showarticletitle{{The ILLIAC IV Computer}}. In
  \bibinfo{booktitle}{\emph{TC}} \bibinfo{year}{1968}\natexlab{}.
\newblock


\bibitem[\protect\citeauthoryear{Benton}{Benton}{2017}]%
        {benton2017ccix}
\bibfield{author}{\bibinfo{person}{Brad Benton}.}
\newblock \showarticletitle{{CCIX}, {G}en-{Z}, {O}pen{CAPI}: {O}verview and
  {C}omparison}. In \bibinfo{booktitle}{\emph{OFA}}
  \bibinfo{year}{2017}\natexlab{}.
\newblock


\bibitem[\protect\citeauthoryear{Besta, Kanakagiri, Kwasniewski,
  Ausavarungnirun, Ber{\'a}nek, Kanellopoulos, Janda, Vonarburg-Shmaria,
  Gianinazzi, Stefan, et~al\mbox{.}}{Besta et~al\mbox{.}}{2021}]%
        {besta2021sisa}
\bibfield{author}{\bibinfo{person}{Maciej Besta}, \bibinfo{person}{Raghavendra
  Kanakagiri}, \bibinfo{person}{Grzegorz Kwasniewski}, \bibinfo{person}{Rachata
  Ausavarungnirun}, \bibinfo{person}{Jakub Ber{\'a}nek},
  \bibinfo{person}{Konstantinos Kanellopoulos}, \bibinfo{person}{Kacper Janda},
  \bibinfo{person}{Zur Vonarburg-Shmaria}, \bibinfo{person}{Lukas Gianinazzi},
  \bibinfo{person}{Ioana Stefan}, {et~al\mbox{.}}}
\newblock \showarticletitle{{SISA: Set-Centric Instruction Set Architecture for
  Graph Mining on Processing-in-Memory Systems}}. In
  \bibinfo{booktitle}{\emph{MICRO}} \bibinfo{year}{2021}\natexlab{}.
\newblock


\bibitem[\protect\citeauthoryear{Bianco, Diamanti, Fuhrer, Gysi, Lapillonne,
  Osuna, and Schulthess}{Bianco et~al\mbox{.}}{2013}]%
        {bianco2013gpu}
\bibfield{author}{\bibinfo{person}{M Bianco}, \bibinfo{person}{T Diamanti},
  \bibinfo{person}{O Fuhrer}, \bibinfo{person}{T Gysi}, \bibinfo{person}{X
  Lapillonne}, \bibinfo{person}{C Osuna}, {and} \bibinfo{person}{T
  Schulthess}.}
\newblock \showarticletitle{{A {GPU C}apable {V}ersion of the {COSMO W}eather
  {M}odel}}. In \bibinfo{booktitle}{\emph{ISC}}
  \bibinfo{year}{2013}\natexlab{}.
\newblock


\bibitem[\protect\citeauthoryear{Bonaventura}{Bonaventura}{2000}]%
        {bonaventura2000semi}
\bibfield{author}{\bibinfo{person}{Luca Bonaventura}.}
\newblock \showarticletitle{{A {S}emi-implicit {S}emi-{L}agrangian {S}cheme
  using the {H}eight {C}oordinate for a {N}onhydrostatic and {F}ully {E}lastic
  {M}odel of {A}tmospheric {F}lows}}. In \bibinfo{booktitle}{\emph{JCP}}
  \bibinfo{year}{2000}\natexlab{}.
\newblock


\bibitem[\protect\citeauthoryear{Boroumand, Ghose, Akin, Narayanaswami,
  Oliveira, Ma, Shiu, and Mutlu}{Boroumand et~al\mbox{.}}{2021}]%
        {boroumand2021google}
\bibfield{author}{\bibinfo{person}{Amirali Boroumand}, \bibinfo{person}{Saugata
  Ghose}, \bibinfo{person}{Berkin Akin}, \bibinfo{person}{Ravi Narayanaswami},
  \bibinfo{person}{Geraldo~F Oliveira}, \bibinfo{person}{Xiaoyu Ma},
  \bibinfo{person}{Eric Shiu}, {and} \bibinfo{person}{Onur Mutlu}.}
\newblock \showarticletitle{{Google Neural Network Models for Edge Devices:
  Analyzing and Mitigating Machine Learning Inference Bottlenecks}}. In
  \bibinfo{booktitle}{\emph{PACT}} \bibinfo{year}{2021}\natexlab{}.
\newblock


\bibitem[\protect\citeauthoryear{Boroumand, Ghose, Kim, Ausavarungnirun, Shiu,
  Thakur, Kim, Kuusela, Knies, Ranganathan, and Mutlu}{Boroumand
  et~al\mbox{.}}{2018}]%
        {googleWorkloads}
\bibfield{author}{\bibinfo{person}{Amirali Boroumand}, \bibinfo{person}{Saugata
  Ghose}, \bibinfo{person}{Youngsok Kim}, \bibinfo{person}{Rachata
  Ausavarungnirun}, \bibinfo{person}{Eric Shiu}, \bibinfo{person}{Rahul
  Thakur}, \bibinfo{person}{Daehyun Kim}, \bibinfo{person}{Aki Kuusela},
  \bibinfo{person}{Allan Knies}, \bibinfo{person}{Parthasarathy Ranganathan},
  {and} \bibinfo{person}{Onur Mutlu}.}
\newblock \showarticletitle{{Google Workloads for Consumer Devices: Mitigating
  Data Movement Bottlenecks}}. In \bibinfo{booktitle}{\emph{ASPLOS}}
  \bibinfo{year}{2018}\natexlab{}.
\newblock


\bibitem[\protect\citeauthoryear{Boroumand, Ghose, Patel, Hassan, Lucia,
  Ausavarungnirun, Hsieh, Hajinazar, Malladi, Zheng, et~al\mbox{.}}{Boroumand
  et~al\mbox{.}}{2019}]%
        {boroumand2019conda}
\bibfield{author}{\bibinfo{person}{Amirali Boroumand}, \bibinfo{person}{Saugata
  Ghose}, \bibinfo{person}{Minesh Patel}, \bibinfo{person}{Hasan Hassan},
  \bibinfo{person}{Brandon Lucia}, \bibinfo{person}{Rachata Ausavarungnirun},
  \bibinfo{person}{Kevin Hsieh}, \bibinfo{person}{Nastaran Hajinazar},
  \bibinfo{person}{Krishna~T Malladi}, \bibinfo{person}{Hongzhong Zheng},
  {et~al\mbox{.}}}
\newblock \showarticletitle{{CoNDA: Efficient Cache Coherence Support for
  Near-Data Accelerators}}. In \bibinfo{booktitle}{\emph{ISCA}}
  \bibinfo{year}{2019}\natexlab{}.
\newblock


\bibitem[\protect\citeauthoryear{Boroumand, Ghose, Patel, Hassan, Lucia, Hsieh,
  Malladi, Zheng, and Mutlu}{Boroumand et~al\mbox{.}}{2016}]%
        {boroumand2016lazypim}
\bibfield{author}{\bibinfo{person}{Amirali Boroumand}, \bibinfo{person}{Saugata
  Ghose}, \bibinfo{person}{Minesh Patel}, \bibinfo{person}{Hasan Hassan},
  \bibinfo{person}{Brandon Lucia}, \bibinfo{person}{Kevin Hsieh},
  \bibinfo{person}{Krishna~T Malladi}, \bibinfo{person}{Hongzhong Zheng}, {and}
  \bibinfo{person}{Onur Mutlu}.}
\newblock \showarticletitle{{LazyPIM: An Efficient Cache Coherence Mechanism
  for Processing-in-Memory}}. In \bibinfo{booktitle}{\emph{CAL}}
  \bibinfo{year}{2016}\natexlab{}.
\newblock


\bibitem[\protect\citeauthoryear{Cali, Kalsi, Bing{\"o}l, Firtina, Subramanian,
  Kim, Ausavarungnirun, Alser, Luna, Boroumand, Nori, Scibisz, Subramoney,
  Alkan, Ghose, and Mutlu}{Cali et~al\mbox{.}}{2020}]%
        {senolcalimicro2020}
\bibfield{author}{\bibinfo{person}{Damla~Senol Cali},
  \bibinfo{person}{Gurpreet~S. Kalsi}, \bibinfo{person}{Z{\"u}lal Bing{\"o}l},
  \bibinfo{person}{Can Firtina}, \bibinfo{person}{Lavanya Subramanian},
  \bibinfo{person}{Jeremie~S. Kim}, \bibinfo{person}{Rachata Ausavarungnirun},
  \bibinfo{person}{Mohammed Alser}, \bibinfo{person}{Juan~G{\'o}mez Luna},
  \bibinfo{person}{Amirali Boroumand}, \bibinfo{person}{Anant Nori},
  \bibinfo{person}{Allison Scibisz}, \bibinfo{person}{Sreenivas Subramoney},
  \bibinfo{person}{Can Alkan}, \bibinfo{person}{Saugata Ghose}, {and}
  \bibinfo{person}{Onur Mutlu}.}
\newblock \showarticletitle{{{GenASM: A High-Performance, Low-Power Approximate
  String Matching Acceleration Framework for Genome Sequence Analysis}}}. In
  \bibinfo{booktitle}{\emph{MICRO}} \bibinfo{year}{2020}\natexlab{}.
\newblock


\bibitem[\protect\citeauthoryear{{Caulfield}, {Chung}, {Putnam}, {Angepat},
  {Fowers}, {Haselman}, {Heil}, {Humphrey}, {Kaur}, {Kim}, {Lo}, {Massengill},
  {Ovtcharov}, {Papamichael}, {Woods}, {Lanka}, {Chiou}, and
  {Burger}}{{Caulfield} et~al\mbox{.}}{2016}]%
        {caulfield2016cloud}
\bibfield{author}{\bibinfo{person}{A.~M. {Caulfield}}, \bibinfo{person}{E.~S.
  {Chung}}, \bibinfo{person}{A. {Putnam}}, \bibinfo{person}{H. {Angepat}},
  \bibinfo{person}{J. {Fowers}}, \bibinfo{person}{M. {Haselman}},
  \bibinfo{person}{S. {Heil}}, \bibinfo{person}{M. {Humphrey}},
  \bibinfo{person}{P. {Kaur}}, \bibinfo{person}{J. {Kim}}, \bibinfo{person}{D.
  {Lo}}, \bibinfo{person}{T. {Massengill}}, \bibinfo{person}{K. {Ovtcharov}},
  \bibinfo{person}{M. {Papamichael}}, \bibinfo{person}{L. {Woods}},
  \bibinfo{person}{S. {Lanka}}, \bibinfo{person}{D. {Chiou}}, {and}
  \bibinfo{person}{D. {Burger}}.}
\newblock \showarticletitle{{A Cloud-Scale Acceleration Architecture}}. In
  \bibinfo{booktitle}{\emph{MICRO}} \bibinfo{year}{2016}\natexlab{}.
\newblock


\bibitem[\protect\citeauthoryear{Chang, G{\'o}mez-Luna, El~Hajj, Huang, Chen,
  and Hwu}{Chang et~al\mbox{.}}{2017}]%
        {chang2017collaborative}
\bibfield{author}{\bibinfo{person}{Li-Wen Chang}, \bibinfo{person}{Juan
  G{\'o}mez-Luna}, \bibinfo{person}{Izzat El~Hajj}, \bibinfo{person}{Sitao
  Huang}, \bibinfo{person}{Deming Chen}, {and} \bibinfo{person}{Wen-mei Hwu}.}
\newblock \showarticletitle{{Collaborative Computing for Heterogeneous
  Integrated Systems}}. In \bibinfo{booktitle}{\emph{ICPE}}
  \bibinfo{year}{2017}\natexlab{}.
\newblock


\bibitem[\protect\citeauthoryear{Chi, Li, Xu, Zhang, Zhao, Liu, Wang, and
  Xie}{Chi et~al\mbox{.}}{2016}]%
        {chi2016prime}
\bibfield{author}{\bibinfo{person}{Ping Chi}, \bibinfo{person}{Shuangchen Li},
  \bibinfo{person}{Cong Xu}, \bibinfo{person}{Tao Zhang},
  \bibinfo{person}{Jishen Zhao}, \bibinfo{person}{Yongpan Liu},
  \bibinfo{person}{Yu Wang}, {and} \bibinfo{person}{Yuan Xie}.}
\newblock \showarticletitle{{PRIME: A Novel Processing-in-memory Architecture
  for Neural Network Computation in ReRAM-based Main Memory}}. In
  \bibinfo{booktitle}{\emph{ISCA}} \bibinfo{year}{2016}\natexlab{}.
\newblock


\bibitem[\protect\citeauthoryear{Chi, Cong, Wei, and Zhou}{Chi
  et~al\mbox{.}}{2018}]%
        {chi2018soda}
\bibfield{author}{\bibinfo{person}{Yuze Chi}, \bibinfo{person}{Jason Cong},
  \bibinfo{person}{Peng Wei}, {and} \bibinfo{person}{Peipei Zhou}.}
\newblock \showarticletitle{{{SODA}: {S}tencil with {O}ptimized {D}ataflow
  {A}rchitecture}}. In \bibinfo{booktitle}{\emph{ICCAD}}
  \bibinfo{year}{2018}\natexlab{}.
\newblock


\bibitem[\protect\citeauthoryear{Choi, Cong, Fang, Hao, Reinman, and Wei}{Choi
  et~al\mbox{.}}{2016}]%
        {10.1145/2897937.2897972}
\bibfield{author}{\bibinfo{person}{Young-kyu Choi}, \bibinfo{person}{Jason
  Cong}, \bibinfo{person}{Zhenman Fang}, \bibinfo{person}{Yuchen Hao},
  \bibinfo{person}{Glenn Reinman}, {and} \bibinfo{person}{Peng Wei}.}
\newblock \showarticletitle{{A Quantitative Analysis on Microarchitectures of
  Modern CPU-FPGA Platforms}}. In \bibinfo{booktitle}{\emph{DAC}}
  \bibinfo{year}{2016}\natexlab{}.
\newblock


\bibitem[\protect\citeauthoryear{Christen, Schenk, and Burkhart}{Christen
  et~al\mbox{.}}{2011}]%
        {christen2011patus}
\bibfield{author}{\bibinfo{person}{Matthias Christen}, \bibinfo{person}{Olaf
  Schenk}, {and} \bibinfo{person}{Helmar Burkhart}.}
\newblock \showarticletitle{{PATUS}: {A} {C}ode {G}eneration and {A}utotuning
  {F}ramework for {P}arallel {I}terative {S}tencil {C}omputations on {M}odern
  {M}icroarchitectures}. In \bibinfo{booktitle}{\emph{IPDPS}}
  \bibinfo{year}{2011}\natexlab{}.
\newblock


\bibitem[\protect\citeauthoryear{Datta, Kamil, Williams, Oliker, Shalf, and
  Yelick}{Datta et~al\mbox{.}}{2009}]%
        {datta2009optimization}
\bibfield{author}{\bibinfo{person}{Kaushik Datta}, \bibinfo{person}{Shoaib
  Kamil}, \bibinfo{person}{Samuel Williams}, \bibinfo{person}{Leonid Oliker},
  \bibinfo{person}{John Shalf}, {and} \bibinfo{person}{Katherine Yelick}.}
\newblock \showarticletitle{{Optimization and {P}erformance {M}odeling of
  {S}tencil {C}omputations on {M}odern {M}icroprocessors}}. In
  \bibinfo{booktitle}{\emph{SIAM review}} \bibinfo{year}{2009}\natexlab{}.
\newblock


\bibitem[\protect\citeauthoryear{de~Fine~Licht, Blott, and
  Hoefler}{de~Fine~Licht et~al\mbox{.}}{2018}]%
        {de2018designing}
\bibfield{author}{\bibinfo{person}{Johannes de Fine~Licht},
  \bibinfo{person}{Michaela Blott}, {and} \bibinfo{person}{Torsten Hoefler}.}
\newblock \showarticletitle{{Designing scalable {FPGA} architectures using
  high-level synthesis}}. In \bibinfo{booktitle}{\emph{PPoPP}}
  \bibinfo{year}{2018}\natexlab{}.
\newblock


\bibitem[\protect\citeauthoryear{de~Fine~Licht, Kuster, De~Matteis, Ben-Nun,
  Hofer, and Hoefler}{de~Fine~Licht et~al\mbox{.}}{2021}]%
        {de2021stencilflow}
\bibfield{author}{\bibinfo{person}{Johannes de Fine~Licht},
  \bibinfo{person}{Andreas Kuster}, \bibinfo{person}{Tiziano De~Matteis},
  \bibinfo{person}{Tal Ben-Nun}, \bibinfo{person}{Dominic Hofer}, {and}
  \bibinfo{person}{Torsten Hoefler}.}
\newblock \showarticletitle{StencilFlow: Mapping large stencil programs to
  distributed spatial computing systems}. In \bibinfo{booktitle}{\emph{CGO}}
  \bibinfo{year}{2021}\natexlab{}.
\newblock


\bibitem[\protect\citeauthoryear{Diamantopoulos, Giefers, and
  Hagleitner}{Diamantopoulos et~al\mbox{.}}{2018}]%
        {8373077}
\bibfield{author}{\bibinfo{person}{Dionysios Diamantopoulos},
  \bibinfo{person}{Heiner Giefers}, {and} \bibinfo{person}{Christoph
  Hagleitner}.}
\newblock \showarticletitle{{ecTALK: Energy Efficient Coherent Transprecision
  Accelerators – The Bidirectional Long Short-Term Memory Neural Network Case
  }}. In \bibinfo{booktitle}{\emph{COOL CHIPS}}
  \bibinfo{year}{2018}\natexlab{}.
\newblock


\bibitem[\protect\citeauthoryear{Diamantopoulos and Hagleitner}{Diamantopoulos
  and Hagleitner}{2018}]%
        {dioFPT}
\bibfield{author}{\bibinfo{person}{Dionysios Diamantopoulos} {and}
  \bibinfo{person}{Christoph Hagleitner}.}
\newblock \showarticletitle{{A {S}ystem-{L}evel {T}ransprecision {FPGA}
  {A}ccelerator for {BLSTM} {U}sing {O}n-chip {M}emory {R}eshaping}}. In
  \bibinfo{booktitle}{\emph{FPT}} \bibinfo{year}{2018}\natexlab{}.
\newblock


\bibitem[\protect\citeauthoryear{Doms and Sch{\"a}ttler}{Doms and
  Sch{\"a}ttler}{1999}]%
        {doms1999nonhydrostatic}
\bibfield{author}{\bibinfo{person}{G Doms} {and} \bibinfo{person}{U
  Sch{\"a}ttler}.}
\newblock \showarticletitle{The {N}onhydrostatic {L}imited-{A}rea {M}odel {LM}
  ({L}okal-model) of the {DWD. P}art {I: S}cientific {D}ocumentation}. In
  \bibinfo{booktitle}{\emph{DWD, GB Forschung und Entwicklung}}
  \bibinfo{year}{1999}\natexlab{}.
\newblock


\bibitem[\protect\citeauthoryear{Drumond, Daglis, Mirzadeh, Ustiugov, Picorel,
  Falsafi, Grot, and Pnevmatikatos}{Drumond et~al\mbox{.}}{2017}]%
        {de2017mondrian}
\bibfield{author}{\bibinfo{person}{Mario Drumond}, \bibinfo{person}{Alexandros
  Daglis}, \bibinfo{person}{Nooshin Mirzadeh}, \bibinfo{person}{Dmitrii
  Ustiugov}, \bibinfo{person}{Javier Picorel}, \bibinfo{person}{Babak Falsafi},
  \bibinfo{person}{Boris Grot}, {and} \bibinfo{person}{Dionisios
  Pnevmatikatos}.}
\newblock \showarticletitle{{The {M}ondrian {D}ata {E}ngine}}. In
  \bibinfo{booktitle}{\emph{ISCA}} \bibinfo{year}{2017}\natexlab{}.
\newblock


\bibitem[\protect\citeauthoryear{Duarte, Han, Harris, Jindariani, Kreinar,
  Kreis, Ngadiuba, Pierini, Rivera, Tran, and Wu}{Duarte et~al\mbox{.}}{2018}]%
        {duarte2018fast}
\bibfield{author}{\bibinfo{person}{Javier Duarte}, \bibinfo{person}{Song Han},
  \bibinfo{person}{Philip Harris}, \bibinfo{person}{Sergo Jindariani},
  \bibinfo{person}{Edward Kreinar}, \bibinfo{person}{Benjamin Kreis},
  \bibinfo{person}{Jennifer Ngadiuba}, \bibinfo{person}{Maurizio Pierini},
  \bibinfo{person}{Ryan Rivera}, \bibinfo{person}{Nhan Tran}, {and}
  \bibinfo{person}{Z Wu}.}
\newblock \showarticletitle{{Fast inference of deep neural networks in FPGAs
  for pinproceedings physics}}. In \bibinfo{booktitle}{\emph{JINST}}
  \bibinfo{year}{2018}\natexlab{}.
\newblock


\bibitem[\protect\citeauthoryear{Fang, Mulder, Hidders, Lee, and Hofstee}{Fang
  et~al\mbox{.}}{2020}]%
        {Fang2020}
\bibfield{author}{\bibinfo{person}{Jian Fang}, \bibinfo{person}{Yvo T.~B.
  Mulder}, \bibinfo{person}{Jan Hidders}, \bibinfo{person}{Jinho Lee}, {and}
  \bibinfo{person}{H.~Peter Hofstee}.}
\newblock \showarticletitle{{In-memory database acceleration on FPGAs: a
  survey}}. In \bibinfo{booktitle}{\emph{VLDB}}
  \bibinfo{year}{2020}\natexlab{}.
\newblock


\bibitem[\protect\citeauthoryear{Farmahini-Farahani, Ahn, Morrow, and
  Kim}{Farmahini-Farahani et~al\mbox{.}}{2015}]%
        {7056040}
\bibfield{author}{\bibinfo{person}{A. Farmahini-Farahani},
  \bibinfo{person}{J.~H. Ahn}, \bibinfo{person}{K. Morrow}, {and}
  \bibinfo{person}{N.~S. Kim}.}
\newblock \showarticletitle{{NDA: Near-DRAM Acceleration Architecture
  Leveraging Commodity DRAM Devices and Standard Memory Modules}}. In
  \bibinfo{booktitle}{\emph{HPCA}} \bibinfo{year}{2015}\natexlab{}.
\newblock


\bibitem[\protect\citeauthoryear{Fernandez, Quislant, Guti{\'e}rrez, Plata,
  Giannoula, Alser, G{\'o}mez-Luna, and Mutlu}{Fernandez et~al\mbox{.}}{2020}]%
        {fernandez2020natsa}
\bibfield{author}{\bibinfo{person}{Ivan Fernandez}, \bibinfo{person}{Ricardo
  Quislant}, \bibinfo{person}{Eladio Guti{\'e}rrez}, \bibinfo{person}{Oscar
  Plata}, \bibinfo{person}{Christina Giannoula}, \bibinfo{person}{Mohammed
  Alser}, \bibinfo{person}{Juan G{\'o}mez-Luna}, {and} \bibinfo{person}{Onur
  Mutlu}.}
\newblock \showarticletitle{{NATSA: A Near-Data Processing Accelerator for Time
  Series Analysis}}. In \bibinfo{booktitle}{\emph{ICCD}}
  \bibinfo{year}{2020}\natexlab{}.
\newblock


\bibitem[\protect\citeauthoryear{Flynn}{Flynn}{1966}]%
        {flynn1966very}
\bibfield{author}{\bibinfo{person}{Michael~J Flynn}.}
\newblock \showarticletitle{{Very High-Speed Computing Systems}}. In
  \bibinfo{booktitle}{\emph{Proceedings of the IEEE}}
  \bibinfo{year}{1966}\natexlab{}.
\newblock


\bibitem[\protect\citeauthoryear{Fu and Clapp}{Fu and Clapp}{2011}]%
        {fu2011eliminating}
\bibfield{author}{\bibinfo{person}{Haohuan Fu} {and} \bibinfo{person}{Robert~G
  Clapp}.}
\newblock \showarticletitle{Eliminating the memory bottleneck: an FPGA-based
  solution for 3D reverse time migration}. In \bibinfo{booktitle}{\emph{FPGA}}
  \bibinfo{year}{2011}\natexlab{}.
\newblock


\bibitem[\protect\citeauthoryear{Gaide, Gaitonde, Ravishankar, and Bauer}{Gaide
  et~al\mbox{.}}{2019}]%
        {gaide2019xilinx}
\bibfield{author}{\bibinfo{person}{Brian Gaide}, \bibinfo{person}{Dinesh
  Gaitonde}, \bibinfo{person}{Chirag Ravishankar}, {and}
  \bibinfo{person}{Trevor Bauer}.}
\newblock \showarticletitle{{Xilinx Adaptive Compute Acceleration Platform:
  Versal™ Architecture}}. In \bibinfo{booktitle}{\emph{FPGA}}
  \bibinfo{year}{2019}\natexlab{}.
\newblock


\bibitem[\protect\citeauthoryear{Gao, Tziantzioulis, and Wentzlaff}{Gao
  et~al\mbox{.}}{2019}]%
        {gao2019computedram}
\bibfield{author}{\bibinfo{person}{Fei Gao}, \bibinfo{person}{Georgios
  Tziantzioulis}, {and} \bibinfo{person}{David Wentzlaff}.}
\newblock \showarticletitle{{ComputeDRAM: In-Memory Compute Using Off-the-Shelf
  DRAMs}}. In \bibinfo{booktitle}{\emph{MICRO}}
  \bibinfo{year}{2019}\natexlab{}.
\newblock


\bibitem[\protect\citeauthoryear{Gao, Ayers, and Kozyrakis}{Gao
  et~al\mbox{.}}{2015}]%
        {gao2015practical}
\bibfield{author}{\bibinfo{person}{Mingyu Gao}, \bibinfo{person}{Grant Ayers},
  {and} \bibinfo{person}{Christos Kozyrakis}.}
\newblock \showarticletitle{{Practical Near-Data Processing for In-Memory
  Analytics Frameworks}}. In \bibinfo{booktitle}{\emph{PACT}}
  \bibinfo{year}{2015}\natexlab{}.
\newblock


\bibitem[\protect\citeauthoryear{Gao and Kozyrakis}{Gao and Kozyrakis}{2016}]%
        {7446059}
\bibfield{author}{\bibinfo{person}{M. Gao} {and} \bibinfo{person}{C.
  Kozyrakis}.}
\newblock \showarticletitle{{HRL: Efficient and Flexible Reconfigurable Logic
  for Near-Data Processing}}. In \bibinfo{booktitle}{\emph{HPCA}}
  \bibinfo{year}{2016}\natexlab{}.
\newblock


\bibitem[\protect\citeauthoryear{Ghose, Boroumand, Kim, G{\'o}mez-Luna, and
  Mutlu}{Ghose et~al\mbox{.}}{2019a}]%
        {ghose2019processing}
\bibfield{author}{\bibinfo{person}{Saugata Ghose}, \bibinfo{person}{Amirali
  Boroumand}, \bibinfo{person}{Jeremie~S Kim}, \bibinfo{person}{Juan
  G{\'o}mez-Luna}, {and} \bibinfo{person}{Onur Mutlu}.}
\newblock \showarticletitle{{Processing-in-memory: A workload-driven
  perspective}}. In \bibinfo{booktitle}{\emph{IBM JRD}}
  \bibinfo{year}{2019}\natexlab{a}.
\newblock


\bibitem[\protect\citeauthoryear{Ghose, Li, Hajinazar, Cali, and Mutlu}{Ghose
  et~al\mbox{.}}{2019b}]%
        {ghose2019demystifying}
\bibfield{author}{\bibinfo{person}{Saugata Ghose}, \bibinfo{person}{Tianshi
  Li}, \bibinfo{person}{Nastaran Hajinazar}, \bibinfo{person}{Damla~Senol
  Cali}, {and} \bibinfo{person}{Onur Mutlu}.}
\newblock \showarticletitle{{Demystifying {C}omplex {W}orkload-{DRAM}
  {I}nteractions: {A}n {E}xperimental {S}tudy}}. In
  \bibinfo{booktitle}{\emph{POMACS}} \bibinfo{year}{2019}\natexlab{b}.
\newblock


\bibitem[\protect\citeauthoryear{Giannoula, Vijaykumar, Papadopoulou,
  Karakostas, Fernandez, G{\'o}mez-Luna, Orosa, Koziris, Goumas, and
  Mutlu}{Giannoula et~al\mbox{.}}{2021}]%
        {giannoula2021syncron}
\bibfield{author}{\bibinfo{person}{Christina Giannoula},
  \bibinfo{person}{Nandita Vijaykumar}, \bibinfo{person}{Nikela Papadopoulou},
  \bibinfo{person}{Vasileios Karakostas}, \bibinfo{person}{Ivan Fernandez},
  \bibinfo{person}{Juan G{\'o}mez-Luna}, \bibinfo{person}{Lois Orosa},
  \bibinfo{person}{Nectarios Koziris}, \bibinfo{person}{Georgios Goumas}, {and}
  \bibinfo{person}{Onur Mutlu}.}
\newblock \showarticletitle{SynCron: Efficient Synchronization Support for
  Near-Data-Processing Architectures}. In \bibinfo{booktitle}{\emph{HPCA}}
  \bibinfo{year}{2021}\natexlab{}.
\newblock


\bibitem[\protect\citeauthoryear{Giefers, Polig, and Hagleitner}{Giefers
  et~al\mbox{.}}{2015}]%
        {giefers2015accelerating}
\bibfield{author}{\bibinfo{person}{Heiner Giefers}, \bibinfo{person}{Raphael
  Polig}, {and} \bibinfo{person}{Christoph Hagleitner}.}
\newblock \showarticletitle{{Accelerating arithmetic kernels with coherent
  attached {FPGA} coprocessors}}. In \bibinfo{booktitle}{\emph{DATE}}
  \bibinfo{year}{2015}\natexlab{}.
\newblock


\bibitem[\protect\citeauthoryear{G{\'{o}}mez{-}Luna, Hajj, Fernandez,
  Giannoula, Oliveira, and Mutlu}{G{\'{o}}mez{-}Luna et~al\mbox{.}}{2021}]%
        {upmem2021}
\bibfield{author}{\bibinfo{person}{Juan G{\'{o}}mez{-}Luna},
  \bibinfo{person}{Izzat~El Hajj}, \bibinfo{person}{Ivan Fernandez},
  \bibinfo{person}{Christina Giannoula}, \bibinfo{person}{Geraldo~F. Oliveira},
  {and} \bibinfo{person}{Onur Mutlu}.}
\newblock \showarticletitle{{Benchmarking a New Paradigm: An Experimental
  Analysis of a Real Processing-in-Memory Architecture}}. In
  \bibinfo{booktitle}{\emph{arXiv}} \bibinfo{year}{2021}\natexlab{}.
\newblock


\bibitem[\protect\citeauthoryear{G{\'o}mez-Luna, Hajj, Fernandez, Giannoula,
  Oliveira, and Mutlu}{G{\'o}mez-Luna et~al\mbox{.}}{2021}]%
        {gomez2021benchmarking}
\bibfield{author}{\bibinfo{person}{Juan G{\'o}mez-Luna},
  \bibinfo{person}{Izzat~El Hajj}, \bibinfo{person}{Ivan Fernandez},
  \bibinfo{person}{Christina Giannoula}, \bibinfo{person}{Geraldo~F Oliveira},
  {and} \bibinfo{person}{Onur Mutlu}.}
\newblock \showarticletitle{{Benchmarking Memory-Centric Computing Systems:
  Analysis of Real Processing-in-Memory Hardware}}. In
  \bibinfo{booktitle}{\emph{CUT}} \bibinfo{year}{2021}\natexlab{}.
\newblock


\bibitem[\protect\citeauthoryear{Gonz{\'a}lez and Gonz{\'a}lez}{Gonz{\'a}lez
  and Gonz{\'a}lez}{1997}]%
        {gonzalez1997speculative}
\bibfield{author}{\bibinfo{person}{Jos{\'e} Gonz{\'a}lez} {and}
  \bibinfo{person}{Antonio Gonz{\'a}lez}.}
\newblock \showarticletitle{{Speculative {E}xecution via {A}ddress {P}rediction
  and {D}ata {P}refetching}}. In \bibinfo{booktitle}{\emph{ICS}}
  \bibinfo{year}{1997}\natexlab{}.
\newblock


\bibitem[\protect\citeauthoryear{Gu, Yoon, Bae, Jo, Lee, Yoon, Kang, Kwon,
  Yoon, Cho, Jeong, and Chang}{Gu et~al\mbox{.}}{2016}]%
        {gu2016biscuit}
\bibfield{author}{\bibinfo{person}{Boncheol Gu}, \bibinfo{person}{Andre~S.
  Yoon}, \bibinfo{person}{Duck-Ho Bae}, \bibinfo{person}{Insoon Jo},
  \bibinfo{person}{Jinyoung Lee}, \bibinfo{person}{Jonghyun Yoon},
  \bibinfo{person}{Jeong-Uk Kang}, \bibinfo{person}{Moonsang Kwon},
  \bibinfo{person}{Chanho Yoon}, \bibinfo{person}{Sangyeun Cho},
  \bibinfo{person}{Jaeheon Jeong}, {and} \bibinfo{person}{Duckhyun Chang}.}
\newblock \showarticletitle{{Biscuit: A Framework for Near-data Processing of
  Big Data Workloads}}. In \bibinfo{booktitle}{\emph{ISCA}}
  \bibinfo{year}{2016}\natexlab{}.
\newblock


\bibitem[\protect\citeauthoryear{Gysi, Grosser, and Hoefler}{Gysi
  et~al\mbox{.}}{2015}]%
        {gysi2015modesto}
\bibfield{author}{\bibinfo{person}{Tobias Gysi}, \bibinfo{person}{Tobias
  Grosser}, {and} \bibinfo{person}{Torsten Hoefler}.}
\newblock \showarticletitle{{{MODESTO}: {D}ata-centric {A}nalytic
  {O}ptimization of {C}omplex {S}tencil {P}rograms on {H}eterogeneous
  {A}rchitectures}}. In \bibinfo{booktitle}{\emph{SC}}
  \bibinfo{year}{2015}\natexlab{}.
\newblock


\bibitem[\protect\citeauthoryear{Hajinazar, Oliveira, Gregorio, Ferreira,
  Ghiasi, Patel, Alser, Ghose, Luna, and Mutlu}{Hajinazar
  et~al\mbox{.}}{2021}]%
        {hajinazar2021simdram}
\bibfield{author}{\bibinfo{person}{Nastaran Hajinazar},
  \bibinfo{person}{Geraldo~F Oliveira}, \bibinfo{person}{Sven Gregorio},
  \bibinfo{person}{Jo{\~a}o Ferreira}, \bibinfo{person}{Nika~Mansouri Ghiasi},
  \bibinfo{person}{Minesh Patel}, \bibinfo{person}{Mohammed Alser},
  \bibinfo{person}{Saugata Ghose}, \bibinfo{person}{Juan~G{\'o}mez Luna}, {and}
  \bibinfo{person}{Onur Mutlu}.}
\newblock \showarticletitle{SIMDRAM: An End-to-End Framework for Bit-Serial
  SIMD Computing in DRAM}. In \bibinfo{booktitle}{\emph{ASPLOS}}
  \bibinfo{year}{2021}\natexlab{}.
\newblock


\bibitem[\protect\citeauthoryear{Hashemi, Ebrahimi, Mutlu, Patt,
  et~al\mbox{.}}{Hashemi et~al\mbox{.}}{2016a}]%
        {hashemi2016accelerating}
\bibfield{author}{\bibinfo{person}{Milad Hashemi}, \bibinfo{person}{Eiman
  Ebrahimi}, \bibinfo{person}{Onur Mutlu}, \bibinfo{person}{Yale~N Patt},
  {et~al\mbox{.}}}
\newblock \showarticletitle{{Accelerating Dependent Cache Misses with an
  Enhanced Memory Controller}}. In \bibinfo{booktitle}{\emph{ISCA}}
  \bibinfo{year}{2016}\natexlab{a}.
\newblock


\bibitem[\protect\citeauthoryear{Hashemi, Mutlu, and Patt}{Hashemi
  et~al\mbox{.}}{2016b}]%
        {hashemi2016continuous}
\bibfield{author}{\bibinfo{person}{Milad Hashemi}, \bibinfo{person}{Onur
  Mutlu}, {and} \bibinfo{person}{Yale~N Patt}.}
\newblock \showarticletitle{{Continuous Runahead: Transparent Hardware
  Acceleration for Memory Intensive Workloads}}. In
  \bibinfo{booktitle}{\emph{MICRO}} \bibinfo{year}{2016}\natexlab{b}.
\newblock


\bibitem[\protect\citeauthoryear{Henretty, Stock, Pouchet, Franchetti,
  Ramanujam, and Sadayappan}{Henretty et~al\mbox{.}}{2011}]%
        {henretty2011data}
\bibfield{author}{\bibinfo{person}{Tom Henretty}, \bibinfo{person}{Kevin
  Stock}, \bibinfo{person}{Louis-No{\"e}l Pouchet}, \bibinfo{person}{Franz
  Franchetti}, \bibinfo{person}{J Ramanujam}, {and} \bibinfo{person}{P
  Sadayappan}.}
\newblock \showarticletitle{{Data {L}ayout {T}ransformation for {S}tencil
  {C}omputations on {S}hort-{V}ector {SIMD} {A}rchitectures}}. In
  \bibinfo{booktitle}{\emph{CC}} \bibinfo{year}{2011}\natexlab{}.
\newblock


\bibitem[\protect\citeauthoryear{Hermosilla, Bermejo, Balaguer, and
  Ruiz}{Hermosilla et~al\mbox{.}}{2008}]%
        {hermosilla2008non}
\bibfield{author}{\bibinfo{person}{Txomin Hermosilla}, \bibinfo{person}{E
  Bermejo}, \bibinfo{person}{A Balaguer}, {and} \bibinfo{person}{Luis~A Ruiz}.}
\newblock \showarticletitle{Non-linear fourth-order image interpolation for
  subpixel edge detection and localization}. In
  \bibinfo{booktitle}{\emph{IMAVIS}} \bibinfo{year}{2008}\natexlab{}.
\newblock


\bibitem[\protect\citeauthoryear{Hsieh, Ebrahimi, Kim, Chatterjee, O'Connor,
  Vijaykumar, Mutlu, and Keckler}{Hsieh et~al\mbox{.}}{2016a}]%
        {7551394}
\bibfield{author}{\bibinfo{person}{Kevin Hsieh}, \bibinfo{person}{Eiman
  Ebrahimi}, \bibinfo{person}{Gwangsun Kim}, \bibinfo{person}{Niladrish
  Chatterjee}, \bibinfo{person}{Mike O'Connor}, \bibinfo{person}{Nandita
  Vijaykumar}, \bibinfo{person}{Onur Mutlu}, {and} \bibinfo{person}{Stephen~W
  Keckler}.}
\newblock \showarticletitle{{Transparent Offloading and Mapping (TOM): Enabling
  Programmer-Transparent Near-Data Processing in GPU Systems}}. In
  \bibinfo{booktitle}{\emph{ISCA}} \bibinfo{year}{2016}\natexlab{a}.
\newblock


\bibitem[\protect\citeauthoryear{Hsieh, Khan, Vijaykumar, Chang, Boroumand,
  Ghose, and Mutlu}{Hsieh et~al\mbox{.}}{2016b}]%
        {hsieh2016accelerating}
\bibfield{author}{\bibinfo{person}{Kevin Hsieh}, \bibinfo{person}{Samira Khan},
  \bibinfo{person}{Nandita Vijaykumar}, \bibinfo{person}{Kevin~K Chang},
  \bibinfo{person}{Amirali Boroumand}, \bibinfo{person}{Saugata Ghose}, {and}
  \bibinfo{person}{Onur Mutlu}.}
\newblock \showarticletitle{{Accelerating Pointer Chasing in 3D-Stacked Memory:
  Challenges, Mechanisms, Evaluation}}. In \bibinfo{booktitle}{\emph{ICCD}}
  \bibinfo{year}{2016}\natexlab{b}.
\newblock


\bibitem[\protect\citeauthoryear{Huang, Chang, El~Hajj, Garcia~de Gonzalo,
  G\'{o}mez-Luna, Chalamalasetti, El-Hadedy, Milojicic, Mutlu, Chen, and
  Hwu}{Huang et~al\mbox{.}}{2019}]%
        {chai_icpe19}
\bibfield{author}{\bibinfo{person}{Sitao Huang}, \bibinfo{person}{Li-Wen
  Chang}, \bibinfo{person}{Izzat El~Hajj}, \bibinfo{person}{Simon Garcia~de
  Gonzalo}, \bibinfo{person}{Juan G\'{o}mez-Luna}, \bibinfo{person}{Sai~Rahul
  Chalamalasetti}, \bibinfo{person}{Mohamed El-Hadedy}, \bibinfo{person}{Dejan
  Milojicic}, \bibinfo{person}{Onur Mutlu}, \bibinfo{person}{Deming Chen},
  {and} \bibinfo{person}{Wen-mei Hwu}.}
\newblock \showarticletitle{{Analysis and Modeling of Collaborative Execution
  Strategies for Heterogeneous {CPU-FPGA} Architectures}}. In
  \bibinfo{booktitle}{\emph{ICPE}} \bibinfo{year}{2019}\natexlab{}.
\newblock


\bibitem[\protect\citeauthoryear{Huynh, Wang, and Vincent}{Huynh
  et~al\mbox{.}}{2014}]%
        {huynh2014high}
\bibfield{author}{\bibinfo{person}{HT Huynh}, \bibinfo{person}{Zhi~J Wang},
  {and} \bibinfo{person}{Peter~E Vincent}.}
\newblock \showarticletitle{High-order methods for computational fluid
  dynamics: A brief review of compact differential formulations on unstructured
  grids}. In \bibinfo{booktitle}{\emph{Computers \& Fluids}}
  \bibinfo{year}{2014}\natexlab{}.
\newblock


\bibitem[\protect\citeauthoryear{Istv{\'a}n, Sidler, and Alonso}{Istv{\'a}n
  et~al\mbox{.}}{2017}]%
        {istvan2017caribou}
\bibfield{author}{\bibinfo{person}{Zsolt Istv{\'a}n}, \bibinfo{person}{David
  Sidler}, {and} \bibinfo{person}{Gustavo Alonso}.}
\newblock \showarticletitle{{Caribou: Intelligent Distributed Storage}}. In
  \bibinfo{booktitle}{\emph{VLDB}} \bibinfo{year}{2017}\natexlab{}.
\newblock


\bibitem[\protect\citeauthoryear{Jiang, Wang, Liu, G\'{o}mez-Luna, Guan, Deng,
  Zhang, and Mutlu}{Jiang et~al\mbox{.}}{2020}]%
        {jiang2020}
\bibfield{author}{\bibinfo{person}{Jiantong Jiang}, \bibinfo{person}{Zeke
  Wang}, \bibinfo{person}{Xue Liu}, \bibinfo{person}{Juan G\'{o}mez-Luna},
  \bibinfo{person}{Nan Guan}, \bibinfo{person}{Qingxu Deng},
  \bibinfo{person}{Wei Zhang}, {and} \bibinfo{person}{Onur Mutlu}.}
\newblock \showarticletitle{{Boyi: A Systematic Framework for Automatically
  Deciding the Right Execution Model of {OpenCL} Applications on {FPGAs}}}. In
  \bibinfo{booktitle}{\emph{FPGA}} \bibinfo{year}{2020}\natexlab{}.
\newblock


\bibitem[\protect\citeauthoryear{Jongerius, Wijnholds, Nijboer, and
  Corporaal}{Jongerius et~al\mbox{.}}{2014}]%
        {6898703}
\bibfield{author}{\bibinfo{person}{R. Jongerius}, \bibinfo{person}{S.
  Wijnholds}, \bibinfo{person}{R. Nijboer}, {and} \bibinfo{person}{H.
  Corporaal}.}
\newblock \showarticletitle{{An End-to-End Computing Model for the Square
  Kilometre Array}}. In \bibinfo{booktitle}{\emph{Computer}}
  \bibinfo{year}{2014}\natexlab{}.
\newblock


\bibitem[\protect\citeauthoryear{Jun, Liu, Lee, Hicks, Ankcorn, King, Xu,
  et~al\mbox{.}}{Jun et~al\mbox{.}}{2015}]%
        {jun2015bluedbm}
\bibfield{author}{\bibinfo{person}{Sang-Woo Jun}, \bibinfo{person}{Ming Liu},
  \bibinfo{person}{Sungjin Lee}, \bibinfo{person}{Jamey Hicks},
  \bibinfo{person}{John Ankcorn}, \bibinfo{person}{Myron King},
  \bibinfo{person}{Shuotao Xu}, {et~al\mbox{.}}}
\newblock \showarticletitle{{Blue{DBM}: {A}n Appliance for {B}ig {D}ata
  Analytics}}. In \bibinfo{booktitle}{\emph{ISCA}}
  \bibinfo{year}{2015}\natexlab{}.
\newblock


\bibitem[\protect\citeauthoryear{Kang, Kee, Miller, and Park}{Kang
  et~al\mbox{.}}{2013}]%
        {kang2013enabling}
\bibfield{author}{\bibinfo{person}{Yangwook Kang}, \bibinfo{person}{Yang-suk
  Kee}, \bibinfo{person}{Ethan~L Miller}, {and} \bibinfo{person}{Chanik Park}.}
\newblock \showarticletitle{{Enabling Cost-effective Data Processing with Smart
  SSD}}. In \bibinfo{booktitle}{\emph{MSST}} \bibinfo{year}{2013}\natexlab{}.
\newblock


\bibitem[\protect\citeauthoryear{Kara, Alistarh, Alonso, Mutlu, and Zhang}{Kara
  et~al\mbox{.}}{2017}]%
        {kara2017fpga}
\bibfield{author}{\bibinfo{person}{Kaan Kara}, \bibinfo{person}{Dan Alistarh},
  \bibinfo{person}{Gustavo Alonso}, \bibinfo{person}{Onur Mutlu}, {and}
  \bibinfo{person}{Ce Zhang}.}
\newblock \showarticletitle{{FPGA-accelerated {D}ense {L}inear {M}achine
  {L}earning: {A} {P}recision-{C}onvergence {T}rade-off}}. In
  \bibinfo{booktitle}{\emph{FCCM}} \bibinfo{year}{2017}\natexlab{}.
\newblock


\bibitem[\protect\citeauthoryear{Kara, Hagleitner, Diamantopoulos, Syrivelis,
  and Alonso}{Kara et~al\mbox{.}}{2020}]%
        {kara2020hbm}
\bibfield{author}{\bibinfo{person}{Kaan Kara}, \bibinfo{person}{Christoph
  Hagleitner}, \bibinfo{person}{Dionysios Diamantopoulos},
  \bibinfo{person}{Dimitris Syrivelis}, {and} \bibinfo{person}{Gustavo
  Alonso}.}
\newblock \showarticletitle{High Bandwidth Memory on FPGAs: A Data Analytics
  Perspective}. In \bibinfo{booktitle}{\emph{FPL}}
  \bibinfo{year}{2020}\natexlab{}.
\newblock


\bibitem[\protect\citeauthoryear{{Ke}, {Gupta}, {Cho}, {Brooks}, {Chandra},
  {Diril}, {Firoozshahian}, {Hazelwood}, {Jia}, {Lee}, {Li}, {Maher},
  {Mudigere}, {Naumov}, {Schatz}, {Smelyanskiy}, {Wang}, {Reagen}, {Wu},
  {Hempstead}, and {Zhang}}{{Ke} et~al\mbox{.}}{2020}]%
        {ke2020recnmp}
\bibfield{author}{\bibinfo{person}{L. {Ke}}, \bibinfo{person}{U. {Gupta}},
  \bibinfo{person}{B.~Y. {Cho}}, \bibinfo{person}{D. {Brooks}},
  \bibinfo{person}{V. {Chandra}}, \bibinfo{person}{U. {Diril}},
  \bibinfo{person}{A. {Firoozshahian}}, \bibinfo{person}{K. {Hazelwood}},
  \bibinfo{person}{B. {Jia}}, \bibinfo{person}{H.~S. {Lee}},
  \bibinfo{person}{M. {Li}}, \bibinfo{person}{B. {Maher}}, \bibinfo{person}{D.
  {Mudigere}}, \bibinfo{person}{M. {Naumov}}, \bibinfo{person}{M. {Schatz}},
  \bibinfo{person}{M. {Smelyanskiy}}, \bibinfo{person}{X. {Wang}},
  \bibinfo{person}{B. {Reagen}}, \bibinfo{person}{C. {Wu}}, \bibinfo{person}{M.
  {Hempstead}}, {and} \bibinfo{person}{X. {Zhang}}.}
\newblock \showarticletitle{{RecNMP: Accelerating personalized recommendation
  with near-memory processing}}. In \bibinfo{booktitle}{\emph{ISCA}}
  \bibinfo{year}{2020}\natexlab{}.
\newblock


\bibitem[\protect\citeauthoryear{Kehler, Hanesiak, Curry, Sills, and
  Taylor}{Kehler et~al\mbox{.}}{2016}]%
        {kehler2016high}
\bibfield{author}{\bibinfo{person}{Scott Kehler}, \bibinfo{person}{John
  Hanesiak}, \bibinfo{person}{Michelle Curry}, \bibinfo{person}{David Sills},
  {and} \bibinfo{person}{Neil Taylor}.}
\newblock \showarticletitle{{High {R}esolution {D}eterministic {P}rediction
  {S}ystem ({HRDPS}) {S}imulations of {M}anitoba {L}ake {B}reezes}}. In
  \bibinfo{booktitle}{\emph{Atmosphere-Ocean}} \bibinfo{year}{2016}\natexlab{}.
\newblock


\bibitem[\protect\citeauthoryear{Kim, Kung, Chai, Yalamanchili, and
  Mukhopadhyay}{Kim et~al\mbox{.}}{2016}]%
        {kim2016neurocube}
\bibfield{author}{\bibinfo{person}{Duckhwan Kim}, \bibinfo{person}{Jaeha Kung},
  \bibinfo{person}{Sek Chai}, \bibinfo{person}{Sudhakar Yalamanchili}, {and}
  \bibinfo{person}{Saibal Mukhopadhyay}.}
\newblock \showarticletitle{{Neurocube: A Programmable Digital Neuromorphic
  Architecture with High-Density 3D Memory}}. In
  \bibinfo{booktitle}{\emph{ISCA}} \bibinfo{year}{2016}\natexlab{}.
\newblock


\bibitem[\protect\citeauthoryear{{Kim}, {Oh}, {Lee}, {Lee}, {Hwang}, {Hwang},
  {Na}, {Moon}, {Kim}, {Park}, {Ryu}, {Park}, {Kang}, {Kim}, {Kim}, {Bang},
  {Cho}, {Jang}, {Han}, {LeeLee}, {Choi}, and {Jun}}{{Kim}
  et~al\mbox{.}}{2012}]%
        {6025219}
\bibfield{author}{\bibinfo{person}{J. {Kim}}, \bibinfo{person}{C.~S. {Oh}},
  \bibinfo{person}{H. {Lee}}, \bibinfo{person}{D. {Lee}},
  \bibinfo{person}{H.~R. {Hwang}}, \bibinfo{person}{S. {Hwang}},
  \bibinfo{person}{B. {Na}}, \bibinfo{person}{J. {Moon}}, \bibinfo{person}{J.
  {Kim}}, \bibinfo{person}{H. {Park}}, \bibinfo{person}{J. {Ryu}},
  \bibinfo{person}{K. {Park}}, \bibinfo{person}{S.~K. {Kang}},
  \bibinfo{person}{S. {Kim}}, \bibinfo{person}{H. {Kim}}, \bibinfo{person}{J.
  {Bang}}, \bibinfo{person}{H. {Cho}}, \bibinfo{person}{M. {Jang}},
  \bibinfo{person}{C. {Han}}, \bibinfo{person}{J. {LeeLee}},
  \bibinfo{person}{J.~S. {Choi}}, {and} \bibinfo{person}{Y. {Jun}}.}
\newblock \showarticletitle{{A 1.2 V 12.8 GB/s 2 Gb Mobile Wide-I/O DRAM With
  4$\times$128 I/Os Using TSV Based Stacking}}. In
  \bibinfo{booktitle}{\emph{JSSC}} \bibinfo{year}{2012}\natexlab{}.
\newblock


\bibitem[\protect\citeauthoryear{Kim, Cali, Xin, Lee, Ghose, Alser, Hassan,
  Ergin, Alkan, and Mutlu}{Kim et~al\mbox{.}}{2018}]%
        {kim2018grim}
\bibfield{author}{\bibinfo{person}{Jeremie~S Kim}, \bibinfo{person}{Damla~Senol
  Cali}, \bibinfo{person}{Hongyi Xin}, \bibinfo{person}{Donghyuk Lee},
  \bibinfo{person}{Saugata Ghose}, \bibinfo{person}{Mohammed Alser},
  \bibinfo{person}{Hasan Hassan}, \bibinfo{person}{Oguz Ergin},
  \bibinfo{person}{Can Alkan}, {and} \bibinfo{person}{Onur Mutlu}.}
\newblock \showarticletitle{{GRIM-Filter: Fast seed location filtering in DNA
  read mapping using processing-in-memory technologies}}. In
  \bibinfo{booktitle}{\emph{BMC Genomics}} \bibinfo{year}{2018}\natexlab{}.
\newblock


\bibitem[\protect\citeauthoryear{Koo, Matam, I, Narra, Li, Tseng, Swanson, and
  Annavaram}{Koo et~al\mbox{.}}{2017}]%
        {koo2017summarizer}
\bibfield{author}{\bibinfo{person}{Gunjae Koo}, \bibinfo{person}{Kiran~Kumar
  Matam}, \bibinfo{person}{Te I}, \bibinfo{person}{H.~V. Krishna~Giri Narra},
  \bibinfo{person}{Jing Li}, \bibinfo{person}{Hung-Wei Tseng},
  \bibinfo{person}{Steven Swanson}, {and} \bibinfo{person}{Murali Annavaram}.}
\newblock \showarticletitle{{Summarizer: Trading Communication with Computing
  Near Storage}}. In \bibinfo{booktitle}{\emph{MICRO}}
  \bibinfo{year}{2017}\natexlab{}.
\newblock


\bibitem[\protect\citeauthoryear{Kwon, Han~Lee, Lee, Kwon, Min~Ryu, Son, O, Yu,
  Lee, Young~Kim, Cho, Guk~Kim, Choi, Shin, Kim, Phuah, Kim, Jun~Song, Choi,
  Kim, Kim, Kim, Wang, Kang, Ro, Seo, Song, Youn, Sohn, and Sung~Kim}{Kwon
  et~al\mbox{.}}{2021}]%
        {fimdram2021ISSCC}
\bibfield{author}{\bibinfo{person}{Young-Cheon Kwon}, \bibinfo{person}{Suk
  Han~Lee}, \bibinfo{person}{Jaehoon Lee}, \bibinfo{person}{Sang-Hyuk Kwon},
  \bibinfo{person}{Je Min~Ryu}, \bibinfo{person}{Jong-Pil Son},
  \bibinfo{person}{Seongil O}, \bibinfo{person}{Hak-Soo Yu},
  \bibinfo{person}{Haesuk Lee}, \bibinfo{person}{Soo Young~Kim},
  \bibinfo{person}{Youngmin Cho}, \bibinfo{person}{Jin Guk~Kim},
  \bibinfo{person}{Jongyoon Choi}, \bibinfo{person}{Hyun-Sung Shin},
  \bibinfo{person}{Jin Kim}, \bibinfo{person}{BengSeng Phuah},
  \bibinfo{person}{HyoungMin Kim}, \bibinfo{person}{Myeong Jun~Song},
  \bibinfo{person}{Ahn Choi}, \bibinfo{person}{Daeho Kim},
  \bibinfo{person}{SooYoung Kim}, \bibinfo{person}{Eun-Bong Kim},
  \bibinfo{person}{David Wang}, \bibinfo{person}{Shinhaeng Kang},
  \bibinfo{person}{Yuhwan Ro}, \bibinfo{person}{Seungwoo Seo},
  \bibinfo{person}{JoonHo Song}, \bibinfo{person}{Jaeyoun Youn},
  \bibinfo{person}{Kyomin Sohn}, {and} \bibinfo{person}{Nam Sung~Kim}.}
\newblock \showarticletitle{{A 20nm 6GB Function-In-Memory DRAM, Based on HBM2
  with a 1.2TFLOPS Programmable Computing Unit Using Bank-Level Parallelism,
  for Machine Learning Applications}}. In \bibinfo{booktitle}{\emph{ISSCC}}
  \bibinfo{year}{2021}\natexlab{}.
\newblock


\bibitem[\protect\citeauthoryear{Lai, Chi, Hu, Wang, Yu, Zhou, Cong, and
  Zhang}{Lai et~al\mbox{.}}{2019}]%
        {lai2019heterocl}
\bibfield{author}{\bibinfo{person}{Yi-Hsiang Lai}, \bibinfo{person}{Yuze Chi},
  \bibinfo{person}{Yuwei Hu}, \bibinfo{person}{Jie Wang},
  \bibinfo{person}{Cody~Hao Yu}, \bibinfo{person}{Yuan Zhou},
  \bibinfo{person}{Jason Cong}, {and} \bibinfo{person}{Zhiru Zhang}.}
\newblock \showarticletitle{{HeteroCL: A Multi-Paradigm Programming
  Infrastructure for Software-Defined Reconfigurable Computing}}. In
  \bibinfo{booktitle}{\emph{FPGA}} \bibinfo{year}{2019}\natexlab{}.
\newblock


\bibitem[\protect\citeauthoryear{Lee, Ghose, Pekhimenko, Khan, and Mutlu}{Lee
  et~al\mbox{.}}{2016}]%
        {lee2016smla}
\bibfield{author}{\bibinfo{person}{Donghyuk Lee}, \bibinfo{person}{Saugata
  Ghose}, \bibinfo{person}{Gennady Pekhimenko}, \bibinfo{person}{Samira Khan},
  {and} \bibinfo{person}{Onur Mutlu}.}
\newblock \showarticletitle{{Simultaneous Multi-Layer Access: Improving
  {3D}-Stacked Memory Bandwidth at Low Cost}}. In \bibinfo{booktitle}{\emph{ACM
  TACO}} \bibinfo{year}{2016}\natexlab{}.
\newblock


\bibitem[\protect\citeauthoryear{{Lee}, {Kim}, {Kim}, {Kim}, {Kim}, {Park},
  {Kim}, {Kim}, {Park}, {Shin}, {Cho}, {Kwon}, {Kim}, {Lee}, {Park}, {Chung},
  and {Hong}}{{Lee} et~al\mbox{.}}{2014}]%
        {6757501}
\bibfield{author}{\bibinfo{person}{D.~U. {Lee}}, \bibinfo{person}{K.~W. {Kim}},
  \bibinfo{person}{K.~W. {Kim}}, \bibinfo{person}{H. {Kim}},
  \bibinfo{person}{J.~Y. {Kim}}, \bibinfo{person}{Y.~J. {Park}},
  \bibinfo{person}{J.~H. {Kim}}, \bibinfo{person}{D.~S. {Kim}},
  \bibinfo{person}{H.~B. {Park}}, \bibinfo{person}{J.~W. {Shin}},
  \bibinfo{person}{J.~H. {Cho}}, \bibinfo{person}{K.~H. {Kwon}},
  \bibinfo{person}{M.~J. {Kim}}, \bibinfo{person}{J. {Lee}},
  \bibinfo{person}{K.~W. {Park}}, \bibinfo{person}{B. {Chung}}, {and}
  \bibinfo{person}{S. {Hong}}.}
\newblock \showarticletitle{{25.2 {A} 1.2{V} 8{G}b 8-{C}hannel 128{GB}/s
  {H}igh-{B}andwidth {M}emory ({HBM}) {S}tacked {DRAM} with {E}ffective
  {M}icrobump {I/O T}est {M}ethods Using 29nm {P}rocess and {TSV}}}. In
  \bibinfo{booktitle}{\emph{ISSCC}} \bibinfo{year}{2014}\natexlab{}.
\newblock


\bibitem[\protect\citeauthoryear{Lee, Kim, Yoo, Choi, Hofstee, Nam, Nutter, and
  Jamsek}{Lee et~al\mbox{.}}{2017}]%
        {10.14778/3137765.3137776}
\bibfield{author}{\bibinfo{person}{Jinho Lee}, \bibinfo{person}{Heesu Kim},
  \bibinfo{person}{Sungjoo Yoo}, \bibinfo{person}{Kiyoung Choi},
  \bibinfo{person}{H.~Peter Hofstee}, \bibinfo{person}{Gi-Joon Nam},
  \bibinfo{person}{Mark~R. Nutter}, {and} \bibinfo{person}{Damir Jamsek}.}
\newblock \showarticletitle{{ExtraV: Boosting Graph Processing near Storage
  with a Coherent Accelerator}}. In \bibinfo{booktitle}{\emph{VLDB}}
  \bibinfo{year}{2017}\natexlab{}.
\newblock


\bibitem[\protect\citeauthoryear{Lee, Sim, and Kim}{Lee et~al\mbox{.}}{2015}]%
        {lee2015bssync}
\bibfield{author}{\bibinfo{person}{Joo~Hwan Lee}, \bibinfo{person}{Jaewoong
  Sim}, {and} \bibinfo{person}{Hyesoon Kim}.}
\newblock \showarticletitle{{BSSync: Processing Near Memory for Machine
  Learning Workloads with Bounded Staleness Consistency Models}}. In
  \bibinfo{booktitle}{\emph{PACT}} \bibinfo{year}{2015}\natexlab{}.
\newblock


\bibitem[\protect\citeauthoryear{Lee, Kang, Lee, Kim, Lee, Seo, Yoon, Lee, Lim,
  Shin, Kim, O, Iyer, Wang, Sohn, and Sung~Kim}{Lee et~al\mbox{.}}{2021}]%
        {fimdram2021ISCA}
\bibfield{author}{\bibinfo{person}{Sukhan Lee}, \bibinfo{person}{Shin-haeng
  Kang}, \bibinfo{person}{Jaehoon Lee}, \bibinfo{person}{Hyeonsu Kim},
  \bibinfo{person}{Eojin Lee}, \bibinfo{person}{Seungwoo Seo},
  \bibinfo{person}{Hosang Yoon}, \bibinfo{person}{Seungwon Lee},
  \bibinfo{person}{Kyounghwan Lim}, \bibinfo{person}{Hyunsung Shin},
  \bibinfo{person}{Jinhyun Kim}, \bibinfo{person}{Seongil O},
  \bibinfo{person}{Anand Iyer}, \bibinfo{person}{David Wang},
  \bibinfo{person}{Kyomin Sohn}, {and} \bibinfo{person}{Nam Sung~Kim}.}
\newblock \showarticletitle{{Hardware Architecture and Software Stack for FIM
  Based on Commercial DRAM Technology}}. In \bibinfo{booktitle}{\emph{ISCA}}
  \bibinfo{year}{2021}\natexlab{}.
\newblock


\bibitem[\protect\citeauthoryear{Lee, Mazumdar, del Mundo, Alaghi, Ceze, and
  Oskin}{Lee et~al\mbox{.}}{2018}]%
        {lee2018application}
\bibfield{author}{\bibinfo{person}{Vincent~T Lee}, \bibinfo{person}{Amrita
  Mazumdar}, \bibinfo{person}{Carlo~C del Mundo}, \bibinfo{person}{Armin
  Alaghi}, \bibinfo{person}{Luis Ceze}, {and} \bibinfo{person}{Mark Oskin}.}
\newblock \showarticletitle{{Application Codesign of Near-Data Processing for
  Similarity Search}}. In \bibinfo{booktitle}{\emph{IPDPS}}
  \bibinfo{year}{2018}\natexlab{}.
\newblock


\bibitem[\protect\citeauthoryear{Li, Chi, and Cong}{Li et~al\mbox{.}}{2020}]%
        {li2020heterohalide}
\bibfield{author}{\bibinfo{person}{Jiajie Li}, \bibinfo{person}{Yuze Chi},
  {and} \bibinfo{person}{Jason Cong}.}
\newblock \showarticletitle{HeteroHalide: From image processing DSL to
  efficient FPGA acceleration}. In \bibinfo{booktitle}{\emph{FPGA}}
  \bibinfo{year}{2020}\natexlab{}.
\newblock


\bibitem[\protect\citeauthoryear{Li, Xu, Zou, Zhao, Lu, and Xie}{Li
  et~al\mbox{.}}{2016}]%
        {li2016pinatubo}
\bibfield{author}{\bibinfo{person}{Shuangchen Li}, \bibinfo{person}{Cong Xu},
  \bibinfo{person}{Qiaosha Zou}, \bibinfo{person}{Jishen Zhao},
  \bibinfo{person}{Yu Lu}, {and} \bibinfo{person}{Yuan Xie}.}
\newblock \showarticletitle{{Pinatubo: A Processing-in-Memory Architecture for
  Bulk Bitwise Operations in Emerging Non-volatile Memories}}. In
  \bibinfo{booktitle}{\emph{DAC}} \bibinfo{year}{2016}\natexlab{}.
\newblock


\bibitem[\protect\citeauthoryear{Liu, Zhao, Ogleari, Li, and Zhao}{Liu
  et~al\mbox{.}}{2018}]%
        {liu2018processing}
\bibfield{author}{\bibinfo{person}{Jiawen Liu}, \bibinfo{person}{Hengyu Zhao},
  \bibinfo{person}{Matheus~A Ogleari}, \bibinfo{person}{Dong Li}, {and}
  \bibinfo{person}{Jishen Zhao}.}
\newblock \showarticletitle{{Processing-in-Memory for Energy-efficient Neural
  Network Training: A Heterogeneous Approach}}. In
  \bibinfo{booktitle}{\emph{MICRO}} \bibinfo{year}{2018}\natexlab{}.
\newblock


\bibitem[\protect\citeauthoryear{Liu, Calciu, Herlihy, and Mutlu}{Liu
  et~al\mbox{.}}{2017}]%
        {liu2017concurrent}
\bibfield{author}{\bibinfo{person}{Zhiyu Liu}, \bibinfo{person}{Irina Calciu},
  \bibinfo{person}{Maurice Herlihy}, {and} \bibinfo{person}{Onur Mutlu}.}
\newblock \showarticletitle{{Concurrent Data Structures for Near-Memory
  Computing}}. In \bibinfo{booktitle}{\emph{SPAA}}
  \bibinfo{year}{2017}\natexlab{}.
\newblock


\bibitem[\protect\citeauthoryear{Mayhew and Krishnan}{Mayhew and
  Krishnan}{2003}]%
        {pcie}
\bibfield{author}{\bibinfo{person}{David Mayhew} {and} \bibinfo{person}{Venkata
  Krishnan}.}
\newblock \showarticletitle{{{PCI} {E}xpress and {A}dvanced {S}witching:
  Evolutionary Path to Building Next Generation Interconnects}}. In
  \bibinfo{booktitle}{\emph{HOTI}} \bibinfo{year}{2003}\natexlab{}.
\newblock


\bibitem[\protect\citeauthoryear{Meng and Skadron}{Meng and Skadron}{2011}]%
        {meng2011performance}
\bibfield{author}{\bibinfo{person}{Jiayuan Meng} {and} \bibinfo{person}{Kevin
  Skadron}.}
\newblock \showarticletitle{{A {P}erformance {S}tudy for {I}terative {S}tencil
  {L}oops on {GPU}s with {G}host {Z}one {O}ptimizations}}. In
  \bibinfo{booktitle}{\emph{IJPP}} \bibinfo{year}{2011}\natexlab{}.
\newblock


\bibitem[\protect\citeauthoryear{Microsoft}{Microsoft}{[n.d.]}]%
        {FPGA_in_azure}
\bibfield{author}{\bibinfo{person}{Microsoft}.}
\newblock \bibinfo{booktitle}{\emph{Deploy ML models to field-programmable gate
  arrays (FPGAs) with Azure Machine Learning,
  \url{https://docs.microsoft.com/en-us/azure/machine-learning/how-to-deploy-fpga-web-service}}}.
\newblock


\bibitem[\protect\citeauthoryear{Morad, Yavits, and Ginosar}{Morad
  et~al\mbox{.}}{2015}]%
        {morad2015gp}
\bibfield{author}{\bibinfo{person}{Amir Morad}, \bibinfo{person}{Leonid
  Yavits}, {and} \bibinfo{person}{Ran Ginosar}.}
\newblock \showarticletitle{{GP-SIMD Processing-in-Memory}}. In
  \bibinfo{booktitle}{\emph{ACM TACO}} \bibinfo{year}{2015}\natexlab{}.
\newblock


\bibitem[\protect\citeauthoryear{Mutlu}{Mutlu}{2021}]%
        {mutlu2021intelligent}
\bibfield{author}{\bibinfo{person}{Onur Mutlu}.}
\newblock \showarticletitle{{Intelligent Architectures for Intelligent
  Computing Systems}}. In \bibinfo{booktitle}{\emph{DATE}}
  \bibinfo{year}{2021}\natexlab{}.
\newblock


\bibitem[\protect\citeauthoryear{Mutlu, Ghose, G{\'o}mez-Luna, and
  Ausavarungnirun}{Mutlu et~al\mbox{.}}{2019a}]%
        {mutlu2019enabling}
\bibfield{author}{\bibinfo{person}{Onur Mutlu}, \bibinfo{person}{Saugata
  Ghose}, \bibinfo{person}{Juan G{\'o}mez-Luna}, {and} \bibinfo{person}{Rachata
  Ausavarungnirun}.}
\newblock \showarticletitle{{Enabling Practical Processing in and near Memory
  for Data-Intensive Computing}}. In \bibinfo{booktitle}{\emph{DAC}}
  \bibinfo{year}{2019}\natexlab{a}.
\newblock


\bibitem[\protect\citeauthoryear{Mutlu, Ghose, G{\'o}mez-Luna, and
  Ausavarungnirun}{Mutlu et~al\mbox{.}}{2019b}]%
        {mutlu2019}
\bibfield{author}{\bibinfo{person}{Onur Mutlu}, \bibinfo{person}{Saugata
  Ghose}, \bibinfo{person}{Juan G{\'o}mez-Luna}, {and} \bibinfo{person}{Rachata
  Ausavarungnirun}.}
\newblock \showarticletitle{{Processing Data Where It Makes Sense: {E}nabling
  In-Memory Computation}}. In \bibinfo{booktitle}{\emph{MicPro}}
  \bibinfo{year}{2019}\natexlab{b}.
\newblock


\bibitem[\protect\citeauthoryear{Mutlu, Ghose, G{\'o}mez-Luna, and
  Ausavarungnirun}{Mutlu et~al\mbox{.}}{2021}]%
        {mutlu2020modern}
\bibfield{author}{\bibinfo{person}{Onur Mutlu}, \bibinfo{person}{Saugata
  Ghose}, \bibinfo{person}{Juan G{\'o}mez-Luna}, {and} \bibinfo{person}{Rachata
  Ausavarungnirun}.}
\newblock \showarticletitle{{A Modern Primer on Processing in Memor}}. In
  \bibinfo{booktitle}{\emph{Emerging Computing: From Devices to Systems-Looking
  Beyond Moore and Von Neumann. Springer}} \bibinfo{year}{2021}\natexlab{}.
\newblock


\bibitem[\protect\citeauthoryear{{Nai}, {Hadidi}, {Sim}, {Kim}, {Kumar}, and
  {Kim}}{{Nai} et~al\mbox{.}}{2017}]%
        {nai2017graphpim}
\bibfield{author}{\bibinfo{person}{L. {Nai}}, \bibinfo{person}{R. {Hadidi}},
  \bibinfo{person}{J. {Sim}}, \bibinfo{person}{H. {Kim}}, \bibinfo{person}{P.
  {Kumar}}, {and} \bibinfo{person}{H. {Kim}}.}
\newblock \showarticletitle{{GraphPIM: Enabling Instruction-Level PIM
  Offloading in Graph Computing Frameworks}}. In
  \bibinfo{booktitle}{\emph{HPCA}} \bibinfo{year}{2017}\natexlab{}.
\newblock


\bibitem[\protect\citeauthoryear{{Nair}, {Antao}, {Bertolli}, {Bose},
  {Brunheroto}, {Chen}, {Cher}, {Costa}, {Doi}, {Evangelinos}, {Fleischer},
  {Fox}, {Gallo}, {Grinberg}, {Gunnels}, {Jacob}, {Jacob}, {Jacobson},
  {Karkhanis}, {Kim}, {Moreno}, {O'Brien}, {Ohmacht}, {Park}, {Prener},
  {Rosenburg}, {Ryu}, {Sallenave}, {Serrano}, {Siegl}, {Sugavanam}, and
  {Sura}}{{Nair} et~al\mbox{.}}{2015}]%
        {nair2015active}
\bibfield{author}{\bibinfo{person}{R. {Nair}}, \bibinfo{person}{S.~F. {Antao}},
  \bibinfo{person}{C. {Bertolli}}, \bibinfo{person}{P. {Bose}},
  \bibinfo{person}{J.~R. {Brunheroto}}, \bibinfo{person}{T. {Chen}},
  \bibinfo{person}{C.~. {Cher}}, \bibinfo{person}{C.~H.~A. {Costa}},
  \bibinfo{person}{J. {Doi}}, \bibinfo{person}{C. {Evangelinos}},
  \bibinfo{person}{B.~M. {Fleischer}}, \bibinfo{person}{T.~W. {Fox}},
  \bibinfo{person}{D.~S. {Gallo}}, \bibinfo{person}{L. {Grinberg}},
  \bibinfo{person}{J.~A. {Gunnels}}, \bibinfo{person}{A.~C. {Jacob}},
  \bibinfo{person}{P. {Jacob}}, \bibinfo{person}{H.~M. {Jacobson}},
  \bibinfo{person}{T. {Karkhanis}}, \bibinfo{person}{C. {Kim}},
  \bibinfo{person}{J.~H. {Moreno}}, \bibinfo{person}{J.~K. {O'Brien}},
  \bibinfo{person}{M. {Ohmacht}}, \bibinfo{person}{Y. {Park}},
  \bibinfo{person}{D.~A. {Prener}}, \bibinfo{person}{B.~S. {Rosenburg}},
  \bibinfo{person}{K.~D. {Ryu}}, \bibinfo{person}{O. {Sallenave}},
  \bibinfo{person}{M.~J. {Serrano}}, \bibinfo{person}{P.~D.~M. {Siegl}},
  \bibinfo{person}{K. {Sugavanam}}, {and} \bibinfo{person}{Z. {Sura}}.}
\newblock \showarticletitle{{Active Memory Cube: A Processing-in-Memory
  Architecture for Exascale Systems}}. In \bibinfo{booktitle}{\emph{IBM JRD}}
  \bibinfo{year}{2015}\natexlab{}.
\newblock


\bibitem[\protect\citeauthoryear{Navarro, Mohsen, Yan, Li, Gu, Meyerson, and
  Gerstein}{Navarro et~al\mbox{.}}{2019}]%
        {navarro2019genomics}
\bibfield{author}{\bibinfo{person}{F{\'a}bio~CP Navarro},
  \bibinfo{person}{Hussein Mohsen}, \bibinfo{person}{Chengfei Yan},
  \bibinfo{person}{Shantao Li}, \bibinfo{person}{Mengting Gu},
  \bibinfo{person}{William Meyerson}, {and} \bibinfo{person}{Mark Gerstein}.}
\newblock \showarticletitle{{Genomics and data science: an application within
  an umbrella}}. In \bibinfo{booktitle}{\emph{BMC}}
  \bibinfo{year}{2019}\natexlab{}.
\newblock


\bibitem[\protect\citeauthoryear{Neale, Chen, Gettelman, Lauritzen, Park,
  Williamson, Conley, Garcia, Kinnison, Lamarque, et~al\mbox{.}}{Neale
  et~al\mbox{.}}{2010}]%
        {neale2010description}
\bibfield{author}{\bibinfo{person}{Richard~B Neale},
  \bibinfo{person}{Chih-Chieh Chen}, \bibinfo{person}{Andrew Gettelman},
  \bibinfo{person}{Peter~H Lauritzen}, \bibinfo{person}{Sungsu Park},
  \bibinfo{person}{David~L Williamson}, \bibinfo{person}{Andrew~J Conley},
  \bibinfo{person}{Rolando Garcia}, \bibinfo{person}{Doug Kinnison},
  \bibinfo{person}{Jean-Francois Lamarque}, {et~al\mbox{.}}}
\newblock \showarticletitle{{Description of the {NCAR} Community Atmosphere
  Model ({CAM} 5.0)}}. In \bibinfo{booktitle}{\emph{NCAR Tech. Note}}
  \bibinfo{year}{2010}\natexlab{}.
\newblock


\bibitem[\protect\citeauthoryear{Oliveira, Gómez-Luna, Orosa, Ghose,
  Vijaykumar, Fernandez, Sadrosadati, and Mutlu}{Oliveira
  et~al\mbox{.}}{2021}]%
        {oliveira2021pimbench}
\bibfield{author}{\bibinfo{person}{Geraldo~Francisco Oliveira},
  \bibinfo{person}{Juan Gómez-Luna}, \bibinfo{person}{Lois Orosa},
  \bibinfo{person}{Saugata Ghose}, \bibinfo{person}{Nandita Vijaykumar},
  \bibinfo{person}{Ivan Fernandez}, \bibinfo{person}{Mohammad Sadrosadati},
  {and} \bibinfo{person}{Onur Mutlu}.}
\newblock \showarticletitle{{DAMOV: A New Methodology and Benchmark Suite for
  Evaluating Data Movement Bottlenecks}}. In \bibinfo{booktitle}{\emph{IEEE
  Access}} \bibinfo{year}{2021}\natexlab{}.
\newblock


\bibitem[\protect\citeauthoryear{{\"O}zi{\c{s}}ik, Orlande, Cola{\c{c}}o, and
  Cotta}{{\"O}zi{\c{s}}ik et~al\mbox{.}}{2017}]%
        {ozicsik2017finite}
\bibfield{author}{\bibinfo{person}{M~Necati {\"O}zi{\c{s}}ik},
  \bibinfo{person}{Helcio~RB Orlande}, \bibinfo{person}{Marcelo~J
  Cola{\c{c}}o}, {and} \bibinfo{person}{Renato~M Cotta}.}
  \bibinfo{year}{2017}\natexlab{}.
\newblock \bibinfo{booktitle}{\emph{Finite difference methods in heat
  transfer}}.
\newblock \bibinfo{publisher}{CRC Press}.
\newblock


\bibitem[\protect\citeauthoryear{Park, Kim, Yun, Lee, Rhu, and Ahn}{Park
  et~al\mbox{.}}{2021}]%
        {park2021trim}
\bibfield{author}{\bibinfo{person}{Jaehyun Park}, \bibinfo{person}{Byeongho
  Kim}, \bibinfo{person}{Sungmin Yun}, \bibinfo{person}{Eojin Lee},
  \bibinfo{person}{Minsoo Rhu}, {and} \bibinfo{person}{Jung~Ho Ahn}.}
\newblock \showarticletitle{{TRiM: Enhancing Processor-Memory Interfaces with
  Scalable Tensor Reduction in Memory}}. In \bibinfo{booktitle}{\emph{MICRO}}
  \bibinfo{year}{2021}\natexlab{}.
\newblock


\bibitem[\protect\citeauthoryear{Pattnaik, Tang, Jog, Kayiran, Mishra,
  Kandemir, Mutlu, and Das}{Pattnaik et~al\mbox{.}}{2016}]%
        {7756764}
\bibfield{author}{\bibinfo{person}{Ashutosh Pattnaik}, \bibinfo{person}{Xulong
  Tang}, \bibinfo{person}{Adwait Jog}, \bibinfo{person}{Onur Kayiran},
  \bibinfo{person}{Asit~K Mishra}, \bibinfo{person}{Mahmut~T Kandemir},
  \bibinfo{person}{Onur Mutlu}, {and} \bibinfo{person}{Chita~R Das}.}
\newblock \showarticletitle{{Scheduling Techniques for GPU Architectures with
  Processing-in-Memory Capabilities}}. In \bibinfo{booktitle}{\emph{PACT}}
  \bibinfo{year}{2016}\natexlab{}.
\newblock


\bibitem[\protect\citeauthoryear{Pawlowski}{Pawlowski}{2011}]%
        {7477494}
\bibfield{author}{\bibinfo{person}{J.~T. Pawlowski}.}
\newblock \showarticletitle{{Hybrid Memory Cube (HMC)}}. In
  \bibinfo{booktitle}{\emph{HCS}} \bibinfo{year}{2011}\natexlab{}.
\newblock


\bibitem[\protect\citeauthoryear{Pohl, Sattler, and Graefe}{Pohl
  et~al\mbox{.}}{2019}]%
        {hbm_joins}
\bibfield{author}{\bibinfo{person}{Constantin Pohl}, \bibinfo{person}{Kai-Uwe
  Sattler}, {and} \bibinfo{person}{Goetz Graefe}.}
\newblock \showarticletitle{Joins on high-bandwidth memory: a new level in the
  memory hierarchy}. In \bibinfo{booktitle}{\emph{VLDB}}
  \bibinfo{year}{2019}\natexlab{}.
\newblock


\bibitem[\protect\citeauthoryear{Pugsley, Jestes, Zhang, Balasubramonian,
  Srinivasan, Buyuktosunoglu, Davis, and Li}{Pugsley et~al\mbox{.}}{2014}]%
        {6844483}
\bibfield{author}{\bibinfo{person}{Seth~H Pugsley}, \bibinfo{person}{Jeffrey
  Jestes}, \bibinfo{person}{Huihui Zhang}, \bibinfo{person}{Rajeev
  Balasubramonian}, \bibinfo{person}{Vijayalakshmi Srinivasan},
  \bibinfo{person}{Alper Buyuktosunoglu}, \bibinfo{person}{Al Davis}, {and}
  \bibinfo{person}{Feifei Li}.}
\newblock \showarticletitle{{NDC: Analyzing the Impact of 3D-Stacked
  Memory+Logic Devices on MapReduce Workloads}}. In
  \bibinfo{booktitle}{\emph{ISPASS}} \bibinfo{year}{2014}\natexlab{}.
\newblock


\bibitem[\protect\citeauthoryear{Rojek et~al\mbox{.}}{Rojek
  et~al\mbox{.}}{2019}]%
        {mpdata}
\bibfield{author}{\bibinfo{person}{Krzysztof Rojek} {et~al\mbox{.}}}
\newblock \showarticletitle{{CFD A}cceleration with {FPGA}}. In
  \bibinfo{booktitle}{\emph{H2RC}} \bibinfo{year}{2019}\natexlab{}.
\newblock


\bibitem[\protect\citeauthoryear{Sadasivam, Thompto, Kalla, and
  Starke}{Sadasivam et~al\mbox{.}}{2017}]%
        {POWER9}
\bibfield{author}{\bibinfo{person}{Satish~Kumar Sadasivam},
  \bibinfo{person}{Brian~W Thompto}, \bibinfo{person}{Ron Kalla}, {and}
  \bibinfo{person}{William~J Starke}.}
\newblock \showarticletitle{{IBM POWER9 Processor Architecture}}. In
  \bibinfo{booktitle}{\emph{IEEE Micro}} \bibinfo{year}{2017}\natexlab{}.
\newblock


\bibitem[\protect\citeauthoryear{Sano, Hatsuda, and Yamamoto}{Sano
  et~al\mbox{.}}{2014}]%
        {sano2014multi}
\bibfield{author}{\bibinfo{person}{Kentaro Sano}, \bibinfo{person}{Yoshiaki
  Hatsuda}, {and} \bibinfo{person}{Satoru Yamamoto}.}
\newblock \showarticletitle{{Multi-{FPGA} {A}ccelerator for {S}calable
  {S}tencil {C}omputation with {C}onstant {M}emory {B}andwidth}}. In
  \bibinfo{booktitle}{\emph{TPDS}} \bibinfo{year}{2014}\natexlab{}.
\newblock


\bibitem[\protect\citeauthoryear{Santos, Oliveira, Tom{\'e}, Alves, Almeida,
  and Carro}{Santos et~al\mbox{.}}{2017}]%
        {7927081}
\bibfield{author}{\bibinfo{person}{Paulo~C Santos}, \bibinfo{person}{Geraldo~F
  Oliveira}, \bibinfo{person}{Diego~G Tom{\'e}}, \bibinfo{person}{Marco~AZ
  Alves}, \bibinfo{person}{Eduardo~C Almeida}, {and} \bibinfo{person}{Luigi
  Carro}.}
\newblock \showarticletitle{{Operand Size Reconfiguration for Big Data
  Processing in Memory}}. In \bibinfo{booktitle}{\emph{DATE}}
  \bibinfo{year}{2017}\natexlab{}.
\newblock


\bibitem[\protect\citeauthoryear{Sch{\"a}r, Fuhrer, Arteaga, Ban, Charpilloz,
  Di~Girolamo, Hentgen, Hoefler, Lapillonne, Leutwyler, Osterried, Panosetti,
  Rudishli, Schlemmer, Schulthess, Sprenger, Ubbiali, and Wernli}{Sch{\"a}r
  et~al\mbox{.}}{2020}]%
        {schar2020kilometer}
\bibfield{author}{\bibinfo{person}{Christoph Sch{\"a}r},
  \bibinfo{person}{Oliver Fuhrer}, \bibinfo{person}{Andrea Arteaga},
  \bibinfo{person}{Nikolina Ban}, \bibinfo{person}{Christophe Charpilloz},
  \bibinfo{person}{Salvatore Di~Girolamo}, \bibinfo{person}{Laureline Hentgen},
  \bibinfo{person}{Torsten Hoefler}, \bibinfo{person}{Xavier Lapillonne},
  \bibinfo{person}{David Leutwyler}, \bibinfo{person}{Katherine Osterried},
  \bibinfo{person}{Davide Panosetti}, \bibinfo{person}{Stefan Rudishli},
  \bibinfo{person}{Linda Schlemmer}, \bibinfo{person}{Thomas~C. Schulthess},
  \bibinfo{person}{Michael Sprenger}, \bibinfo{person}{Stefano Ubbiali}, {and}
  \bibinfo{person}{Heini Wernli}.}
\newblock \showarticletitle{{Kilometer-scale Climate Models: Prospects and
  Challenges}}. In \bibinfo{booktitle}{\emph{BAMS}}
  \bibinfo{year}{2020}\natexlab{}.
\newblock


\bibitem[\protect\citeauthoryear{Seshadri, Hsieh, Boroum, Lee, Kozuch, Mutlu,
  Gibbons, and Mowry}{Seshadri et~al\mbox{.}}{2015a}]%
        {seshadri2015fast}
\bibfield{author}{\bibinfo{person}{Vivek Seshadri}, \bibinfo{person}{Kevin
  Hsieh}, \bibinfo{person}{Amirali Boroum}, \bibinfo{person}{Donghyuk Lee},
  \bibinfo{person}{Michael~A Kozuch}, \bibinfo{person}{Onur Mutlu},
  \bibinfo{person}{Phillip~B Gibbons}, {and} \bibinfo{person}{Todd~C Mowry}.}
\newblock \showarticletitle{{Fast Bulk Bitwise AND and OR in DRAM}}. In
  \bibinfo{booktitle}{\emph{CAL}} \bibinfo{year}{2015}\natexlab{a}.
\newblock


\bibitem[\protect\citeauthoryear{Seshadri, Kim, Fallin, Lee, Ausavarungnirun,
  Pekhimenko, Luo, Mutlu, Gibbons, Kozuch, et~al\mbox{.}}{Seshadri
  et~al\mbox{.}}{2013}]%
        {seshadri2013rowclone}
\bibfield{author}{\bibinfo{person}{Vivek Seshadri}, \bibinfo{person}{Yoongu
  Kim}, \bibinfo{person}{Chris Fallin}, \bibinfo{person}{Donghyuk Lee},
  \bibinfo{person}{Rachata Ausavarungnirun}, \bibinfo{person}{Gennady
  Pekhimenko}, \bibinfo{person}{Yixin Luo}, \bibinfo{person}{Onur Mutlu},
  \bibinfo{person}{Phillip~B Gibbons}, \bibinfo{person}{Michael~A Kozuch},
  {et~al\mbox{.}}}
\newblock \showarticletitle{{RowClone: Fast and Energy-Efficient In-DRAM Bulk
  Data Copy and Initialization}}. In \bibinfo{booktitle}{\emph{MICRO}}
  \bibinfo{year}{2013}\natexlab{}.
\newblock


\bibitem[\protect\citeauthoryear{Seshadri, Lee, Mullins, Hassan, Boroumand,
  Kim, Kozuch, Mutlu, Gibbons, and Mowry}{Seshadri et~al\mbox{.}}{2016}]%
        {seshadri2016buddy}
\bibfield{author}{\bibinfo{person}{Vivek Seshadri}, \bibinfo{person}{Donghyuk
  Lee}, \bibinfo{person}{Thomas Mullins}, \bibinfo{person}{Hasan Hassan},
  \bibinfo{person}{Amirali Boroumand}, \bibinfo{person}{Jeremie Kim},
  \bibinfo{person}{Michael~A Kozuch}, \bibinfo{person}{Onur Mutlu},
  \bibinfo{person}{Phillip~B Gibbons}, {and} \bibinfo{person}{Todd~C Mowry}.}
\newblock \showarticletitle{{Buddy-RAM: Improving the Performance and
  Efficiency of Bulk Bitwise Operations Using DRAM}}. In
  \bibinfo{booktitle}{\emph{arXiv}} \bibinfo{year}{2016}\natexlab{}.
\newblock


\bibitem[\protect\citeauthoryear{Seshadri, Lee, Mullins, Hassan, Boroumand,
  Kim, Kozuch, Mutlu, Gibbons, and Mowry}{Seshadri et~al\mbox{.}}{2017}]%
        {seshadri2017ambit}
\bibfield{author}{\bibinfo{person}{Vivek Seshadri}, \bibinfo{person}{Donghyuk
  Lee}, \bibinfo{person}{Thomas Mullins}, \bibinfo{person}{Hasan Hassan},
  \bibinfo{person}{Amirali Boroumand}, \bibinfo{person}{Jeremie Kim},
  \bibinfo{person}{Michael~A Kozuch}, \bibinfo{person}{Onur Mutlu},
  \bibinfo{person}{Phillip~B Gibbons}, {and} \bibinfo{person}{Todd~C Mowry}.}
\newblock \showarticletitle{{Ambit: In-Memory Accelerator for Bulk Bitwise
  Operations Using Commodity DRAM Technology}}. In
  \bibinfo{booktitle}{\emph{MICRO}} \bibinfo{year}{2017}\natexlab{}.
\newblock


\bibitem[\protect\citeauthoryear{Seshadri, Mullins, Boroumand, Mutlu, Gibbons,
  Kozuch, and Mowry}{Seshadri et~al\mbox{.}}{2015b}]%
        {seshadri2015gather}
\bibfield{author}{\bibinfo{person}{Vivek Seshadri}, \bibinfo{person}{Thomas
  Mullins}, \bibinfo{person}{Amirali Boroumand}, \bibinfo{person}{Onur Mutlu},
  \bibinfo{person}{Phillip~B Gibbons}, \bibinfo{person}{Michael~A Kozuch},
  {and} \bibinfo{person}{Todd~C Mowry}.}
\newblock \showarticletitle{{Gather-Scatter DRAM: In-DRAM Address Translation
  to Improve the Spatial Locality of Non-unit Strided Accesses}}. In
  \bibinfo{booktitle}{\emph{MICRO}} \bibinfo{year}{2015}\natexlab{b}.
\newblock


\bibitem[\protect\citeauthoryear{Seshadri and Mutlu}{Seshadri and
  Mutlu}{2017}]%
        {seshadri2017simple}
\bibfield{author}{\bibinfo{person}{Vivek Seshadri} {and} \bibinfo{person}{Onur
  Mutlu}.} \bibinfo{year}{2017}\natexlab{}.
\newblock \showarticletitle{Simple operations in memory to reduce data
  movement}.
\newblock In \bibinfo{booktitle}{\emph{Advances in Computers}}.
\newblock


\bibitem[\protect\citeauthoryear{Seshadri and Mutlu}{Seshadri and
  Mutlu}{2019}]%
        {seshadri2019dram}
\bibfield{author}{\bibinfo{person}{Vivek Seshadri} {and} \bibinfo{person}{Onur
  Mutlu}.}
\newblock \showarticletitle{In-DRAM bulk bitwise execution engine}. In
  \bibinfo{booktitle}{\emph{arXiv}} \bibinfo{year}{2019}\natexlab{}.
\newblock


\bibitem[\protect\citeauthoryear{Sharma}{Sharma}{2019}]%
        {sharma2019compute}
\bibfield{author}{\bibinfo{person}{DD Sharma}.}
\newblock \showarticletitle{{Compute Express Link}}. In
  \bibinfo{booktitle}{\emph{CXL Consortium White Paper}}
  \bibinfo{year}{2019}\natexlab{}.
\newblock


\bibitem[\protect\citeauthoryear{Simon, Qureshi, Rios, Levisse, Zapater, and
  Atienza}{Simon et~al\mbox{.}}{2020}]%
        {simon2020blade}
\bibfield{author}{\bibinfo{person}{William~Andrew Simon},
  \bibinfo{person}{Yasir~Mahmood Qureshi}, \bibinfo{person}{Marco Rios},
  \bibinfo{person}{Alexandre Levisse}, \bibinfo{person}{Marina Zapater}, {and}
  \bibinfo{person}{David Atienza}.}
\newblock \showarticletitle{{BLADE: An in-Cache Computing Architecture for Edge
  Devices}}. In \bibinfo{booktitle}{\emph{TC}} \bibinfo{year}{2020}\natexlab{}.
\newblock


\bibitem[\protect\citeauthoryear{Singh et~al\mbox{.}}{Singh
  et~al\mbox{.}}{2019d}]%
        {NAPEL}
\bibfield{author}{\bibinfo{person}{Gagandeep Singh} {et~al\mbox{.}}}
\newblock \showarticletitle{{{NAPEL: N}ear-{M}emory {C}omputing {A}pplication
  {P}erformance {P}rediction via {E}nsemble {L}earning}}. In
  \bibinfo{booktitle}{\emph{DAC}} \bibinfo{year}{2019}\natexlab{d}.
\newblock


\bibitem[\protect\citeauthoryear{Singh, Alser, Cali, Diamantopoulos,
  G{\'o}mez-Luna, Corporaal, and Mutlu}{Singh et~al\mbox{.}}{2021a}]%
        {singh2021fpga}
\bibfield{author}{\bibinfo{person}{Gagandeep Singh}, \bibinfo{person}{Mohammed
  Alser}, \bibinfo{person}{Damla~Senol Cali}, \bibinfo{person}{Dionysios
  Diamantopoulos}, \bibinfo{person}{Juan G{\'o}mez-Luna}, \bibinfo{person}{Henk
  Corporaal}, {and} \bibinfo{person}{Onur Mutlu}.}
\newblock \showarticletitle{FPGA-based Near-Memory Acceleration of Modern
  Data-Intensive Applications}. In \bibinfo{booktitle}{\emph{IEEE Micro}}
  \bibinfo{year}{2021}\natexlab{a}.
\newblock


\bibitem[\protect\citeauthoryear{Singh, Chelini, Corda, Awan, Stuijk, Jordans,
  Corporaal, and Boonstra}{Singh et~al\mbox{.}}{2018}]%
        {singh2018review}
\bibfield{author}{\bibinfo{person}{Gagandeep Singh}, \bibinfo{person}{Lorenzo
  Chelini}, \bibinfo{person}{Stefano Corda}, \bibinfo{person}{Ahsan~Javed
  Awan}, \bibinfo{person}{Sander Stuijk}, \bibinfo{person}{Roel Jordans},
  \bibinfo{person}{Henk Corporaal}, {and} \bibinfo{person}{Albert-Jan
  Boonstra}.}
\newblock \showarticletitle{{A {R}eview of {N}ear-{M}emory {C}omputing
  {A}rchitectures: {O}pportunities and {C}hallenges}}. In
  \bibinfo{booktitle}{\emph{DSD}} \bibinfo{year}{2018}\natexlab{}.
\newblock


\bibitem[\protect\citeauthoryear{Singh, Chelini, Corda, Awan, Stuijk, Jordans,
  Corporaal, and Boonstra}{Singh et~al\mbox{.}}{2019a}]%
        {singh2019near}
\bibfield{author}{\bibinfo{person}{Gagandeep Singh}, \bibinfo{person}{Lorenzo
  Chelini}, \bibinfo{person}{Stefano Corda}, \bibinfo{person}{Ahsan~Javed
  Awan}, \bibinfo{person}{Sander Stuijk}, \bibinfo{person}{Roel Jordans},
  \bibinfo{person}{Henk Corporaal}, {and} \bibinfo{person}{Albert-Jan
  Boonstra}.}
\newblock \showarticletitle{{Near-Memory Computing: Past, Present, and
  Future}}. In \bibinfo{booktitle}{\emph{MicPro}}
  \bibinfo{year}{2019}\natexlab{a}.
\newblock


\bibitem[\protect\citeauthoryear{Singh, Diamantopolous, G{\'o}mez-Luna, Stuijk,
  Mutlu, and Corporaal}{Singh et~al\mbox{.}}{2021b}]%
        {singh2021modeling}
\bibfield{author}{\bibinfo{person}{Gagandeep Singh}, \bibinfo{person}{Dionysios
  Diamantopolous}, \bibinfo{person}{Juan G{\'o}mez-Luna},
  \bibinfo{person}{Sander Stuijk}, \bibinfo{person}{Onur Mutlu}, {and}
  \bibinfo{person}{Henk Corporaal}.}
\newblock \showarticletitle{{Modeling FPGA-Based Systems via Few-Shot
  Learning}}. In \bibinfo{booktitle}{\emph{FPGA}}
  \bibinfo{year}{2021}\natexlab{b}.
\newblock


\bibitem[\protect\citeauthoryear{Singh, Diamantopoulos, Hagleitner,
  G{\'o}mez-Luna, Stuijk, Mutlu, and Corporaal}{Singh et~al\mbox{.}}{2020}]%
        {singh2020nero}
\bibfield{author}{\bibinfo{person}{Gagandeep Singh}, \bibinfo{person}{Dionysios
  Diamantopoulos}, \bibinfo{person}{Christoph Hagleitner},
  \bibinfo{person}{Juan G{\'o}mez-Luna}, \bibinfo{person}{Sander Stuijk},
  \bibinfo{person}{Onur Mutlu}, {and} \bibinfo{person}{Henk Corporaal}.}
\newblock \showarticletitle{{NERO: A Near High-Bandwidth Memory Stencil
  Accelerator for Weather Prediction Modeling}}. In
  \bibinfo{booktitle}{\emph{FPL}} \bibinfo{year}{2020}\natexlab{}.
\newblock


\bibitem[\protect\citeauthoryear{Singh, Diamantopoulos, Hagleitner, Stuijk, and
  Corporaal}{Singh et~al\mbox{.}}{2019b}]%
        {narmada}
\bibfield{author}{\bibinfo{person}{Gagandeep Singh}, \bibinfo{person}{Dionysios
  Diamantopoulos}, \bibinfo{person}{Christoph Hagleitner},
  \bibinfo{person}{Sander Stuijk}, {and} \bibinfo{person}{Henk Corporaal}.}
\newblock \showarticletitle{{NARMADA}: {N}ear-memory horizontal diffusion
  accelerator for scalable stencil computations}. In
  \bibinfo{booktitle}{\emph{FPL}} \bibinfo{year}{2019}\natexlab{b}.
\newblock


\bibitem[\protect\citeauthoryear{Singh, Diamantopoulos, Stuijk, Hagleitner, and
  Corporaal}{Singh et~al\mbox{.}}{2019c}]%
        {singh2019low}
\bibfield{author}{\bibinfo{person}{Gagandeep Singh}, \bibinfo{person}{Dionysios
  Diamantopoulos}, \bibinfo{person}{Sander Stuijk}, \bibinfo{person}{Christoph
  Hagleitner}, {and} \bibinfo{person}{Henk Corporaal}.}
\newblock \showarticletitle{{Low Precision Processing for High Order Stencil
  Computations}}. In \bibinfo{booktitle}{\emph{Springer LNCS}}
  \bibinfo{year}{2019}\natexlab{c}.
\newblock


\bibitem[\protect\citeauthoryear{Strzodka, Shaheen, Pajak, and Seidel}{Strzodka
  et~al\mbox{.}}{2010}]%
        {strzodka2010cache}
\bibfield{author}{\bibinfo{person}{Robert Strzodka}, \bibinfo{person}{Mohammed
  Shaheen}, \bibinfo{person}{Dawid Pajak}, {and} \bibinfo{person}{Hans-Peter
  Seidel}.}
\newblock \showarticletitle{{Cache {O}blivious {P}arallelograms in {I}terative
  {S}tencil {C}omputations}}. In \bibinfo{booktitle}{\emph{ICS}}
  \bibinfo{year}{2010}\natexlab{}.
\newblock


\bibitem[\protect\citeauthoryear{Stuecheli et~al\mbox{.}}{Stuecheli
  et~al\mbox{.}}{2018}]%
        {openCAPI}
\bibfield{author}{\bibinfo{person}{Jeffrey Stuecheli} {et~al\mbox{.}}}
\newblock \showarticletitle{{I{BM POWER9} opens up a new era of acceleration
  enablement: Open{CAPI}}}. In \bibinfo{booktitle}{\emph{IBM JRD}}
  \bibinfo{year}{2018}\natexlab{}.
\newblock


\bibitem[\protect\citeauthoryear{Stuecheli, Blaner, Johns, and
  Siegel}{Stuecheli et~al\mbox{.}}{2015}]%
        {stuecheli2015capi}
\bibfield{author}{\bibinfo{person}{Jeffrey Stuecheli}, \bibinfo{person}{Bart
  Blaner}, \bibinfo{person}{CR Johns}, {and} \bibinfo{person}{MS Siegel}.}
\newblock \showarticletitle{{CAPI: A Coherent Accelerator Processor
  Interface}}. In \bibinfo{booktitle}{\emph{IBM JRD}}
  \bibinfo{year}{2015}\natexlab{}.
\newblock


\bibitem[\protect\citeauthoryear{{Sukhwani}, {Roewer}, {Haymes}, {Kim},
  {McPadden}, {Dreps}, {Sanner}, {Lunteren}, and {Asaad}}{{Sukhwani}
  et~al\mbox{.}}{2017}]%
        {ConTutto_2017_MICRO}
\bibfield{author}{\bibinfo{person}{B. {Sukhwani}}, \bibinfo{person}{T.
  {Roewer}}, \bibinfo{person}{C.~L. {Haymes}}, \bibinfo{person}{K. {Kim}},
  \bibinfo{person}{A.~J. {McPadden}}, \bibinfo{person}{D.~M. {Dreps}},
  \bibinfo{person}{D. {Sanner}}, \bibinfo{person}{J.~V. {Lunteren}}, {and}
  \bibinfo{person}{S. {Asaad}}.}
\newblock \showarticletitle{ConTutto – A Novel FPGA-based Prototyping
  Platform Enabling Innovation in the Memory Subsystem of a Server Class
  Processor}. In \bibinfo{booktitle}{\emph{MICRO}}
  \bibinfo{year}{2017}\natexlab{}.
\newblock


\bibitem[\protect\citeauthoryear{Szustak, Rojek, and Gepner}{Szustak
  et~al\mbox{.}}{2013}]%
        {szustak2013using}
\bibfield{author}{\bibinfo{person}{Lukasz Szustak}, \bibinfo{person}{Krzysztof
  Rojek}, {and} \bibinfo{person}{Pawel Gepner}.}
\newblock \showarticletitle{{Using {I}ntel {X}eon {P}hi Coprocessor to
  Accelerate Computations in {MPDATA} Algorithm}}. In
  \bibinfo{booktitle}{\emph{PPAM}} \bibinfo{year}{2013}\natexlab{}.
\newblock


\bibitem[\protect\citeauthoryear{Tang, Chowdhury, Kuszmaul, Luk, and
  Leiserson}{Tang et~al\mbox{.}}{2011}]%
        {tang2011pochoir}
\bibfield{author}{\bibinfo{person}{Yuan Tang}, \bibinfo{person}{Rezaul~Alam
  Chowdhury}, \bibinfo{person}{Bradley~C Kuszmaul}, \bibinfo{person}{Chi-Keung
  Luk}, {and} \bibinfo{person}{Charles~E Leiserson}.}
\newblock \showarticletitle{{The {P}ochoir {S}tencil {C}ompiler}}. In
  \bibinfo{booktitle}{\emph{SPAA}} \bibinfo{year}{2011}\natexlab{}.
\newblock


\bibitem[\protect\citeauthoryear{Thaler, Moosbrugger, Osuna, Bianco, Vogt,
  Afanasyev, Mosimann, Fuhrer, Schulthess, and Hoefler}{Thaler
  et~al\mbox{.}}{2019}]%
        {cosmo_knl}
\bibfield{author}{\bibinfo{person}{Felix Thaler}, \bibinfo{person}{Stefan
  Moosbrugger}, \bibinfo{person}{Carlos Osuna}, \bibinfo{person}{Mauro Bianco},
  \bibinfo{person}{Hannes Vogt}, \bibinfo{person}{Anton Afanasyev},
  \bibinfo{person}{Lukas Mosimann}, \bibinfo{person}{Oliver Fuhrer},
  \bibinfo{person}{Thomas~C Schulthess}, {and} \bibinfo{person}{Torsten
  Hoefler}.}
\newblock \showarticletitle{Porting the {COSMO W}eather {M}odel to {M}anycore
  {CPU}s}. In \bibinfo{booktitle}{\emph{PASC}} \bibinfo{year}{2019}\natexlab{}.
\newblock


\bibitem[\protect\citeauthoryear{Thomas}{Thomas}{1949}]%
        {thomas}
\bibfield{author}{\bibinfo{person}{Llewellyn Thomas}.}
\newblock \showarticletitle{{Elliptic Problems in {L}inear {D}ifferential
  {E}quations over a {N}etwork}}. In \bibinfo{booktitle}{\emph{Watson Sci.
  Comput. Lab. Rept., Columbia University}} \bibinfo{year}{1949}\natexlab{}.
\newblock


\bibitem[\protect\citeauthoryear{Tsai et~al\mbox{.}}{Tsai
  et~al\mbox{.}}{2017}]%
        {jenga}
\bibfield{author}{\bibinfo{person}{Po-An Tsai} {et~al\mbox{.}}}
\newblock \showarticletitle{{Jenga: {S}oftware-{D}efined {C}ache
  {H}ierarchies}}. In \bibinfo{booktitle}{\emph{ISCA}}
  \bibinfo{year}{2017}\natexlab{}.
\newblock


\bibitem[\protect\citeauthoryear{Tullsen, Eggers, and Levy}{Tullsen
  et~al\mbox{.}}{1995}]%
        {smt}
\bibfield{author}{\bibinfo{person}{Dean~M Tullsen}, \bibinfo{person}{Susan~J
  Eggers}, {and} \bibinfo{person}{Henry~M Levy}.}
\newblock \showarticletitle{{Simultaneous Multithreading: Maximizing On-Chip
  Parallelism}}. In \bibinfo{booktitle}{\emph{ISCA}}
  \bibinfo{year}{1995}\natexlab{}.
\newblock


\bibitem[\protect\citeauthoryear{van Lunteren, Luijten, Diamantopoulos,
  Auernhammer, Hagleitner, Chelini, Corda, and Singh}{van Lunteren
  et~al\mbox{.}}{2019}]%
        {chris}
\bibfield{author}{\bibinfo{person}{Jan van Lunteren}, \bibinfo{person}{Ronald
  Luijten}, \bibinfo{person}{Dionysios Diamantopoulos},
  \bibinfo{person}{Florian Auernhammer}, \bibinfo{person}{Christoph
  Hagleitner}, \bibinfo{person}{Lorenzo Chelini}, \bibinfo{person}{Stefano
  Corda}, {and} \bibinfo{person}{Gagandeep Singh}.}
\newblock \showarticletitle{Coherently {A}ttached {P}rogrammable
  {N}ear-{M}emory {A}cceleration {P}latform and its application to {S}tencil
  {P}rocessing}. In \bibinfo{booktitle}{\emph{DATE}}
  \bibinfo{year}{2019}\natexlab{}.
\newblock


\bibitem[\protect\citeauthoryear{Volkov and Demmel}{Volkov and Demmel}{2008}]%
        {volkov2008benchmarking}
\bibfield{author}{\bibinfo{person}{Vasily Volkov} {and}
  \bibinfo{person}{James~W Demmel}.}
\newblock \showarticletitle{Benchmarking GPUs to tune dense linear algebra}. In
  \bibinfo{booktitle}{\emph{SC}} \bibinfo{year}{2008}\natexlab{}.
\newblock


\bibitem[\protect\citeauthoryear{Wahib and Maruyama}{Wahib and
  Maruyama}{2014}]%
        {wahib2014scalable}
\bibfield{author}{\bibinfo{person}{Mohamed Wahib} {and} \bibinfo{person}{Naoya
  Maruyama}.}
\newblock \showarticletitle{{Scalable {K}ernel {F}usion for {M}emory-{B}ound
  {GPU} {A}pplications}}. In \bibinfo{booktitle}{\emph{SC}}
  \bibinfo{year}{2014}\natexlab{}.
\newblock


\bibitem[\protect\citeauthoryear{Waidyasooriya and Hariyama}{Waidyasooriya and
  Hariyama}{2019}]%
        {waidyasooriya2019multi}
\bibfield{author}{\bibinfo{person}{Hasitha~Muthumala Waidyasooriya} {and}
  \bibinfo{person}{Masanori Hariyama}.}
\newblock \showarticletitle{Multi-FPGA accelerator architecture for stencil
  computation exploiting spacial and temporal scalability}. In
  \bibinfo{booktitle}{\emph{IEEE Access}} \bibinfo{year}{2019}\natexlab{}.
\newblock


\bibitem[\protect\citeauthoryear{{Waidyasooriya}, {Takei}, {Tatsumi}, and
  {Hariyama}}{{Waidyasooriya} et~al\mbox{.}}{2017}]%
        {7582502}
\bibfield{author}{\bibinfo{person}{H.~M. {Waidyasooriya}}, \bibinfo{person}{Y.
  {Takei}}, \bibinfo{person}{S. {Tatsumi}}, {and} \bibinfo{person}{M.
  {Hariyama}}.}
\newblock \showarticletitle{{Open{CL-B}ased {FPGA-P}latform for {S}tencil
  {C}omputation and {I}ts {O}ptimization {M}ethodology}}. In
  \bibinfo{booktitle}{\emph{TPDS}} \bibinfo{year}{2017}\natexlab{}.
\newblock


\bibitem[\protect\citeauthoryear{Wang and Liang}{Wang and Liang}{2017}]%
        {wang2017comprehensive}
\bibfield{author}{\bibinfo{person}{Shuo Wang} {and} \bibinfo{person}{Yun
  Liang}.}
\newblock \showarticletitle{A comprehensive framework for synthesizing stencil
  algorithms on FPGAs using OpenCL model}. In \bibinfo{booktitle}{\emph{DAC}}
  \bibinfo{year}{2017}\natexlab{}.
\newblock


\bibitem[\protect\citeauthoryear{Wang, Huang, Zhang, and Alonso}{Wang
  et~al\mbox{.}}{2020}]%
        {wang2020}
\bibfield{author}{\bibinfo{person}{Zeke Wang}, \bibinfo{person}{Hongjing
  Huang}, \bibinfo{person}{Jie Zhang}, {and} \bibinfo{person}{Gustavo Alonso}.}
\newblock \showarticletitle{{Shuhai: {B}enchmarking {H}igh {B}andwidth {M}emory
  on {FPGA}s}}. In \bibinfo{booktitle}{\emph{FCCM}}
  \bibinfo{year}{2020}\natexlab{}.
\newblock


\bibitem[\protect\citeauthoryear{Wenzel, Schmid, Martin, Plauth, Eberhardt, and
  Polze}{Wenzel et~al\mbox{.}}{2018}]%
        {wenzel2018getting}
\bibfield{author}{\bibinfo{person}{Lukas Wenzel}, \bibinfo{person}{Robert
  Schmid}, \bibinfo{person}{Balthasar Martin}, \bibinfo{person}{Max Plauth},
  \bibinfo{person}{Felix Eberhardt}, {and} \bibinfo{person}{Andreas Polze}.}
\newblock \showarticletitle{{Getting {S}tarted with {CAPI} {SNAP}: {H}ardware
  {D}evelopment for {S}oftware {E}ngineers}}. In
  \bibinfo{booktitle}{\emph{Euro-Par}} \bibinfo{year}{2018}\natexlab{}.
\newblock


\bibitem[\protect\citeauthoryear{Williams, Waterman, and Patterson}{Williams
  et~al\mbox{.}}{2009}]%
        {williams2009roofline}
\bibfield{author}{\bibinfo{person}{Samuel Williams}, \bibinfo{person}{Andrew
  Waterman}, {and} \bibinfo{person}{David Patterson}.}
\newblock \showarticletitle{{Roofline: {A}n {I}nsightful {V}isual {P}erformance
  {M}odel for {M}ulticore architectures}}. In \bibinfo{booktitle}{\emph{CACM}}
  \bibinfo{year}{2009}\natexlab{}.
\newblock


\bibitem[\protect\citeauthoryear{Wu, Sharifi, Lenjani, Skadron, and Venkat}{Wu
  et~al\mbox{.}}{2021}]%
        {wu2021sieve}
\bibfield{author}{\bibinfo{person}{Lingxi Wu}, \bibinfo{person}{Rasool
  Sharifi}, \bibinfo{person}{Marzieh Lenjani}, \bibinfo{person}{Kevin Skadron},
  {and} \bibinfo{person}{Ashish Venkat}.}
\newblock \showarticletitle{{Sieve: Scalable In-situ DRAM-based Accelerator
  Designs for Massively Parallel k-mer Matching}}. In
  \bibinfo{booktitle}{\emph{ISCA}} \bibinfo{year}{2021}\natexlab{}.
\newblock


\bibitem[\protect\citeauthoryear{Xu, Fu, Shi, Gan, Li, Luk, and Yang}{Xu
  et~al\mbox{.}}{2018}]%
        {xu2018performance}
\bibfield{author}{\bibinfo{person}{Jingheng Xu}, \bibinfo{person}{Haohuan Fu},
  \bibinfo{person}{Wen Shi}, \bibinfo{person}{Lin Gan}, \bibinfo{person}{Yuxuan
  Li}, \bibinfo{person}{Wayne Luk}, {and} \bibinfo{person}{Guangwen Yang}.}
\newblock \showarticletitle{{Performance {T}uning and {A}nalysis for
  {S}tencil-{B}ased {A}pplications on {POWER}8 {P}rocessor}}. In
  \bibinfo{booktitle}{\emph{ACM TACO}} \bibinfo{year}{2018}\natexlab{}.
\newblock


\bibitem[\protect\citeauthoryear{Zhang, Jayasena, Lyashevsky, Greathouse, Xu,
  and Ignatowski}{Zhang et~al\mbox{.}}{2014}]%
        {zhang2014top}
\bibfield{author}{\bibinfo{person}{Dongping Zhang}, \bibinfo{person}{Nuwan
  Jayasena}, \bibinfo{person}{Alexander Lyashevsky}, \bibinfo{person}{Joseph~L
  Greathouse}, \bibinfo{person}{Lifan Xu}, {and} \bibinfo{person}{Michael
  Ignatowski}.}
\newblock \showarticletitle{{TOP-PIM: Throughput-Oriented Programmable
  Processing in Memory}}. In \bibinfo{booktitle}{\emph{HPDC}}
  \bibinfo{year}{2014}\natexlab{}.
\newblock


\bibitem[\protect\citeauthoryear{Zhang, Marks, Sippel, Rogers, Zhang,
  Gopalakrishnan, Zhang, and Tallapragada}{Zhang et~al\mbox{.}}{2018}]%
        {doi:10.1175/WAF-D-17-0097.1}
\bibfield{author}{\bibinfo{person}{Jun~A. Zhang}, \bibinfo{person}{Frank~D.
  Marks}, \bibinfo{person}{Jason~A. Sippel}, \bibinfo{person}{Robert~F.
  Rogers}, \bibinfo{person}{Xuejin Zhang}, \bibinfo{person}{Sundararaman~G.
  Gopalakrishnan}, \bibinfo{person}{Zhan Zhang}, {and} \bibinfo{person}{Vijay
  Tallapragada}.}
\newblock \showarticletitle{{Evaluating the {I}mpact of {I}mprovement in the
  {H}orizontal {D}iffusion {P}arameterization on {H}urricane {P}rediction in
  the {O}perational {H}urricane {W}eather {R}esearch and {F}orecast ({HWRF})
  {M}odel}}. In \bibinfo{booktitle}{\emph{Weather and Forecasting}}
  \bibinfo{year}{2018}\natexlab{}.
\newblock


\bibitem[\protect\citeauthoryear{Zhu, Zhuo, Wang, Chen, and Xie}{Zhu
  et~al\mbox{.}}{2018}]%
        {hbm_gpu_data_intensive}
\bibfield{author}{\bibinfo{person}{Maohua Zhu}, \bibinfo{person}{Youwei Zhuo},
  \bibinfo{person}{Chao Wang}, \bibinfo{person}{Wenguang Chen}, {and}
  \bibinfo{person}{Yuan Xie}.}
\newblock \showarticletitle{{Performance Evaluation and Optimization of
  {HBM}-Enabled {GPU} for Data-intensive Applications}}. In
  \bibinfo{booktitle}{\emph{VLSI}} \bibinfo{year}{2018}\natexlab{}.
\newblock


\bibitem[\protect\citeauthoryear{Zohouri, Podobas, and Matsuoka}{Zohouri
  et~al\mbox{.}}{2018}]%
        {zohouri2018combined}
\bibfield{author}{\bibinfo{person}{Hamid~Reza Zohouri}, \bibinfo{person}{Artur
  Podobas}, {and} \bibinfo{person}{Satoshi Matsuoka}.}
\newblock \showarticletitle{Combined spatial and temporal blocking for
  high-performance stencil computation on FPGAs using OpenCL}. In
  \bibinfo{booktitle}{\emph{FPGA}} \bibinfo{year}{2018}\natexlab{}.
\newblock


\end{thebibliography}
